%% file: main.tex
\documentclass[aps,prd,notitlepage,twocolumn,preprintnumbers,superscriptaddress,nofootinbib,balancelastpage,longbibliography]{revtex4-2}

\input{preamble/packages}
\input{preamble/formatting}
\input{preamble/commands}

\begin{document}

\preprint{MIT-CTP/5972}

\title{Exploring the dark axion portal in the LUXE-NPOD experiment}

\author{Noam Ness}
\email{noam.ness@campus.technion.ac.il}
\affiliation{Physics Department, Technion -- Israel Institute of Technology, Haifa 3200003, Israel}

\author{Barry~E. Cimring}
\email{bcimring@mit.edu}
\affiliation{Center for Theoretical Physics -- a Leinweber Institute,
Massachusetts Institute of Technology, Cambridge, Massachusetts 02139, USA}

\begin{abstract}
\input{sections/0_Abstract}
\end{abstract}

\maketitle

\section{Introduction}
\label{sec:Introduction}
\input{sections/1_Introduction}

\section{Portals to hidden sectors}
\label{sec:PortalstoHiddenSectors}
\input{sections/2_Portals_to_Hidden_Sectors}

\section{Dark axion portal at LUXE-NPOD}
\label{sec:DarkAxionPortalatLUXENPOD}
\input{sections/3_The_Dark_Axion_Portal_at_LUXE-NPOD}

\section{Projected sensitivity}
\label{sec:ProjectedSensitivity}
\input{sections/4_Projected_Sensitivity}

\section{Conclusions}
\label{sec:Conclusions}
\input{sections/5_Conclusions}

\section*{Acknowledgments}
\label{sec:Acknowledgments}
\input{sections/6_Acknowledgments}

\appendix

\section{Cascade equations in 1D}
\label{apx:CascadeEquations}
\input{sections/Appendix_A-Cascade_Equations_in_1D}

\section{Secondary NP differential cross sections}
\label{apx:SecondaryNPDifferentialCrossSections}
\input{sections/Appendix_B-Secondary_NP_Differential_Cross_Sections}

\input{main.bbl}

\end{document}

%% file: preamble/packages.tex
\usepackage{amsmath} 

\usepackage{xspace} 

\usepackage{graphicx} 

\usepackage{subcaption} 
\usepackage{hyperref} 

\usepackage[capitalize]{cleveref} 

\usepackage{xcolor} 

\usepackage{slashed} 

%% file: preamble/formatting.tex
\hypersetup{
	colorlinks=true,       
	linkcolor=black,        
	citecolor=blue,        
	filecolor=blue,     
	urlcolor=blue         
}

%% file: preamble/commands.tex
\newcommand{\mtxt}[1]{{\rm \textnormal{#1}}} 

\newcommand{\numrange}[2]{#1{\rm\textnormal{--}}#2} 

\newcommand{\eV}{{\rm eV}\xspace}
\newcommand{\keV}{{\rm keV}\xspace}
\newcommand{\MeV}{{\rm MeV}\xspace}
\newcommand{\GeV}{{\rm GeV}\xspace}
\newcommand{\TeV}{{\rm TeV}\xspace}
\newcommand{\PeV}{{\rm PeV}\xspace}
\newcommand{\meter}{{\rm m}\xspace}
\newcommand{\cm}{{\rm cm}\xspace}
\newcommand{\micrometer}{\mu{\rm m}\xspace}
\newcommand{\nm}{{\rm nm}\xspace}
\newcommand{\gram}{{\rm g}\xspace}
\newcommand{\TW}{{\rm TW}\xspace}
\newcommand{\mrad}{{\rm mrad}\xspace}
\newcommand{\fs}{{\rm fs}\xspace}
\newcommand{\ps}{{\rm ps}\xspace}

\newcommand{\lag}{\mathcal{L}} 

\renewcommand{\d}{\partial} 

\renewcommand{\O}{\mathcal{O}} 

\newcommand{\lumi}{\mathcal{L}} 

\newcommand{\prob}{\mathcal{P}} 

\newcommand{\BR}{\mathrm{BR}} 

\newcommand{\abs}[1]{\left|#1\right|} 

\newcommand{\Poiss}{\Pi} 

\newcommand{\Binom}{\mathcal{B}} 

%% file: sections/0_Abstract.tex
The optical dump at the LUXE experiment has the potential to create a large flux of $\O(\GeV)$ photons that can be used to look for new physics when directed at a solid material dump. The LUXE-NPOD extension of LUXE, which focuses the hard photons onto a slab of tungsten, offers two interaction points (laser-electron and photon-tungsten), making it well suited to test theories containing two or more new particles. In this paper, we examine the dark axion portal, a scenario involving both dark photons~(DPs) and axionlike particles~(ALPs) and their mutual interactions, and its implications on the phenomenology at LUXE-NPOD\@. We focus on the case of strictly massive DPs. To simulate the spectra of particle populations generated at the electron-laser interaction point, we solve a set of extended 1D cascade equations. We recover a photon spectrum consistent with previous analyses and present previously unconsidered DP and ALP spectra. Further, we derive the overall sensitivity of LUXE-NPOD to various parameters of the new particles and show that it is capable of probing potentially uncharted regions in the dark axion parameter space. For ALPs in the $\numrange{10}{500}\,\MeV$ mass range, and DPs that are either heavier ($\sim\GeV$) or significantly lighter $\O(\numrange{10^{-18}}{10^{-13}})\,\eV$, we obtain promising constraints on DP kinetic mixing parameters smaller than $10^{-2}$ and on $\O(\numrange{10^{-4}}{10^{-3}})\,\GeV^{-1}$ ALP--photon couplings. We find that restrictions on $\O\left(10^{-3}\right)$ kinetic mixing can be extracted for arbitrarily small DP masses. Our discussion aims to be systematic and demonstrates a practical method of analyzing constraints on multidimensional parameter spaces.

%% file: sections/1_Introduction.tex
The Standard Model~(SM) of particle physics, though remarkably successful, falls short of explaining neutrino oscillations, the observed baryon asymmetry in the Universe, and dark matter. These gaps have motivated extensive theoretical and experimental efforts to explore extensions of the SM.

Many proposed extensions introduce a ``hidden'' or ``dark'' sector, composed of new particles that are neutral under the SM gauge group. Interactions between this sector and the SM can arise via effective operators known as ``portals''. Two well-studied cases are the dark photon (DP), which couples to the SM photon through a vector portal~\cite{Paraphotons,KineticMixing,DPBook}, and the axionlike particle~(ALP), which couples to SM gauge bosons via an axion portal~\cite{PecceiQuinnAxion,PecceiQuinnAxion2,WilczekAxion,WeinbergAxion,KSVZ,KSVZ2,AxionsandALPsReview}. Both scenarios have been constrained by numerous experiments (see e.g.\@ Refs.~\cite{DPLimitsCavendish,DPLimitsPlimptonLawton&AtomicForceMicroscopy,DPLimitSpectroscopy,DPLimitsEarth,DPLimitsJupiter,COBE/FIRAS,DP/ALPLimitALPS,DPLimitSPring-8,DPLimitsUWA,DPLimitsADMX,DPLimitsCROWS,DPLimitsCAST,DPLimitsSHIPS,DPLimitTEXONO,DPLimitCHARM,DPLimitEBeamDumps&FormFactors,DPLimitE137,DPLimitE141,DPLimitE774,DPLimitMuonG-2,DPLimitLSND,DPLimitNA48/2,DPLimitWASA,DPLimitAPEX,DPLimitMAMI,DPLimitBaBar,DPLimitKLOE,DPLimitFASER,ALPLimitCAST,ALPLimitCAST2,ALPLimitSolarNu,ALPLimitGlobularClusters,ALPLimitGlobularClusters2,ALPLimitPVLAS,ALPLimitEuXFL,ALPLimitSAPPHIRES,ALPLimitRevisedOldBeamDump/RevisedColliderInvisibleFinalStates,ALPLimitNOMAD,ALPLimitLEP,ALPLimitBelleII,ALPLimitBESIII,ALPLimitOPAL/LHCpp,ALPLimitCMS,ALPLimitATLAS,ALPLimitOldBeamDump1,ALPLimitOldBeamDump2,ALPLimitNA64,ALPLimitCHARM,ALPLimitMiniBooNE,ALPLimitPrimEx}).

While most studies focus on either the DP or the ALP in isolation, Ref.~\cite{DarkAxionPortal} showed that interactions between DPs, ALPs, and the SM photon can emerge from a unique mechanism that is not merely a combination of vector and axion portals. In some experiments, this so-called dark axion portal enables novel production and decay channels that may dominate the phenomenology in certain regions of parameter space \cite{DAP2,DAP3,DAP4,DAP5,DAP6,DAP7,DAP8,DAP9,DAP10,DAP11,DAP12,DAP13,DAP14,DAP15,DAP16}.

We investigate the implications of the dark axion portal for the proposed LUXE-NPOD experiment~\cite{LUXENPOD,LUXENPOD2}, which uses a stream of hard photons generated at the LUXE experiment~\cite{LUXECDR,LUXETDR,LUXE} to probe physics beyond the SM by placing a material dump and forward detector downstream of the electron-laser interaction point. The original NPOD proposal demonstrated that this configuration is sensitive to new regions of the ALP parameter space. Here we extend the previous analysis to include DPs and the dark axion portal, examining the case in which all three portals (vector, axion, and dark axion) are active. The dual interaction sites---electron-laser and photon-dump---make LUXE-NPOD particularly suitable for testing models with multiple new particles. We present a systematic phenomenological study of this setup using methods that can be applied to other multiportal frameworks.

Our results corroborate previous computations of the hard photon spectrum at LUXE and demonstrate the possibility of producing additional DP and ALP spectra in new physics (NP) scenarios. Moreover, they indicate that LUXE-NPOD is sensitive to regions of dark axion portal parameter space that were not covered by existing single-particle analyses; namely, combinations of kinetic mixings in the $\numrange{10^{-4}}{10^{-2}}$ range, ALP-photons/DPs couplings in the $\O(\numrange{10^{-4}}{10^{-3}})\,\GeV^{-1}$ range, an ALP mass of $\numrange{10}{500}\,\MeV$, and a much heavier ($\sim\GeV$) or lighter $\O(\numrange{10^{-18}}{10^{-13}})\,\eV$ DP.

This paper is organized as follows. In \cref{sec:PortalstoHiddenSectors} we introduce the vector, axion, and dark axion portals as a model for potential NP probed at LUXE-NPOD. \Cref{sec:DarkAxionPortalatLUXENPOD} describes the LUXE-NPOD setup and presents the framework used to analyze and compute the expected NP signals. In \cref{sec:ProjectedSensitivity} we present the projected sensitivity of LUXE-NPOD across various subplanes of the model's parameter space, highlighting regimes where the dark axion portal leads to novel phenomenology. \Cref{sec:Conclusions} concludes with a summary of our results. Technical details of cascade simulations and rate calculations are provided in the appendices.  

%% file: sections/2_Portals_to_Hidden_Sectors.tex
The vector portal, axion portal, and dark axion portal terms can be written as

\begin{align}
\label{eq:VectorPortal}
    \lag_\mtxt{vector portal} &= \frac{\varepsilon}{2} F^{\mu\nu} {F'}_{\mu\nu}
    \,,
    \\\nonumber\\
\label{eq:AxionPortal}
    \lag_\mtxt{axion portal} &= \frac{g_{a\gamma\gamma}}{4}aF^{\mu\nu}\Tilde{F}_{\mu\nu}
    \,,
    \\\nonumber\\
\label{eq:DarkAxionPortal}
    \lag_\mtxt{dark axion portal} &= \frac{g_{a\gamma\gamma'}}{2}aF^{\mu\nu}{\Tilde{F}'}_{\mu\nu}
    \,,
\end{align}
where ${F'}_{\mu\nu}$ is the DP field strength and $\Tilde{F}_{\mu\nu}=\frac{1}{2}\epsilon_{\mu\nu\rho\sigma}F^{\rho\sigma}$. In the above, $\varepsilon$ is the kinetic mixing parameter, $g_{a\gamma\gamma}$ is the ALP-photon-photon coupling, and $g_{a\gamma\gamma'}$ is the ALP-photon-DP coupling. The dark axion portal operator in \cref{eq:DarkAxionPortal} typically arises from a Kim--Shifman--Vainshtein--Zakharov (KSVZ)-type mechanism \cite{DarkAxionPortal}, which introduces an additional ALP-DP-DP term $\frac{g_{a\gamma'\gamma'}}{4}a{F'}^{\mu\nu}{\Tilde{F}'}_{\mu\nu}$ that also affects our phenomenological analysis, as we explain in \cref{sec:DarkAxionPortalatLUXENPOD}.

It is well known that if the DP is massless it completely decouples from the SM and gives rise to so-called millicharged particles (see, e.g., Ref.~\cite{DPBook}). In this work, we consider strictly massive DPs and do not focus on millicharged particles. Accordingly, \cref{eq:VectorPortal,eq:AxionPortal,eq:DarkAxionPortal} are written in the basis where the DP carries a canonical diagonal mass term. Aside from this restriction, we do not assume a dominant coupling or particular mass hierarchy between the ALP and the DP.

%% file: sections/3_The_Dark_Axion_Portal_at_LUXE-NPOD.tex
Complete information regarding the configuration of LUXE can be found in Refs.~\cite{LUXECDR,LUXETDR}, and details of the proposed NPOD extension are given in Ref.~\cite{LUXENPOD}. We briefly recount the relevant points for our purposes. 

LUXE plans to probe QED beyond the Schwinger limit in the strong-field domain where nonperturbative, nonlinear phenomena are relevant~
\cite{LightByLightScattering,LightByLightScattering2,SchwingerPairProduction,NonLinearBreitWheeler,NonlinearCompton,NonlinearCompton2}, by aiming an energetic beam of electrons generated at the European X-ray Free-Electron Laser (XFEL) at a high-intensity laser. The beam-laser collisions result in a collimated flux of $\sim\!\GeV$ photons emerging from the interaction point.

\begin{figure}[b]
    \centering
    \includegraphics[width=0.48\textwidth]{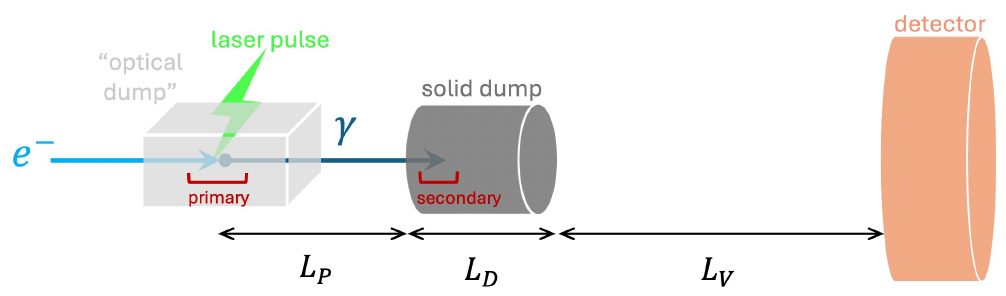}
    \caption{Schematic of the LUXE configuration and the NPOD extension (inspired by Fig.~2 in Ref.~\cite{LUXENPOD}). Electrons collide with a laser pulse (``optical dump'') in the LUXE experiment, generating hard photons which the NPOD extension directs onto a solid dump. A forward detector is stationed after the dump.}
\label{fig:LUXE&LUXENPOD}
\end{figure}

The NPOD proposal suggests directing the stream of photons onto a solid material dump, with a forward detector placed beyond the dump (a depiction of this setup is presented in \cref{fig:LUXE&LUXENPOD}).
In this setup, ALPs can be produced in an elastic Primakoff-like process $\gamma + N \to N + a$ where photons $\gamma$ interact with dump nuclei $N$. After the $\gamma\!\to\! a$ conversion, ALPs that decay into two photons outside the dump and before the detector can be identified by a diphoton signal in the detector. The expected number of signals is computed in Eq.~(2) in Ref.~\cite{LUXENPOD}.

Besides enabling efficient ALP production, the dump is also effective at blocking the SM background from reaching the detector. This fact, combined with appropriate detector technology and signal selection criteria, implies that $<\!1$ diphoton signals from non-ALP sources are expected in one year of data collection. We therefore consider the ALP search to be background free.

A projection of LUXE-NPOD's sensitivity to the ALP parameters is plotted in Fig.~4 of Ref.~\cite{LUXENPOD}. This result demonstrates that LUXE-NPOD has the power to exclude regions of ALP parameter space corresponding to masses of $\O(\numrange{30}{300})\,\MeV$ and couplings of $\O(\numrange{1}{300})\,\PeV^{-1}$.

In what follows, we perform a similar analysis of potential NP signals at LUXE-NPOD for a model including DPs, ALPs, and the dark axion portal. We assume the same experimental configuration and technical specifications that are detailed in Ref.~\cite{LUXENPOD}. This setup has two separate interaction points---one in the optical dump and another in the solid dump---and we are considering two new particles that interact with the SM as well as with each other. Consequently, numerous different NP processes may arise. To make sense of the implications on the observed signals, we attempt to work systematically.

\subsection{Allowed processes}

We begin by defining some useful terminology. We call the electron-laser collision the ``primary'' interaction point and the interaction in the solid dump the ``secondary'' interaction point (see \cref{fig:LUXE&LUXENPOD}). The production of particles at these points will be referred to as ``primary/secondary production.''

\begin{figure}[t]
    \centering
    \begin{subfigure}{0.48\textwidth}
        \centering
        \includegraphics[width=0.45\textwidth]{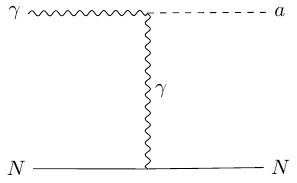}
        \hspace{0.02\textwidth}
        \includegraphics[width=0.45\textwidth]{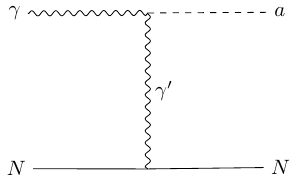}
        \caption{$\mtxt{optical dump} \;\longrightarrow\; \gamma \;\longrightarrow\; \mtxt{solid dump} \;\longrightarrow\; a$}
    \label{subfig:DarkAxionExampleDiagrams}
    \end{subfigure}
   
    \vspace{0.5cm}
   
    \begin{subfigure}{0.48\textwidth}
        \centering
        \hspace{-0.1\textwidth}
        \raisebox{0.045\textwidth}{\includegraphics[width=0.4\textwidth]{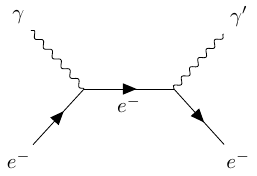}}
        \hspace{0.08\textwidth}
        \includegraphics[width=0.275\textwidth]{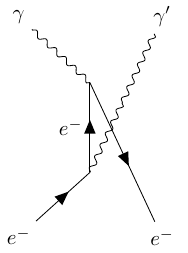}
        \caption{$\mtxt{optical dump} \;\longrightarrow\; \gamma \;\longrightarrow\; \mtxt{solid dump} \;\longrightarrow\; \gamma'$}
    \end{subfigure}

    \vspace{0.5cm}

    \begin{subfigure}{0.48\textwidth}
        \centering
        \includegraphics[width=0.45\textwidth]{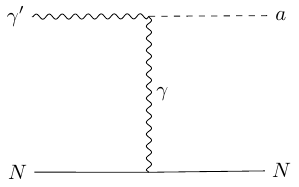}
        \hspace{0.02\textwidth}
        \includegraphics[width=0.45\textwidth]{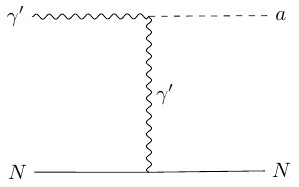}
        \caption{$\mtxt{optical dump} \;\longrightarrow\; \gamma' \;\longrightarrow\; \mtxt{solid dump} \;\longrightarrow\; a$}
    \end{subfigure}

    \vspace{0.5cm}
   
    \begin{subfigure}{0.48\textwidth}
        \centering
        \includegraphics[width=0.45\textwidth]{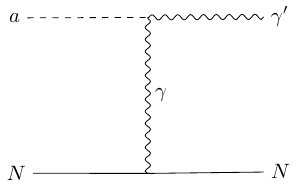}
        \hspace{0.02\textwidth}
        \includegraphics[width=0.45\textwidth]{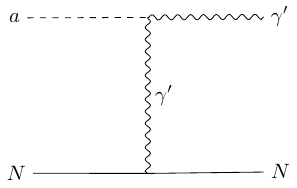}
        \caption{$\mtxt{optical dump} \;\longrightarrow\; a \;\longrightarrow\; \mtxt{solid dump} \;\longrightarrow\; \gamma'$}
    \end{subfigure}
    \caption{Tree-level Feynman diagrams for secondary NP production of DPs/ALPs at LUXE-NPOD\@. The processes are facilitated through interactions with electrons $e^-$ or nuclei $N$ in the solid dump. Each case receives contributions from two mediation channels that interfere with one another. Panels (a)--(d) correspond to the four processes of \crefrange{eq:secNPprod_gammatoa}{eq:secNPprod_atogamma'}, respectively: (a)~secondary ALP production from a  primary photon; (b)~secondary DP production from a primary photon; (c)~secondary ALP production from a primary DP; (d)~secondary DP production from a primary ALP.}
\label{fig:secNPprod}
\end{figure}

Recall that the ALP-only case studied in Ref.~\cite{LUXENPOD} considered the conversion of primary photons to secondary ALPs via interactions with the dump nuclei. The tree-level Feynman diagram for this Primakoff-like process can be seen in \cref{subfig:DarkAxionExampleDiagrams} (left).
The inclusion of DPs and the dark axion portal allows for an analogous $\gamma\to a$ conversion channel mediated by a DP instead of a photon---see \cref{subfig:DarkAxionExampleDiagrams} (right). Additionally, the optical dump now generates not only photons but primary DPs, which can undergo similar $\gamma'\to a$ Primakoff-like conversion involving $\gamma/\gamma'$-mediated tree-level diagrams, or they can pass through the entire dump. There are also various possibilities involving primary ALPs.

\begin{figure}[t]
    \centering
    \begin{subfigure}[b]{0.225\textwidth}
        \centering
        \raisebox{0.2\textwidth}{\includegraphics[width=\textwidth]{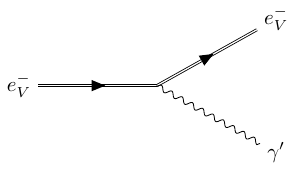}}
        \caption{$\mtxt{optical dump} \;\longrightarrow\; \gamma'$}
    \end{subfigure}
    \hspace{0.015\textwidth}
    \begin{subfigure}[b]{0.225\textwidth}
        \centering
        \includegraphics[width=\textwidth]{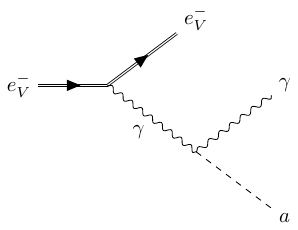}
        \caption{$\mtxt{optical dump} \;\longrightarrow\; a$}
    \label{subfig:primALPproduction}
    \end{subfigure}

    \begin{subfigure}[b]{0.225\textwidth}
        \centering
        \includegraphics[width=\textwidth]{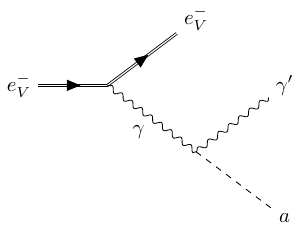}
        \caption{$\mtxt{optical dump} \;\longrightarrow\; \gamma'\! + a$}
    \label{subfig:primALP&DPproduction}
    \end{subfigure}
    \caption{Tree-level Feynman diagrams for primary NP production of DPs/ALPs at LUXE-NPOD\@. The double line represents a dressed Volkov-state electron \cite{VolkovStates}. (a)~DP emission, $e_V^- \to e_V^- \gamma$; (b)~off-shell emission of an ALP together with a photon, $e_V^- \to e_V^- \gamma a$; (c)~off-shell emission of an ALP together with a DP, $e_V^- \to e_V^- \gamma' a$.}
\label{fig:primNPprod}
\end{figure}

We list all the elastic tree-level processes that generate NP particles at LUXE-NPOD to leading order in $\varepsilon$ and the ALP couplings. For completeness, we also include DPs/ALPs that are generated from interactions of primary particles with electrons in the dump (instead of just nuclear interactions). Secondary NP production is generated by
\begin{subequations}
\label{eq:secNPprod}
\begin{alignat}{2}
\label{eq:secNPprod_gammatoa}
    \mtxt{optical dump} &\;\longrightarrow\; \gamma &&\;\longrightarrow\; \mtxt{solid dump} \;\longrightarrow\; a
    \,,
    \\
\label{eq:secNPprod_gammatogamma'}
    \mtxt{optical dump} &\;\longrightarrow\; \gamma &&\;\longrightarrow\; \mtxt{solid dump} \;\longrightarrow\; \gamma'
    \,,
    \\
\label{eq:secNPprod_gamma'toa}
    \mtxt{optical dump} &\;\longrightarrow\; \gamma' &&\;\longrightarrow\; \mtxt{solid dump} \;\longrightarrow\; a
    \,,
    \\
\label{eq:secNPprod_atogamma'}
    \mtxt{optical dump} &\;\longrightarrow\; a &&\;\longrightarrow\; \mtxt{solid dump} \;\longrightarrow\; \gamma'
\,,
\end{alignat}
\end{subequations}
and for primary NP production, we have
\begin{subequations}
\label{eq:primNPprod}
\begin{alignat}{2}
\label{eq:primNPprod_gamma'}
    \mtxt{optical dump} &\;\longrightarrow\; \gamma' &&\;\rlap{\;\;\;\;\;\;\;\;\mtxt{\textcolor{gray}{solid dump}}}{\relbar\joinrel\relbar\joinrel\relbar\joinrel\relbar\joinrel\relbar\joinrel\relbar\joinrel\relbar\joinrel\relbar\joinrel\relbar\joinrel\relbar\joinrel\relbar\joinrel\relbar\joinrel\longrightarrow}\; \gamma'
    \,,
    \\
\label{eq:primNPprod_a}
    \mtxt{optical dump} &\;\longrightarrow\; a &&\;\rlap{\;\;\;\;\;\;\;\;\mtxt{\textcolor{gray}{solid dump}}}{\relbar\joinrel\relbar\joinrel\relbar\joinrel\relbar\joinrel\relbar\joinrel\relbar\joinrel\relbar\joinrel\relbar\joinrel\relbar\joinrel\relbar\joinrel\relbar\joinrel\relbar\joinrel\longrightarrow}\; a
\,.
\end{alignat}
\end{subequations}
Feynman diagrams corresponding to all of the above processes can be found in \cref{fig:secNPprod,fig:primNPprod}.

After NP production, the propagating DPs/ALPs can decay in the volume between the dump and the detector. A DP may decay to an ALP and a photon (if $m_{\gamma'}>m_a$), registering a monophoton signal. An ALP may decay to two photons or (if $m_{\gamma'}<m_a$) to a DP and a photon, registering either a diphoton or a monophoton signal.

\subsection{Expected number of signals}

To find the number of signals we expect at LUXE-NPOD due to NP, we generalize the expression of Eq.~(4) in Ref.~\cite{LUXENPOD} such that for any particular NP process
\begin{equation}
\label{eq:YtoXprocess}
    \mtxt{optical dump} \;\longrightarrow\; Y \;\longrightarrow\; \mtxt{solid dump} \;\longrightarrow\; X,
\end{equation}
as in \crefrange{eq:secNPprod}{eq:primNPprod},\footnote{We accommodate the primary NP production processes in this expression by allowing $Y=X$.} the number of $s$-photon signals ($s=1$, $2$) seen in the detector is given by 
\begin{equation}
\label{eq:NumberOfYtoX}
    \begin{alignedat}[b]{1}
        N_{s\gamma}^{Y\to X} = \int\! dE_Y \; &\frac{d\mathcal{N}_{Y\to X}}{dE_Y} \prob_\mtxt{\ensuremath{X}-decay} \!\left(\substack{\mtxt{after dump} \\ \mtxt{before detector}}\right) 
        \\
        &\times \BR(X\!\to\! s\gamma) \, \Theta\!\left(E_{s\gamma}\!-\!E_\mtxt{thresh}\right) \mathcal{A}
    \,,
    \end{alignedat}
\end{equation}
where $\int\! dE_Y$ integrates over all energies of primary $Y$'s generated in the electron-laser collisions, $d\mathcal{N}_{Y\to X}/dE_Y$ is the differential number of  ${Y\to X}$ events as a function of the $Y$ energy, $\prob_\mtxt{\ensuremath{X}-decay}\!\left(\substack{\mtxt{after dump} \\ \mtxt{before detector}}\right)$ is the probability that $X$ decays in the detectable volume, $\BR(X\to s\gamma)$ is the branching fraction of $X$-decay to an $s$-photon final state, $\Theta\!\left(E_{s\gamma}\!-\!E_\mtxt{thresh}\right)$ is a step function keeping only photon signals with energies above the detection threshold, and $\mathcal{A}$ is the angular acceptance of the detector. We elaborate on each of these elements below.

The differential number of $Y\to X$ events is
\begin{equation}
\label{eq:DiffNumYtoX}
    \frac{d\mathcal{N}_{Y\to X}}{dE_Y} = \lumi_Y \frac{dN_Y}{dE_Y} \sigma_{Y\to X}
\,.
\end{equation}
In the above we have the effective $Y$ luminosity, given by $\lumi_Y = N_e N_\mtxt{BX} \frac{\rho_W}{A_W m_0} \ell_\mtxt{eff}$---where $N_e=1.5\times10^9$ is the XFEL electron bunch population, $N_\mtxt{BX}=10^7$ is the total number of bunch crossings in a one-year run of the experiment, $\rho_W=19.3\,\gram/\cm^3$ and $A_W=183.84$ are the density and mass number of tungsten respectively (taken from Ref.~\cite{PDGreview}), and $\ell_\mtxt{eff}$ is the effective luminosity length for $Y$-dump interactions. For $Y=\gamma$, we take $\ell_\mtxt{eff}=\frac{9}{7}X_0$ (since photons lose most of their energy in the first $9/7$ tungsten radiation lengths $X_0=0.3456\,\cm$ --- see Ref.~\cite{PDGreview}), but for $Y=\gamma'$, $a$, we have $\ell_\mtxt{eff}=L_D=1\,\meter$ (as DPs/ALPs are weakly interacting and not substantially stopped by the dump).

We compute the differential $Y$ flux per incident electron $dN_Y/dE_Y$ using the extended cascade equations of QED~\cite{CascadeEqs} (see Appendix~\ref{apx:CascadeEquations} for details). Schematically, the cascade equations are a set of coupled first-order partial differential equations describing the evolution of the spectrum $I_Y(E_Y,t)$ of population $Y$ at time $t$, i.e.,
\begin{equation}
\label{eq:CascadeSchematic}
\begin{aligned}
    \frac{\d I_Y}{\d t} &= \sum_{X,E_X} r_{X\to Y}(E_Y, E_X,t)I_X(E_X,t)
    \\
    &\phantom{=} - \sum_{Z,E_Z} r_{Y\to Z}(E_Z, E_Y,t)I_Z(E_Z,t)
\,,
\end{aligned}
\end{equation}
where $r_{X\to Y}(E_Y,E_X,t)$ [similarly $r_{Y\to Z}(E_Z,E_Y,t)$] describes the rate to produce particle $Y$ ($Z$) at energy $E_Y$ ($E_Z$) from processes undergone by particle $X$ ($Y$) at energy $E_X$ ($E_Y$). \Cref{eq:CascadeSchematic} approximates the cascade to occur in a single spatial dimension. This approximation is justified by the near collinearity of the $Y$ particles produced by the electrons and the lack of straggling due to feedback of the nonelectron species. 
We solve Eqs.~(\ref{eq:CascadeSchematic}) numerically to find the spectra of particles $dN_Y/dE_Y = I_Y(E_Y,\infty)$ produced by a beam of $16.5\,\GeV$ electrons propagating through the optical dump. For the beams and lasers of interest at LUXE, the spatial extent of the beam and the laser must be taken into account. The electron beam at LUXE is a 3D Gaussian pulse with deviation $5\,\micrometer$ in directions transverse to propagation and deviation of $24\,\micrometer$ in the direction of propagation. The square of the potential describing the laser profile is spatially distributed according to 
\begin{equation}
\label{eq:LaserIntensity}
    \abs{A}^2_\mtxt{rms} = \frac{\abs{A}^2f^2(k\cdot x)}{1 + (z/z_R)} \exp\left(-\frac{2r_\perp^2}{w_0^2(1 + (z/z_R)^2)}\right)
\,,
\end{equation}
where $\abs{A}$ is the magnitude of the laser intensity, $w_0$ is the beam waist (taken to be $10\,\micrometer$), $z_R=\pi w_0^2/\lambda$ is the Rayleigh range, $r_\perp$ is the distance perpendicular to the axis of propagation, and $z$ is the coordinate along the axis of propagation. The pulse shape $f(k\cdot x)$ is Gaussian with width $25\,\fs$ in LUXE's $45\,\TW$ phase-0 run and $120\,\fs$ in LUXE's $350\,\TW$ phase-1 run. The laser is oriented at $17.2^\circ$ relative to the electron beam. The optical dump is therefore wide enough for the locally monochromatic approximation to hold~\cite{LMAapproximation}. To capture the spatial extent of the beam-laser collision, we discretize the incident beam distribution, simulate the propagation of each point through the spatially varying pulse, and sum the resulting spectra.

\begin{figure}
    \centering
    \includegraphics[width=0.48\textwidth]{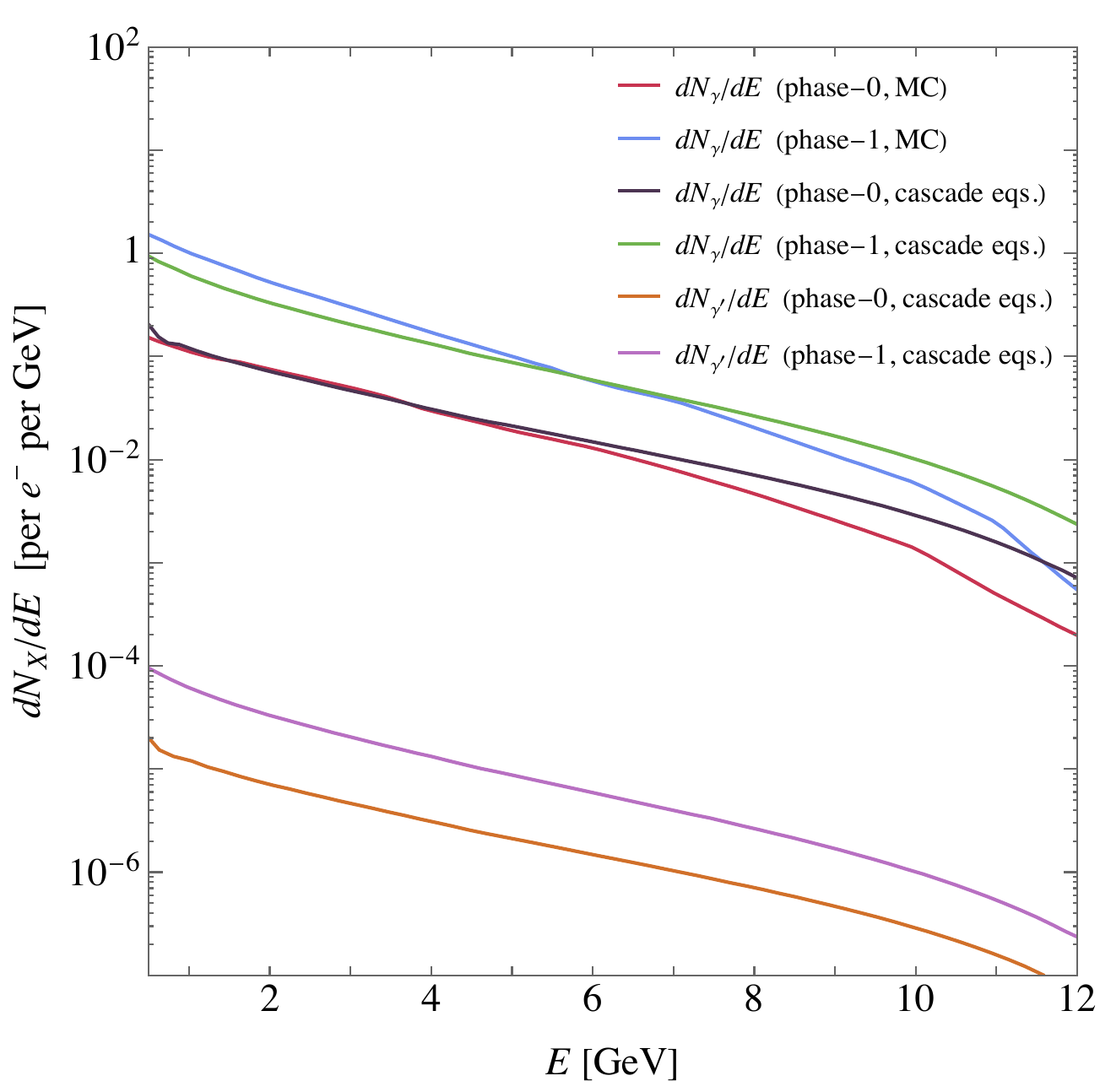}
    \caption{Particle spectra generated in the electron-laser interaction.
    Numerically computed spectra of visible and dark photons produced during phase-0 (black and orange, respectively) and phase-1 (green and pink, respectively) of the LUXE experiment after collision with a high-intensity laser, assuming kinetic mixing $\varepsilon=0.01$ and a DP mass much smaller than that of the electron $m_{\gamma'}\ll m_e$. The photon spectra computed using Monte-Carlo methods are also plotted. }
\label{fig:shower_spectra}
\end{figure}

\Cref{fig:shower_spectra} shows the results of our computation for phase~0 and phase~1, along with the photon spectrum computed in Ref.~\cite{LUXENPOD} using strong-field QED Monte-Carlo simulations.
We see that for small kinetic mixing $\varepsilon$, the straggling of electrons due to DP production is small compared to the straggling due to photons, so the primary DP flux is approximately equal to the photon spectrum suppressed by the kinetic mixing squared $dN_{\gamma'}/dE \approx \varepsilon^2 dN_{\gamma}/dE_\gamma$. The primary production of ALPs is dominated by the $e_V^-\to e_V^-\gamma^{(\prime)}a$ processes depicted in \mbox{\cref{subfig:primALPproduction,subfig:primALP&DPproduction}} (see Appendix~\ref{apx:CascadeEquations} for details); we therefore take a similar crude estimate for the ALP spectrum using the production rate ratio $\Gamma_{e_V^-\to e_V^-a}/\Gamma_\gamma$ calculated in Ref.~\cite{LUXENPOD} (Fig.~10), i.e., $dN_a/dE_a\approx 4\times10^{-16}\left[(g_{a\gamma\gamma}/\TeV^{-1})^2+(g_{a\gamma\gamma'}/\TeV^{-1})^2\right]$. Note that both of these spectra are only present for sufficiently light DPs/ALPs $m_{\gamma',a}\lesssim \O(100)\,\keV$, since the primary production of NP particles is greatly suppressed above this mass---see \cref{fig:rates} (bottom).

Finally, $\sigma_{Y\to X}$ is the inclusive cross section, computed via numerical integration over phase space. The differential cross sections $d\sigma_{Y\to X}/dt$ for each process in \crefrange{eq:secNPprod}{eq:primNPprod}, where $t$ is the square of 4-momentum transfer, can be found in Appendix~\ref{apx:SecondaryNPDifferentialCrossSections}.

Next, the probability that $X$ decays in the detectable volume depends on the point $z$ where it is produced. Assuming the distance from the optical dump to the solid dump is $L_P=13\,\meter$ and the distance from the end of the dump to the detector is $L_V=2.5\,\meter$ (recall the tungsten dump has a length $L_D=1\,\meter$), we may write
\begin{equation}
\label{eq:XDecayProbability}
    \begin{alignedat}[b]{1}
        \prob_\mtxt{\ensuremath{X}-decay} = \! \int \limits_{0}^{L_P+L_D+L_V}\!\! &dz \; p_\mtxt{\ensuremath{X}-prod}(z) 
        \\
        \times& \left(e^{-\frac{L_P+L_D-z}{L_X}} \!-\! e^{-\frac{L_P+L_D+L_V-z}{L_X}} \right)
    \,,
    \end{alignedat}
\end{equation}
where the $X$ decay length $L_X$ is defined by $L_X=\tau_X\sqrt{E_X^2-m_X^2}/m_X$ ($\tau_X$ being the $X$ lifetime, $E_X$ being its energy, and $m_X$ being its mass) and $p_\mtxt{\ensuremath{X}-prod}(z)$ is the probability density that $X$ is produced at a point $z\in\left[0,L_P+L_D+L_V\right]$.

If $X$ is generated in primary production, i.e., at the primary interaction point $z=0$, we take\footnote{We use the convention $\int_0^\infty\!dz\, \delta(z)=1/2$.} $p_\mtxt{\textcolor{gray}{prim}\ensuremath{X}-prod}(z)=2\delta(z)$. If $X$ is generated in secondary production triggered by a photon, we take $p_\mtxt{\textcolor{gray}{\ensuremath{\gamma\!\to}}\ensuremath{X}-prod}(z)=\delta\left(z-L_P\right)$. This is again due to the fact that photons lose most of their energy within $\frac{9}{7}X_0\approx0.44\,\cm$, which is orders of magnitude shorter than the full length of the tungsten dump $L_D=1\,\meter$. Therefore, any secondary NP particles originating from a primary photon are effectively created at a single point at the start of the dump. Lastly, if $X$ is generated in secondary production triggered by a DP or an ALP---which are weakly interacting---we take a simple uniform distribution over the length of the dump $p_\mtxt{\textcolor{gray}{NP\ensuremath{\to}}\ensuremath{X}-prod}(z)=\Theta\left(z-L_P\right)\Theta\left(L_P+L_D-z\right)/L_D$.

We note that in computing the particle lifetime $\tau_X$ we only take into consideration couplings and interactions that were explicitly presented in \cref{sec:PortalstoHiddenSectors} (namely, SM content and DP/ALP interactions through the vector, axion, and dark axion portals). Contributions of DP/ALP interactions purely within the hidden sector are ignored.

The branching fractions depend on the mass hierarchy in our dark axion model. ALPs can always decay to two photons $a\to\gamma\gamma$ and, depending on their mass, may also decay to a photon and DP $a\to\gamma\gamma'$ or to two DPs $a\to\gamma'\gamma'$. The widths of these decays are given, for example, in Ref.~\cite{DarkAxionPortal}, and we can use them to compute $\BR(a\to s\gamma)$ for $s=1$, $2$. DPs, depending on their mass, may decay to charged leptons, hadrons, or a photon and an ALP. We use the decay widths for $\gamma'\to\mtxt{SM}$ and $\gamma'\to a\gamma$ in Refs.~\cite{DPBook} and \cite{DarkAxionPortal}, respectively, to compute $\BR(\gamma'\to1\gamma)$. Here, we neglect any decay channels of DPs/ALPs into pure hidden sector states.

For the detection factors in \cref{eq:NumberOfYtoX}, we assume a photon energy detection threshold of $E_\mtxt{thresh}=0.5\,\GeV$ and take an angular acceptance of $\mathcal{A}=1$ for simplicity. Total angular acceptance may seem unjustified; very light $X$s should be highly boosted and produce very collinear decay remnants that are hard to resolve spatially, and very heavy $X$s might not be boosted enough so their decay remnants could be ejected at an angle too wide to land inside the disk-shaped detector with a radius of $R=1\,\meter$. However, as we will see, NP events at LUXE-NPOD are dominated by $\O(1)\,\GeV$ ALPs ($X=a$) with masses of $\O(\numrange{10}{350})\,\MeV$ coupled to DPs that are either heavier than the ALP or in the ultralight regime $m_{\gamma'}\ll m_a$. This implies angular differences of $\O(\numrange{10}{300})\,\mrad$ for two final photons (or a photon and an ultralight DP). Such angular resolution is achievable with existing detector concepts \cite{AngularResolution}, and the widest angles are roughly compatible with the proposed geometry of the LUXE-NPOD configuration [which has a detection coverage of $\arctan(R/L_V)\approx400\,\mrad$]. We therefore claim that the angular aspect of detection has an $\O(1)$ effect in most regions of the parameter space, with potentially stronger effects near the high-mass boundary. In this regime, a dedicated angular acceptance simulation may modify our projected reach.

Putting the pieces together, the total number of $s$-photon signals ($s=1$, $2$) we expect at LUXE-NPOD is computed via
\begin{equation}
\label{eq:TotalSignals}
\begin{alignedat}[b]{2}
    &N_{s\gamma} = \sum_{Y,X} N_{s\gamma}^{Y\to X} &&
    \\
    &= \sum_{Y,X} \int\!\!dE_Y \! \int\!\!dt \! \int&&\!dz \;  \lumi_Y \frac{dN_Y}{dE_Y}  \frac{d\sigma_{Y\to X}}{dt} p_\mtxt{\ensuremath{X}-prod}(z)
    \\
    & &&\times \left(e^{-\frac{L_P+L_D-z}{L_X}} \!-\! e^{-\frac{L_P+L_D+L_V-z}{L_X}}\right)
    \\ 
    & &&\times \BR(X\!\to\! s\gamma) \, \Theta\!\left(E_{s\gamma}\!-\!E_\mtxt{thresh}\right) \mathcal{A}
\,.
\end{alignedat}
\end{equation}
It is worth stressing that the momentum transfer $t$ depends on the primary energy $E_Y$, which is why this expression is not merely a simple product as suggested in \cref{eq:NumberOfYtoX}---rather, it is a kind of convolution of all the factors we described.

\subsection{Background estimation}
\label{subsec:Background_Estimation}

Background signals from SM processes are identical to those presented in Ref.~\cite{LUXENPOD}. There are two main sources of background\footnote{This does not include background of charged particles, as these can be deflected away from the detector with a magnetic field.}---photons (from EM interactions near the end of the dump) and neutrons (produced in EM showers in the dump) that may be misidentified by the detector as photons.

The total number of $s\gamma$ background signals can be estimated by
\begin{equation}
\label{eq:Nbkg}
\begin{alignedat}[b]{1}
    N_{s\gamma}^\mtxt{bkg} &= \sum_b N_{b\to s\gamma}^\mtxt{bkg} 
    \\
    &= \sum_b N_\mtxt{BX} \prob_{b\to s\gamma}^\mtxt{bkg} R_\mtxt{sel}^{b\to s\gamma}
\,,
\end{alignedat}
\end{equation}
where the sum is over all background sources that can trigger an $s\gamma$ signal ($b=\gamma$, $n$, $2\gamma$, $\gamma+n$, $2n$), $N_{b\to s\gamma}^\mtxt{bkg}$ is the number of $s\gamma$ background signals resulting from the specific source $b$, $\prob_{s\gamma}^\mtxt{bkg}$ is the probability that process $b$ is detected as an $s\gamma$ signal, and $R_\mtxt{sel}^{b\to s\gamma}$ is a rejection factor achieved by implementing a set of selection criteria on the background processes $b$ that trigger an $s\gamma$ signal. Note that the probability depends on $\mu_{\gamma(n)}$, the rate of background photons (neutrons) that arrive at the detector with energy above the detection threshold, and also on the neutron misidentification rate $f_{n\to\gamma}$ (which is highly sensitive to the detector technology used). 

There are two possibilities to obtain a monophoton background signal:
\begin{itemize}
    \item detector triggered by a single background photon and 0 neutrons, with probability $\prob_{\gamma\to1\gamma}^\mtxt{bkg} = \Poiss(\mu_\gamma,1)\times \sum_{k_n=0}^\infty \Poiss(\mu_n,k_n)\Binom(0,k_n,f_{n\to\gamma}) =
    \left(\mu_\gamma e^{-\mu_\gamma}\right) \left(e^{-\mu_nf_{n\to\gamma}}\right)$, or
    \item detector triggered by a single neutron and 0 background photons, with probability $\prob_{n\to1\gamma}^\mtxt{bkg} = \Poiss(\mu_\gamma,0)\times\sum_{k_n=1}^\infty \Poiss(\mu_n,k_n)\Binom(1,k_n,f_{n\to\gamma}) = \left(e^{-\mu_\gamma}\right) \left(\mu_nf_{n\to\gamma}e^{-\mu_nf_{n\to\gamma}}\right)$,
\end{itemize}
where $\Poiss$ and $\Binom$ are the probability mass functions of the Poisson and Binomial distributions respectively.
The background rates $\mu_\gamma$, $\mu_n$ were estimated in Ref.~\cite{LUXENPOD} with a \textsc{Geant\,4} simulation. These rates are smaller and softer during phase-0 of LUXE, but phase-1 levels are conservatively assumed also for phase-0---that is, we use the values $\mu_\gamma=0.013$ and $\mu_n=10$ for both phases.
The resulting estimates for monophoton background events are
\begin{align}
\label{eq:monophoton_bkg}
    N_{\gamma\to1\gamma}^\mtxt{bkg} &= 1.3 \times 10^5 e^{-10 f_{n\to\gamma}} R_\mtxt{sel}^{\gamma\to1\gamma}
    \\
    N_{n\to1\gamma}^\mtxt{bkg} &= 9.9\times 10^7 f_{n\to\gamma} e^{-10 f_{n\to\gamma}} R_\mtxt{sel}^{n\to1\gamma}
\,.
\end{align}
Estimates for the number of diphoton background events are the same as presented in Eqs.~(3--5) of Ref.~\cite{LUXENPOD}.

As noted above, the achievable neutron misidentification rate, $f_{n\to\gamma}$, is determined primarily by the detector technology. Adopting detector specifications of Ref.~\cite{LUXENPOD}, we expect $f_{n\to\gamma}\leq\numrange{10^{-3}}{10^{-4}}$. In addition, time resolutions of $\O(\numrange{10}{100})\,\ps$, position resolutions of $\O(100)\,\micrometer$, angular resolutions of $\O(100)\,\mrad$, and energy resolutions at the level of a few percent---all achievable with existing technologies \cite{Detector1,Detector2,Detector3,Detector4,AngularResolution}---enable precise reconstruction of diphoton kinematics; this, in turn, allows for strong rejection of diphoton background events $R_\mtxt{sel}^{\to2\gamma}\leq\numrange{10^{-3}}{10^{-4}}$. Overall, we are able to reduce the number of diphoton background events to $<1$ (see Eqs.~(3-5) in Ref.~\cite{LUXENPOD}). 

For monophoton signals the situation is more challenging. Although time-of-flight information can be used to suppress single-neutron backgrounds, it does not provide discrimination against irreducible single-photon backgrounds. Moreover, in contrast to the diphoton case, there are no additional kinematic handles to exploit. As a result, the selection efficiency $R_\mtxt{sel}^{\gamma\to1\gamma}$ does not yield a meaningful reduction in event count, and the monophoton background is therefore set by \cref{eq:monophoton_bkg}, $N_{1\gamma}^\mtxt{bkg}\sim\O(10^5)$.

\subsection{Existing constraints}
\label{subsec:Existing_Bounds}

To properly assess the promise of LUXE-NPOD in probing the dark-axion parameter space, our calculations should be compared to existing bounds on DP and ALP parameters. The rigorous and correct way to do so requires implementation of a proper recasting scheme, such as that developed in Ref.~\cite{DPLimitsHeavy_DarkCast}. However, as demonstrated in this manuscript, quantifying the effects of a nonminimal SM extension (that involves both DPs and ALPs) on a given experiment requires a thorough NP treatment and extensive computations, meaning that recasting is much less straightforward. A proper treatment would be akin to replicating the entire scope of the present work dozens of times. While analyzing the dark axion portal in other experiments is an intriguing possibility for future work, for now we simply use the existing bounds ``as is'', without recasting. Although this does not provide an accurate indication of LUXE-NPOD's absolute improvement over existing experiments, it does provide some baseline against which to compare our results.

We consult Ref.~\cite{DPLimitsLight_Handbook} for existing constraints on DP masses below $\sim\!100\,\keV$ (with updated versions of these in Ref.~\cite{CairanOhareGithub}). For limits on DP masses above $\sim\!2\,\MeV$, we refer to Ref.~\cite{DPLimitsHeavy_DarkCast}. Explicitly, we plot the LUXE-NPOD sensitivity against limits from tests of the Coulomb interaction, such as Cavendish-like experiments \cite{DPLimitsCavendish}, the Plimpton--Lawton experiment \cite{DPLimitsPlimptonLawton&AtomicForceMicroscopy}, atomic spectroscopy \cite{DPLimitSpectroscopy}, atomic force microscopy \cite{DPLimitsPlimptonLawton&AtomicForceMicroscopy}, and the magnetic fields of Earth \cite{DPLimitsEarth} and Jupiter \cite{DPLimitsJupiter}; constraints on distortions to the cosmic microwave background obtained by COBE and FIRAS \cite{COBE/FIRAS}; light-shining-through-walls experiments at ALPS \cite{DP/ALPLimitALPS}, SPring-8 \cite{DPLimitSPring-8}, UWA \cite{DPLimitsUWA}, ADMX \cite{DPLimitsADMX}, and CROWS \cite{DPLimitsCROWS}; the helioscopes CAST \cite{DPLimitsCAST} and SHIPS \cite{DPLimitsSHIPS}; the neutrino reactor experiment TEXONO \cite{DPLimitTEXONO}; beam dump experiments such as CHARM \cite{DPLimitCHARM}, E137, E141, and E774 \cite{DPLimitEBeamDumps&FormFactors,DPLimitE137,DPLimitE141,DPLimitE774}; anomalous magnetic moment discrepancies \cite{DPLimitMuonG-2}; the fixed target experiments LSND \cite{DPLimitLSND}, NA48/2 \cite{DPLimitNA48/2}, WASA \cite{DPLimitWASA}, APEX \cite{DPLimitAPEX}, and MAMI \cite{DPLimitMAMI}; and the collider experiments BABAR \cite{DPLimitBaBar}, KLOE \cite{DPLimitKLOE}, and FASER \cite{DPLimitFASER}. Note that we ignore limits that rely on DP dark matter properties such as a particular relic abundance, or those obtained from astrophysical sources, e.g., by DP saturation of energy losses.

For ALPs, we similarly draw from Ref.~\cite{CairanOhareGithub}, again omitting bounds that attribute dark matter properties to ALPs and also some that are obtained from stellar cooling constraints. Explicitly, we plot the LUXE-NPOD sensitivity against limits from the CAST helioscope \cite{ALPLimitCAST,ALPLimitCAST2}; fits of solar data \cite{ALPLimitSolarNu}; analyses of globular clusters \cite{ALPLimitGlobularClusters,ALPLimitGlobularClusters2}; magnetic birefringence measurements at PVLAS \cite{ALPLimitPVLAS}; shining-through-wall experiments by ALPS \cite{DP/ALPLimitALPS} and EuXFEL \cite{ALPLimitEuXFL}; the SAPPHIRES collaboration photon-photon scattering probe \cite{ALPLimitSAPPHIRES}; collider experiments including those recast in Ref.~\cite{ALPLimitRevisedOldBeamDump/RevisedColliderInvisibleFinalStates}, NOMAD \cite{ALPLimitNOMAD}, LEP \cite{ALPLimitLEP}, Belle II \cite{ALPLimitBelleII}, BES III \cite{ALPLimitBESIII}, OPAL \cite{ALPLimitOPAL/LHCpp}, CMS \cite{ALPLimitCMS}, ATLAS \cite{ALPLimitATLAS}, and LHC ($pp$) \cite{ALPLimitOPAL/LHCpp}; beam dump experiments such as those in Refs.~\cite{ALPLimitOldBeamDump1,ALPLimitOldBeamDump2,ALPLimitRevisedOldBeamDump/RevisedColliderInvisibleFinalStates}, NA64 \cite{ALPLimitNA64}, CHARM \cite{ALPLimitCHARM}, and MiniBooNE \cite{ALPLimitMiniBooNE}; and the fixed-target experiment at PrimEx \cite{ALPLimitPrimEx}. Several additional restrictions on the ALP-photon-photon coupling have been derived from astrophysical sources for $\sim\mtxt{MeV}$ ALPs \cite{ALPLimitAstro1,ALPLimitAstro2,ALPLimitAstro3,ALPLimitAstro4,ALPLimitAstro5,ALPLimitAstro6,ALPLimitAstro7,ALPLimitAstro8,ALPLimitAstro9,ALPLimitAstro10}, but their robustness depends on details of the dark sector interactions, and they cease to apply in certain chameleon-like models (see, e.g., the discussion in Ref.~\cite{ChameleonModels}).

Lastly, we note that some studies have also derived bounds on the dark axion portal \cite{DAP2,DAP3,DAP4,DAP5,DAP6,DAP7,DAP8,DAP9,DAP10,DAP11,DAP12,DAP13,DAP14,DAP15,DAP16}. However, most of these extract limits on $g_{a\gamma\gamma'}$ by investigating simplifying benchmark scenarios (for example, Refs.~\cite{DAP2,DAP6,DAP11} derive constraints on the $\left(m_{\gamma'},g_{a\gamma\gamma'}\right)$ plane under the assumption that the ALP is effectively massless and that the dark axion portal is dominant over the other portals), so we do not strictly adhere to them.

%% file: sections/4_Projected_Sensitivity.tex
From \cref{subsec:Background_Estimation}, we see that the diphoton signals are effectively background free. For the monophoton channel, the large background count $N_{1\gamma}^\mtxt{bkg}\sim\O(10^5)$ places us in the Gaussian regime where the asymptotic limit of a profile likelihood ratio test yields a $2\sigma$ exclusion criterion of $N_{1\gamma}=2\sqrt{N_{1\gamma}^\mtxt{bkg}+(\delta N_{1\gamma}^\mtxt{bkg})^2}$, where $\delta$ is the fractional systematic uncertainty on the monophoton background. The statistics-only ($\delta=0$) 95\% C.L.\@ thresholds are
\begin{equation}
\label{eq:95CLregions}
    N_{1\gamma}|_{\delta=0}=\O(700) \,, \quad N_{2\gamma} = 3
\,,
\end{equation}
for monophoton signals and diphoton signals, respectively. Including a benchmark $\delta=1\%\,(5\%)$ does not affect the diphoton condition, but raises the monophoton criterion to $N_{1\gamma}\approx 2\times10^3\,(10^4)$. The exclusion regions we present below are statistics-only projections; a full nuisance parameter treatment with dedicated detector-level background simulation is beyond the scope of this work, and estimating the deterioration of LUXE-NPOD's dark axion reach with varying $\delta$ is nontrivial. We nonetheless verified numerically that a $\delta=5\%$ fractional systematic uncertainty degrades our projected reach by at most a factor of $\O(2)$.

We use the above thresholds to impose limits on the 6 dark axion parameters---two masses $m_a$, $m_{\gamma'}$, a kinetic mixing strength $\varepsilon$, and three ALP couplings $g_{a\gamma\gamma}$, $g_{a\gamma\gamma'}$, $g_{a\gamma'\gamma'}$. In what follows, we methodically evaluate the sensitivity of LUXE-NPOD to the dark axion portal in all two-dimensional parameter subplanes. Since the ALP-DP-DP coupling is a pure hidden sector parameter, we will not plot $g_{a\gamma'\gamma'}$. This leaves $10$ unique pairs of parameters. For each choice, we fix the remaining four parameters and numerically solve Eqs.~(\ref{eq:95CLregions}) for the chosen pair. 
We repeat this process while varying the fixed parameters to test a wide array of regimes; for instance, we always test (at least) the three mass hierarchies of heavier DPs $m_{\gamma'}>m_a$, heavier ALPs $m_{\gamma'}<m_a$, and comparable masses $m_{\gamma'}\approx m_a$.

Unless otherwise specified, the plots in this section display merged results from both diphoton and monophoton signals. We implement a consistent color scheme throughout, where the reach of the experiment is indicated by a solid blue line for LUXE's phase~0 and by a solid black line for LUXE's phase~1 (the 95\% C.L.\@ regions are emphasized with a transparent coloring). Gray regions indicate parameter space excluded by single-particle analyses (see the discussion in \cref{subsec:Existing_Bounds}).

Our results, after exhaustively testing a range of parameter values, indicate that LUXE-NPOD can probe previously unexplored regions of the parameter space when $m_a=\O(\numrange{10}{500})\,\MeV$, in two distinct regimes---heavier DPs ($m_{\gamma'}>m_a$) and ultralight DPs ($m_{\gamma'}\ll m_a$).

\subsection{Heavier dark photons \texorpdfstring{$m_{\gamma'}>m_a$}{Mg'>Ma}}

\begin{figure*}
    \centering
    \includegraphics[width=0.3825\textwidth]{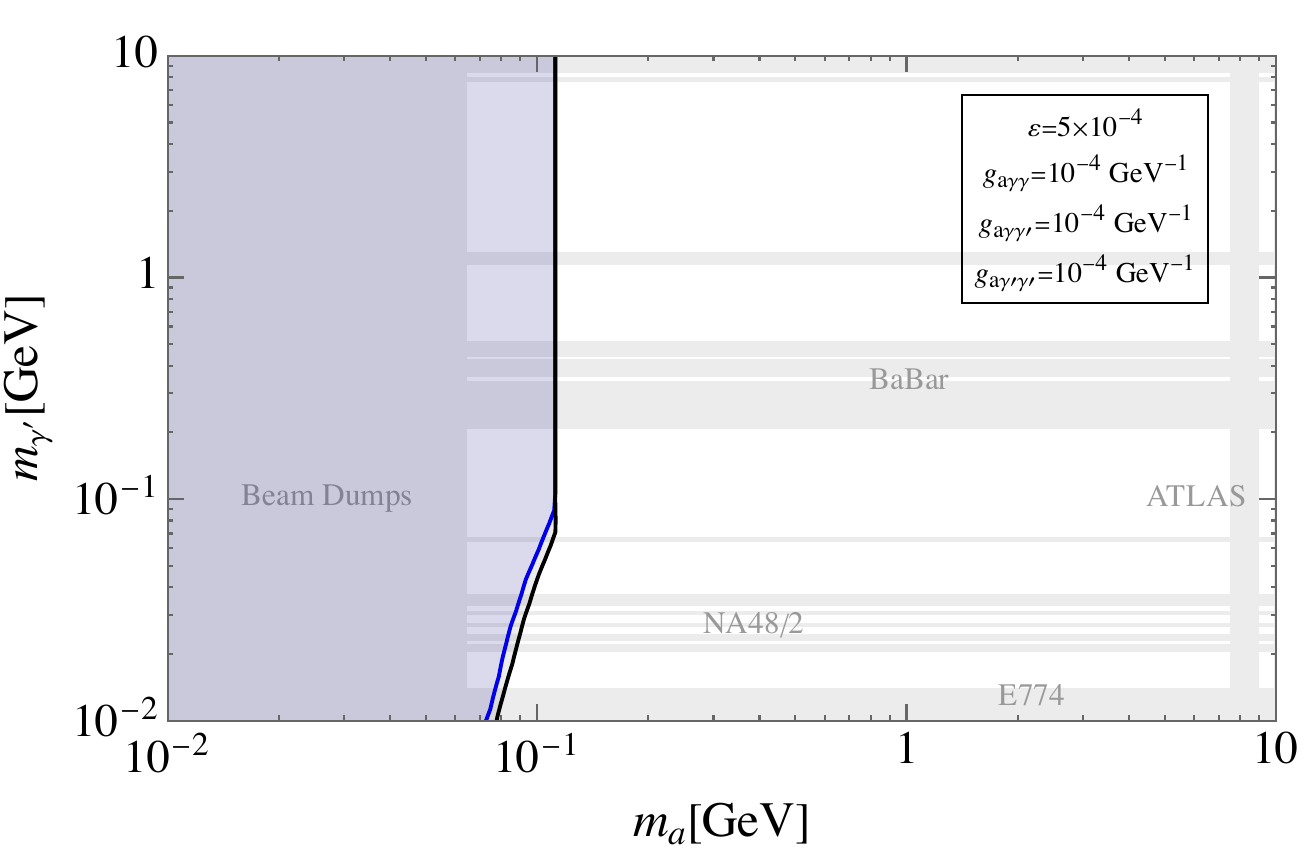}
    \hspace{0.075\textwidth}
    \includegraphics[width=0.3825\textwidth]{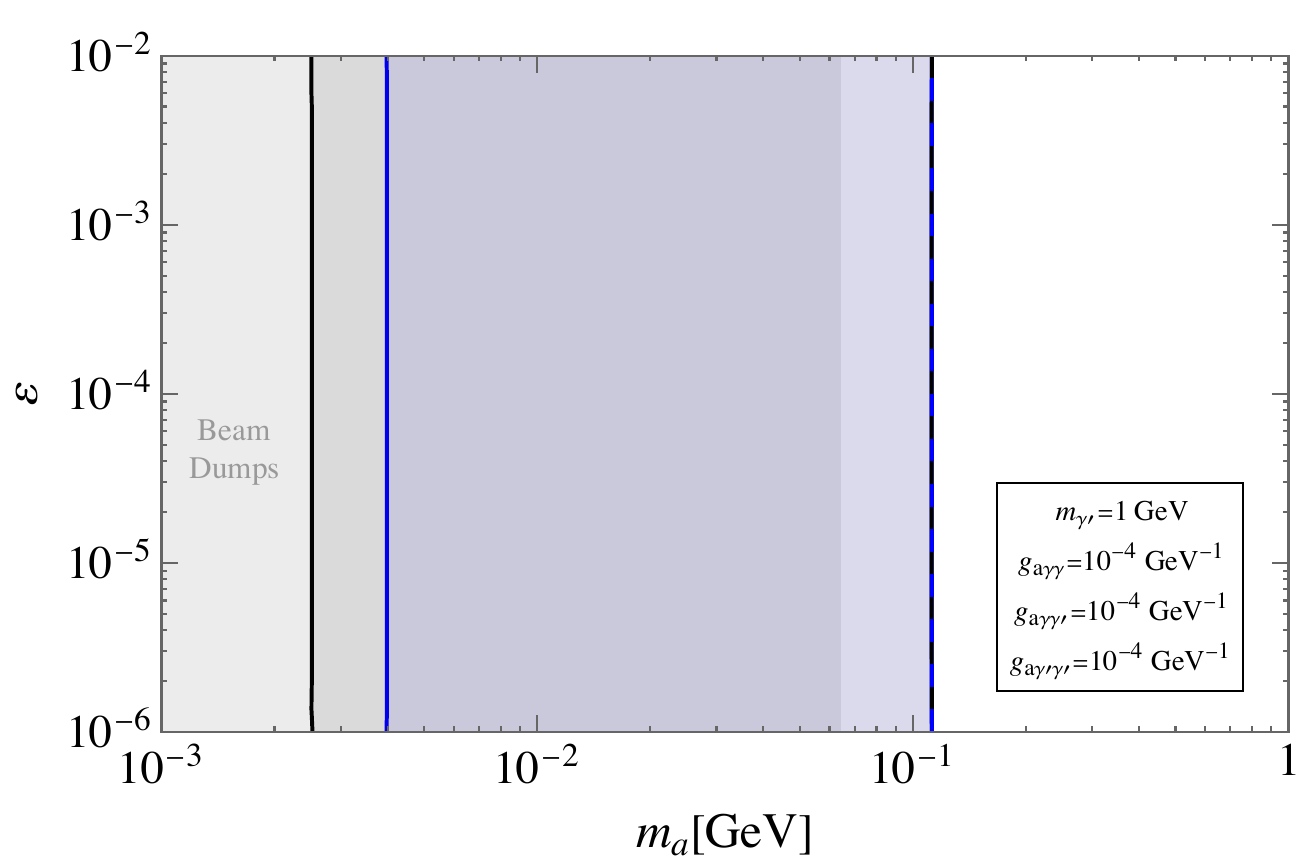}

    \includegraphics[width=0.3825\textwidth]{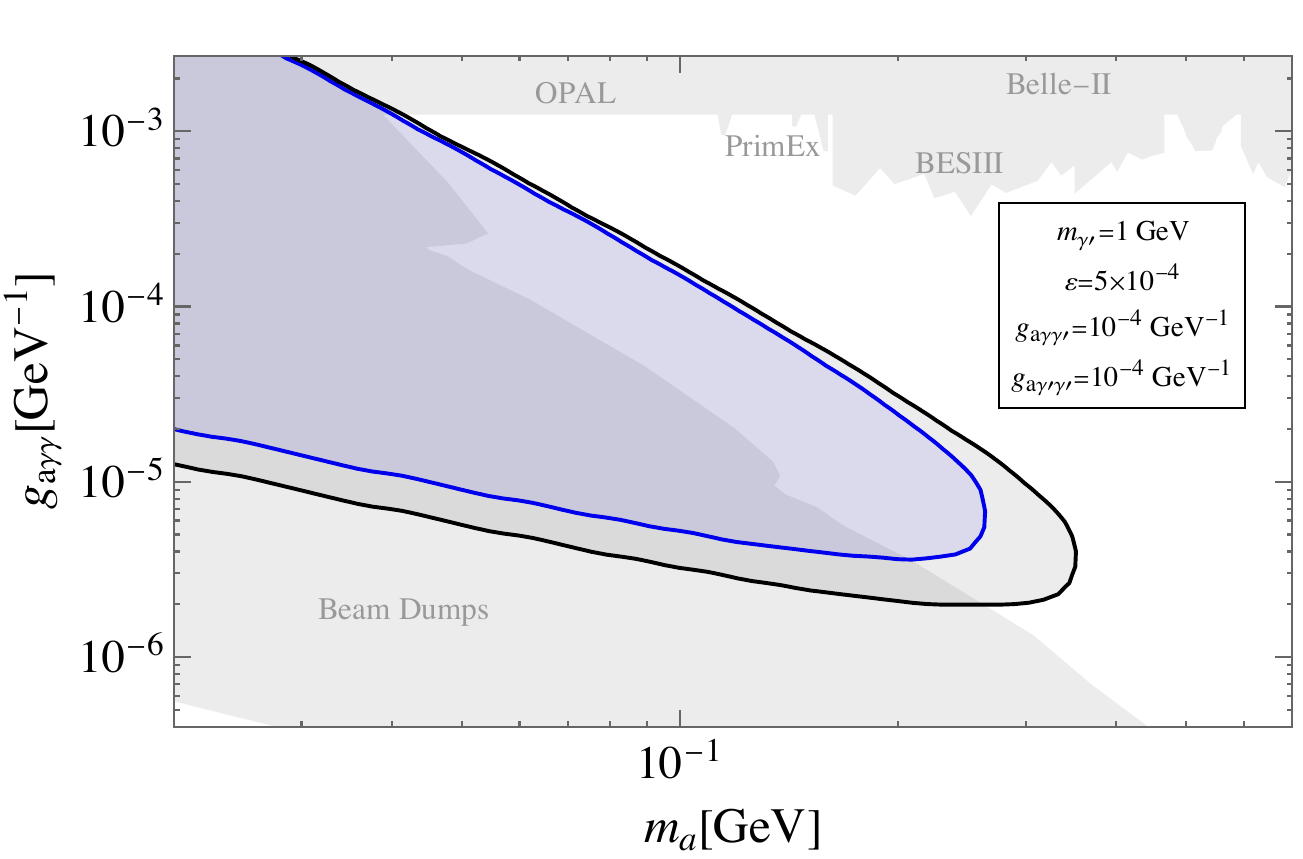}
    \hspace{0.075\textwidth}
    \includegraphics[width=0.3825\textwidth]{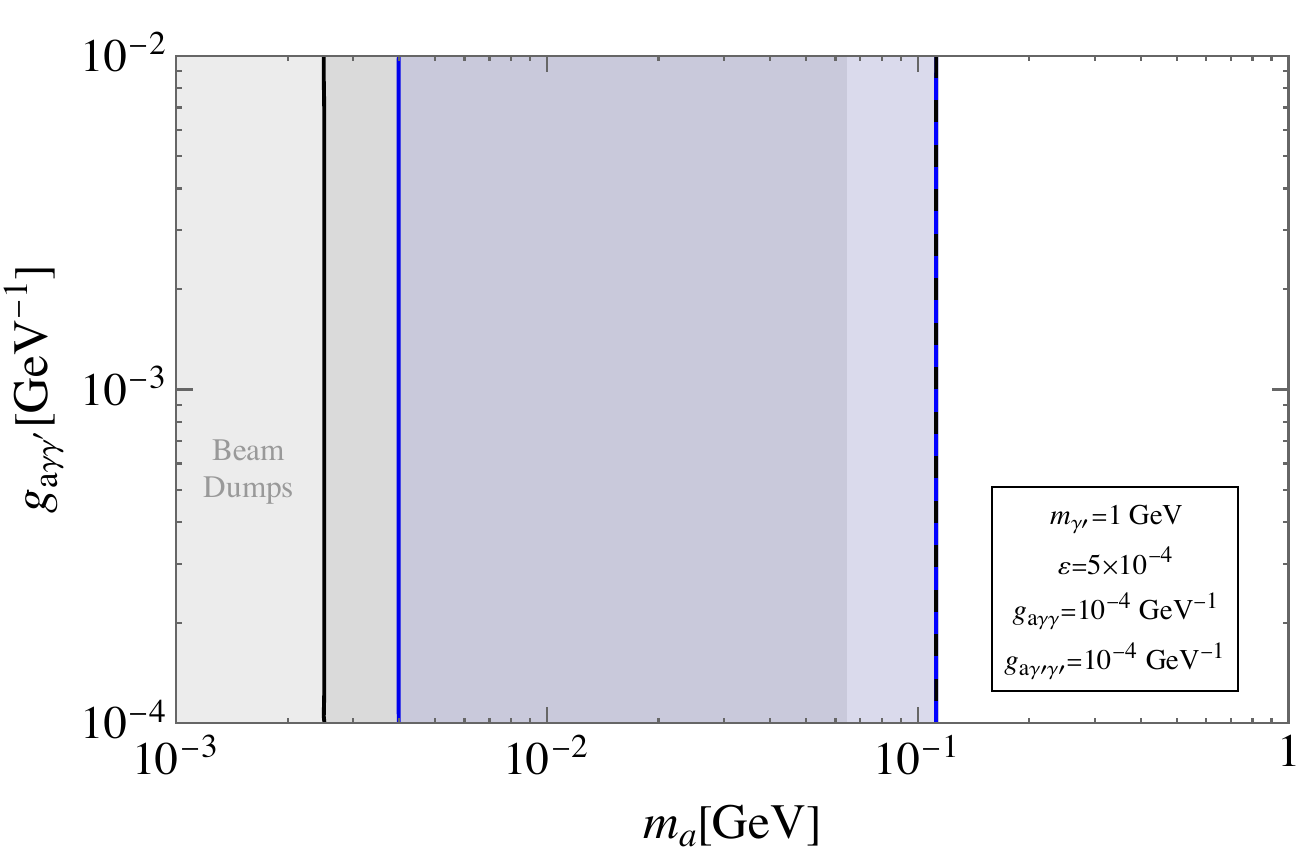}
    
    \includegraphics[width=0.3825\textwidth]{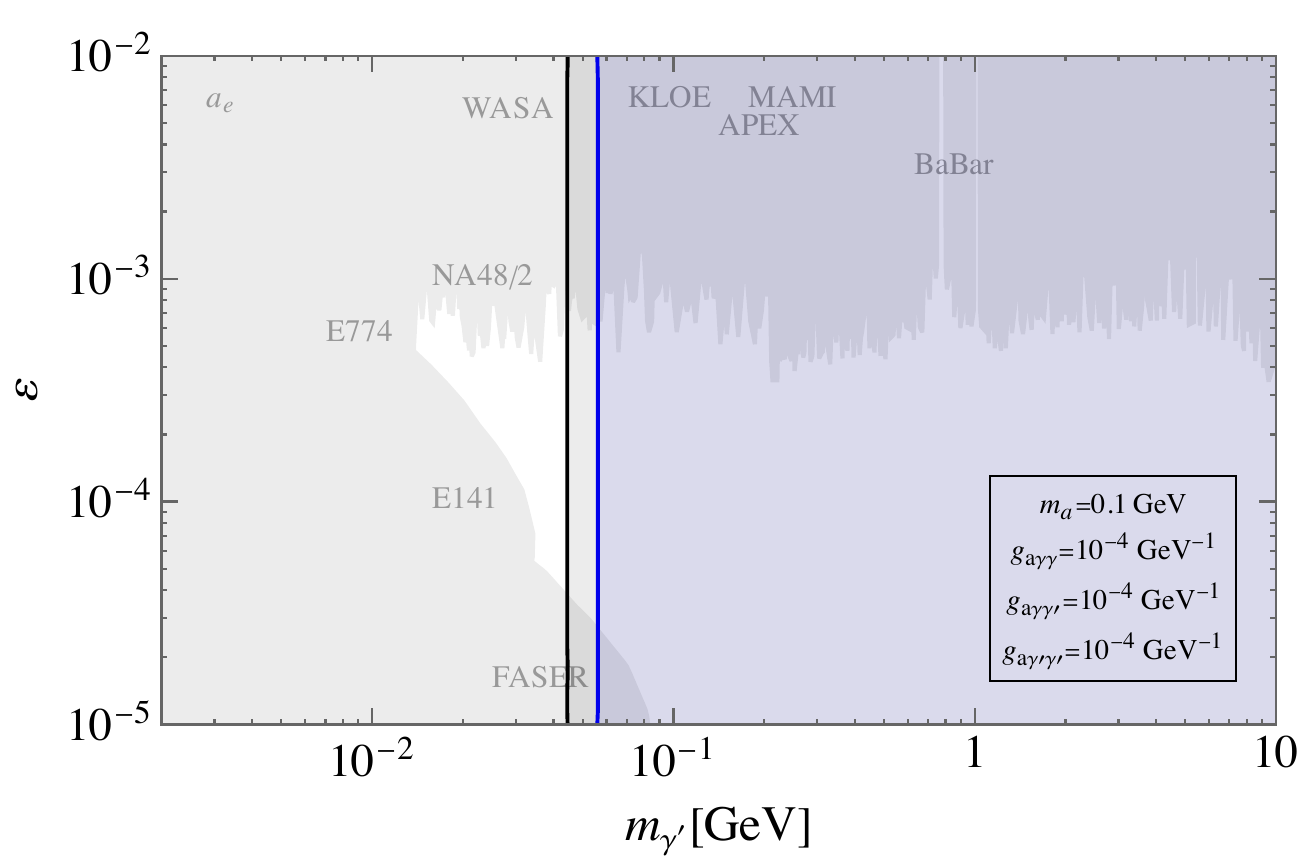}
    \hspace{0.075\textwidth}
    \includegraphics[width=0.3825\textwidth]{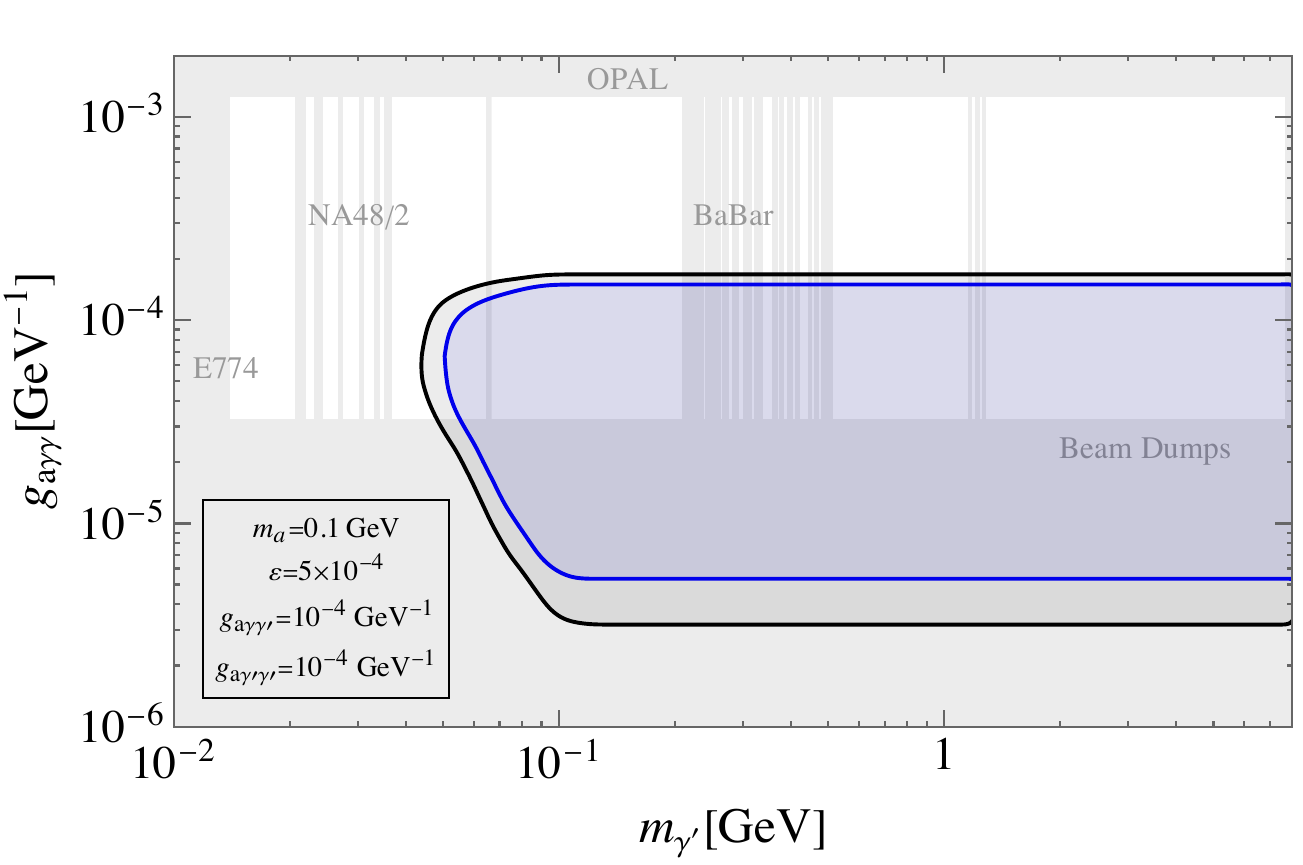}
    
    \includegraphics[width=0.3825\textwidth]{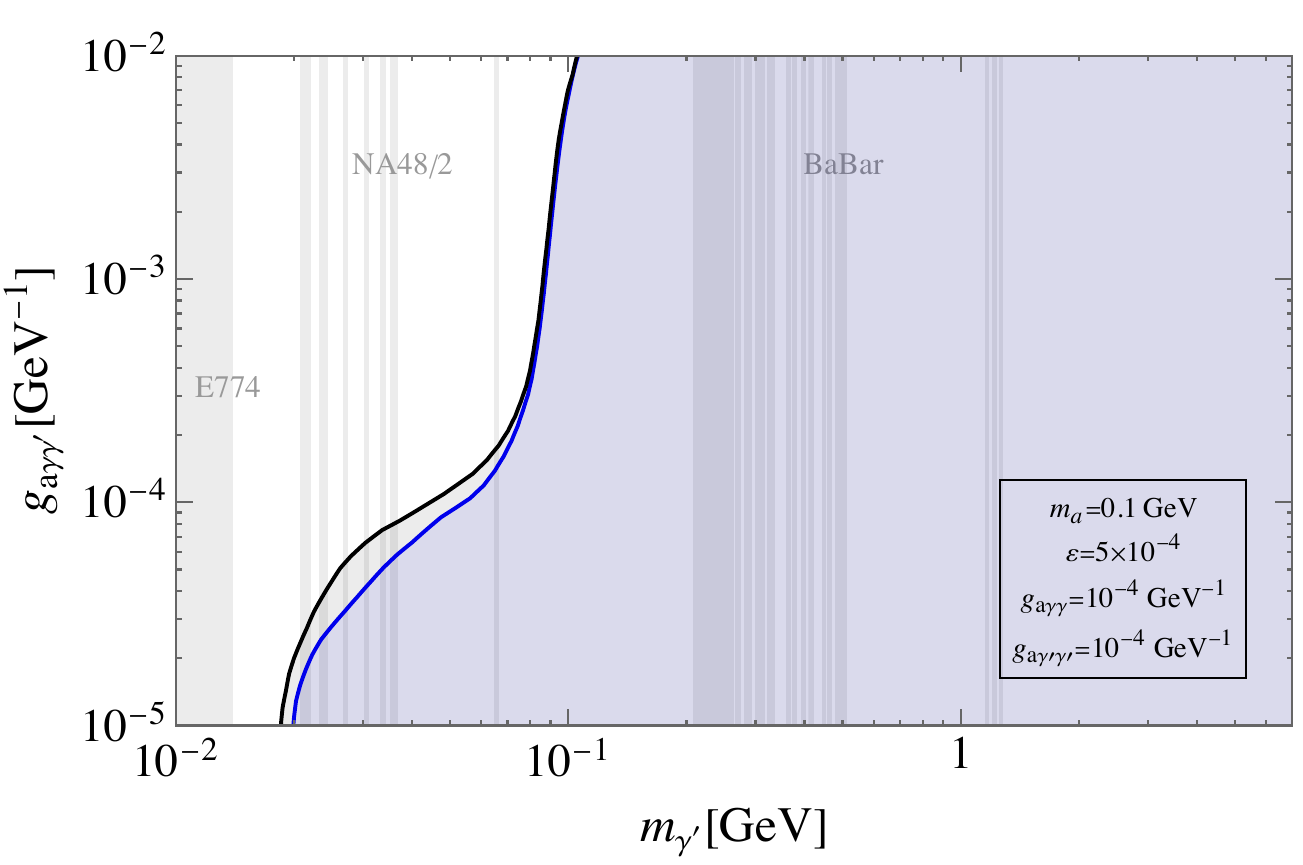}
    \hspace{0.075\textwidth}
    \includegraphics[width=0.3825\textwidth]{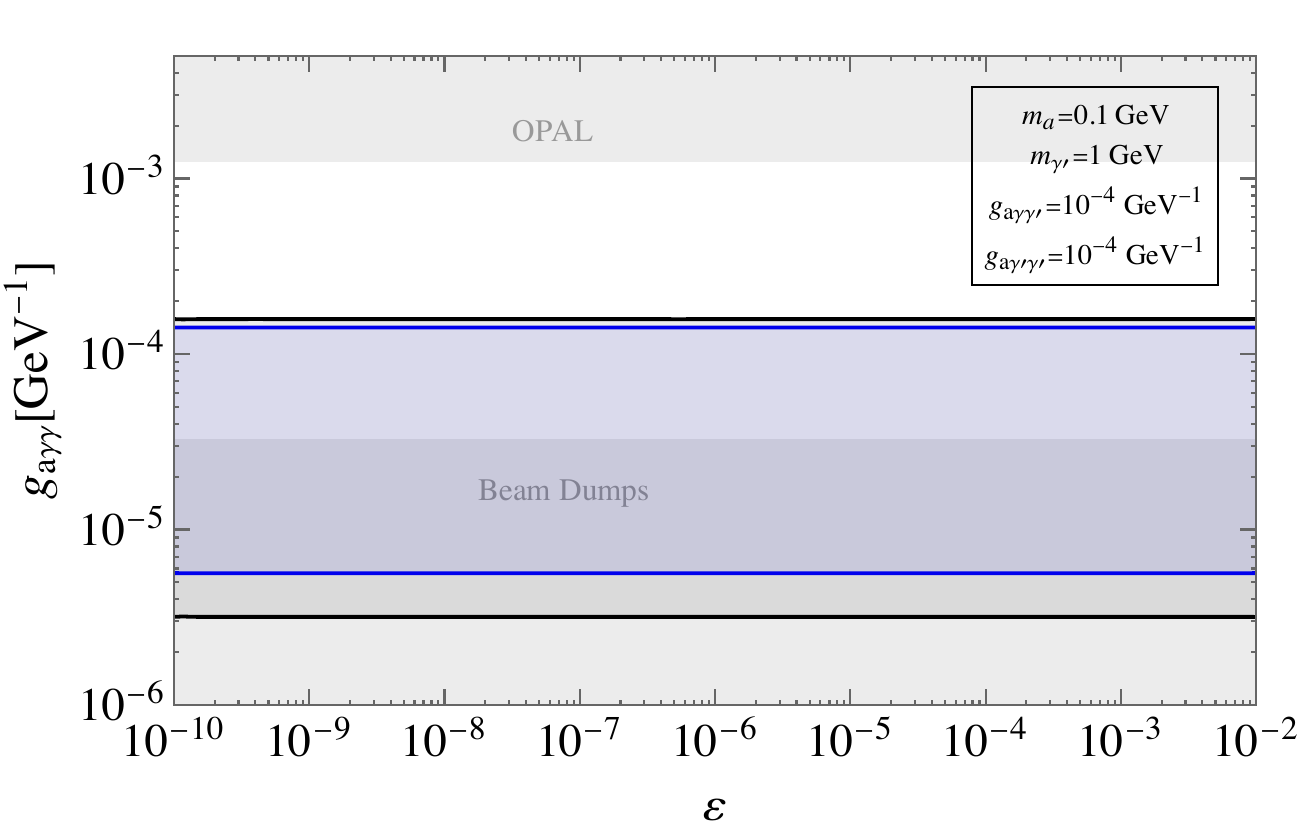}

    \includegraphics[width=0.3825\textwidth]{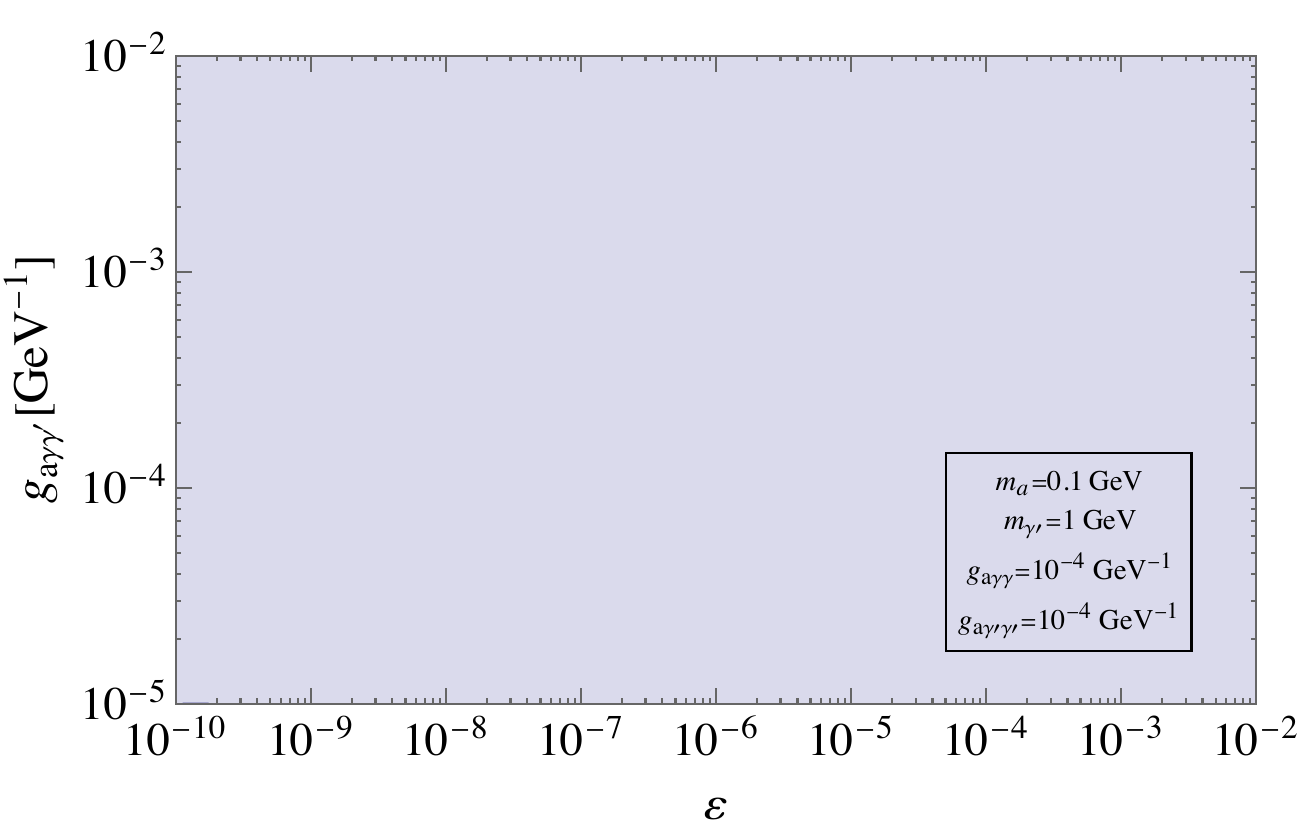}
    \hspace{0.075\textwidth}
    \includegraphics[width=0.3825\textwidth]{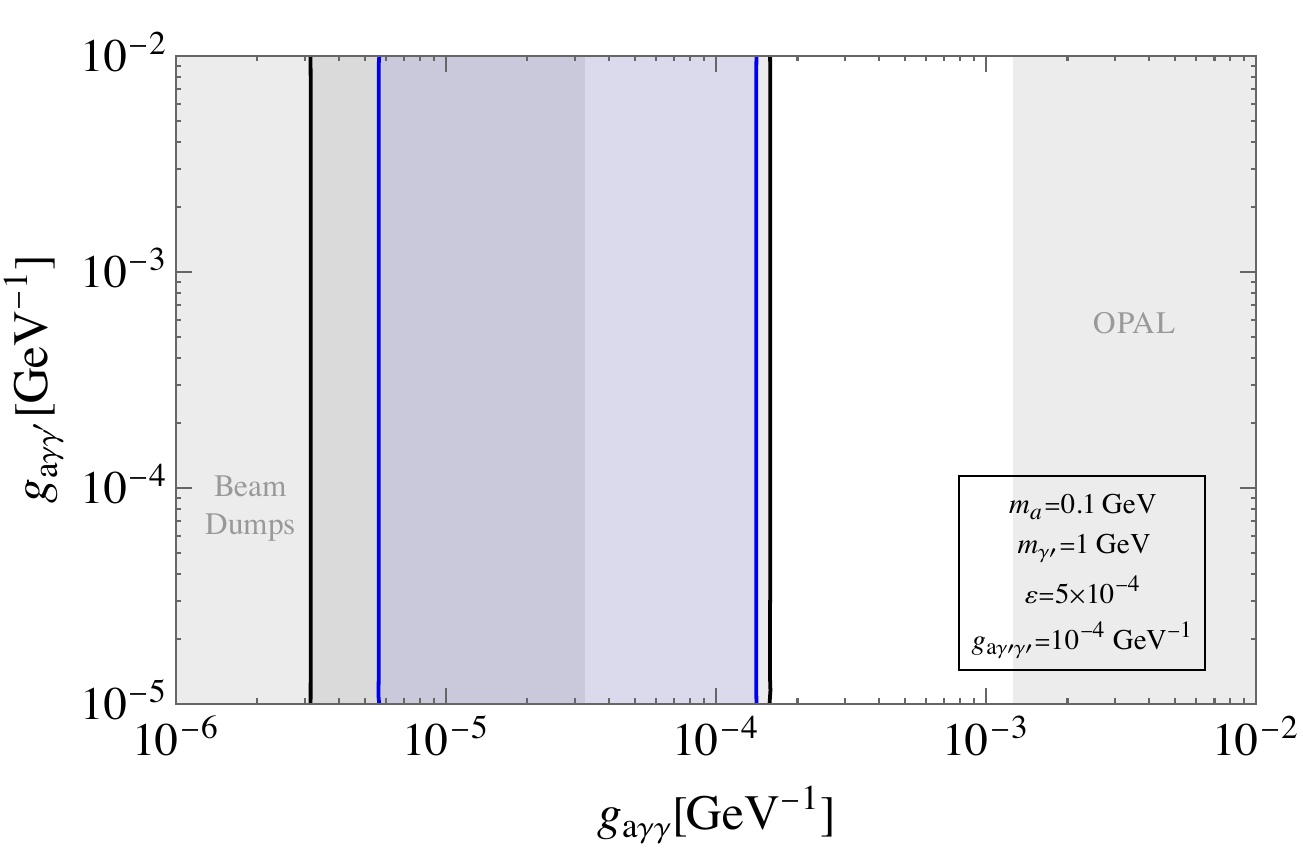}
    
    \caption{Projected reach of LUXE-NPOD in the dark axion parameter space for heavy DPs $m_{\gamma'}>m_a$.}
\label{fig:HeavyDPs}
\end{figure*}

As a consistency check, we consider the DP decoupling limit. Indeed, when we decouple the DP from the ALP we recover the same sensitivity obtained in Ref.~\cite{LUXENPOD}. Moving away from this limit, we introduce a nonzero dark axion portal $g_{a\gamma\gamma'}>0$. Despite this additional interaction, LUXE-NPOD's reach in the $\left(m_a,g_{a\gamma\gamma}\right)$ plane remains essentially unchanged, as shown in \cref{fig:HeavyDPs} (second row, left).
This behavior is expected: the ALP-photon-DP interaction only adds a $\gamma'$ mediation channel for Primakoff-like ALP production (see \cref{subfig:DarkAxionExampleDiagrams}), yet this channel is subdominant due to the large DP mass.

Beyond reproducing the same $\left(m_a,g_{a\gamma\gamma}\right)$ sensitivity as in Ref.~\cite{LUXENPOD}, we obtain novel reach in other parameter subplanes. One example of this, for the $\left(m_{\gamma'},g_{a\gamma\gamma}\right)$ plane, is presented in \cref{fig:HeavyDPs} (third row, right). After fixing the ALP mass $m_a=100\,\MeV$, we see that the experiment is sensitive to $g_{a\gamma\gamma}\sim\O\left(\numrange{1}{100}\right)\,\PeV^{-1}$. This is consistent with our discussion above---heavier DPs do not have much impact on the phenomenology, so the result does not exhibit strong $m_{\gamma'}$ dependence. Indeed, the extremal $g_{a\gamma\gamma}$ values in this curve are exactly equal to the intersections of a constant $m_a=100\,\MeV$ line with the previous curve we discussed (second row, left). When the DP becomes lighter than the ALP $m_{\gamma'}<m_a$, the ALP lifetime sharply decreases as more ALP decay channels (like $a\to\gamma\gamma'$ or $a\to\gamma'\gamma'$) become possible; hence, ALPs with the same $m_a$ and $g_{a\gamma\gamma}$ that previously decayed outside the tungsten dump are now too short lived to escape it, and all sensitivity is lost.

A representative sample of plots for all 10 parameter pairings, in the case of heavier DPs, is shown in \cref{fig:HeavyDPs}. All plots exhibit similar characteristics to the two examples discussed above. Larger $m_a$, $g_{a\gamma\gamma}$ describe short-lived ALPs that decay in the dump, smaller $m_a$, $g_{a\gamma\gamma}$ describe long-lived ALPs that escape the detector, and LUXE-NPOD is sensitive to a particular band of intermediate values. Sensitivity to $\varepsilon$ and $g_{a\gamma\gamma'}$ is minimal to none, since the DP is essentially decoupled from the ALP when $m_{\gamma'}>m_a$.

\subsection{Ultralight dark photons \texorpdfstring{$m_{\gamma'}\ll m_a$}{Mg'<<Ma}}

The second case where LUXE-NPOD exhibits competitive reach is when DPs are ``ultralight.'' We refer to this limit symbolically as $m_{\gamma'}\ll m_a$, though it is important to stress that we are referring here specifically to DPs that are light enough to be produced in the electron-laser collisions $m_{\gamma'}\ll\O(150)\,\keV$ [which incidentally means they are also much lighter than $\O(\numrange{10}{500})\,\MeV$ ALPs].

Three distinctions from the previous case affect the phenomenology. One is that the $\gamma'$-mediated Primakoff-like ALP production from \cref{subfig:DarkAxionExampleDiagrams} is no longer suppressed by a large DP mass; another is that we have the possibility of generating primary DPs, which may contribute to both primary and secondary NP production; the third is that ALPs can now decay into DPs and not just photons (eliciting all the implications this has on the ALP lifetime and branching fractions).

\begin{figure*}
    \centering
    \includegraphics[width=0.3825\textwidth]{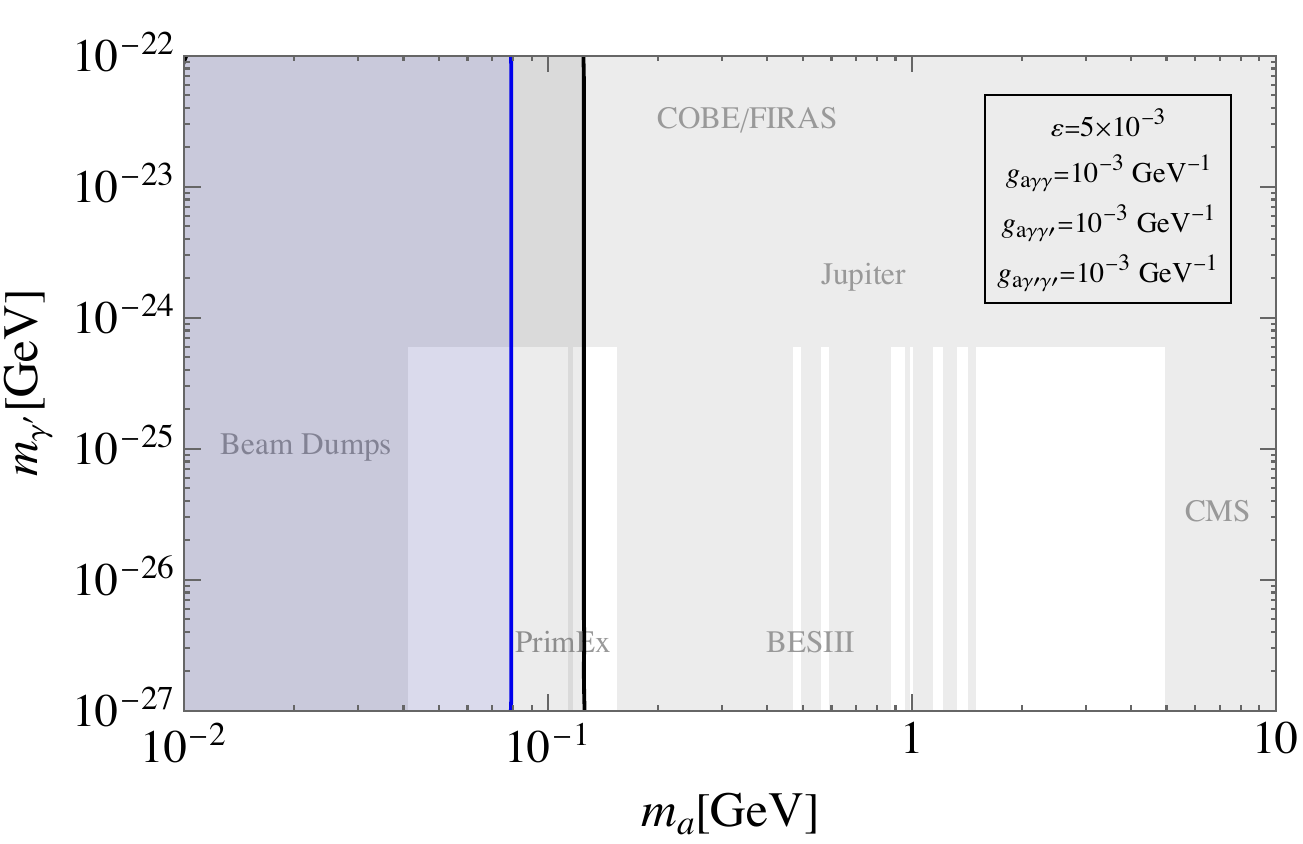}
    \hspace{0.075\textwidth}
    \includegraphics[width=0.3825\textwidth]{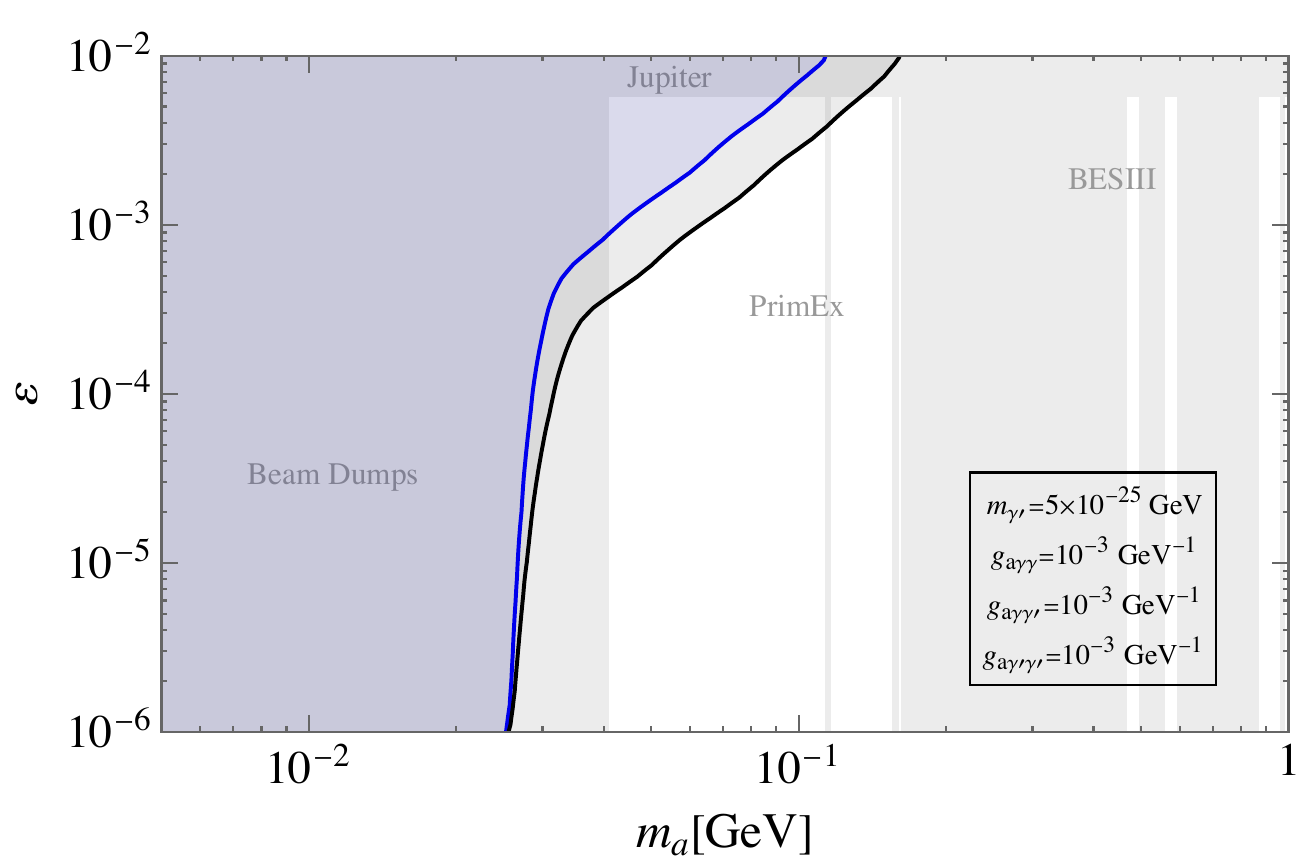}

    \includegraphics[width=0.3825\textwidth]{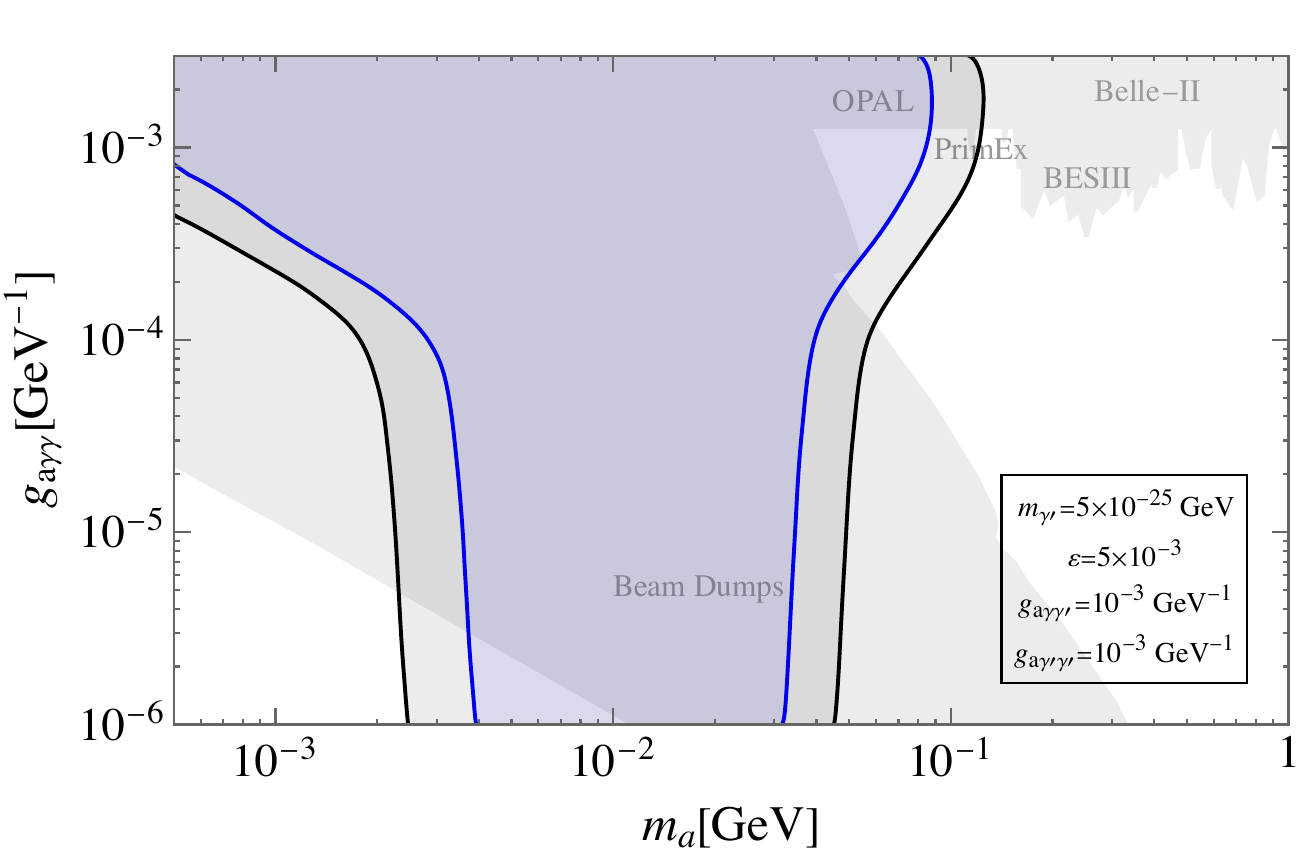}
    \hspace{0.075\textwidth}
    \includegraphics[width=0.3825\textwidth]{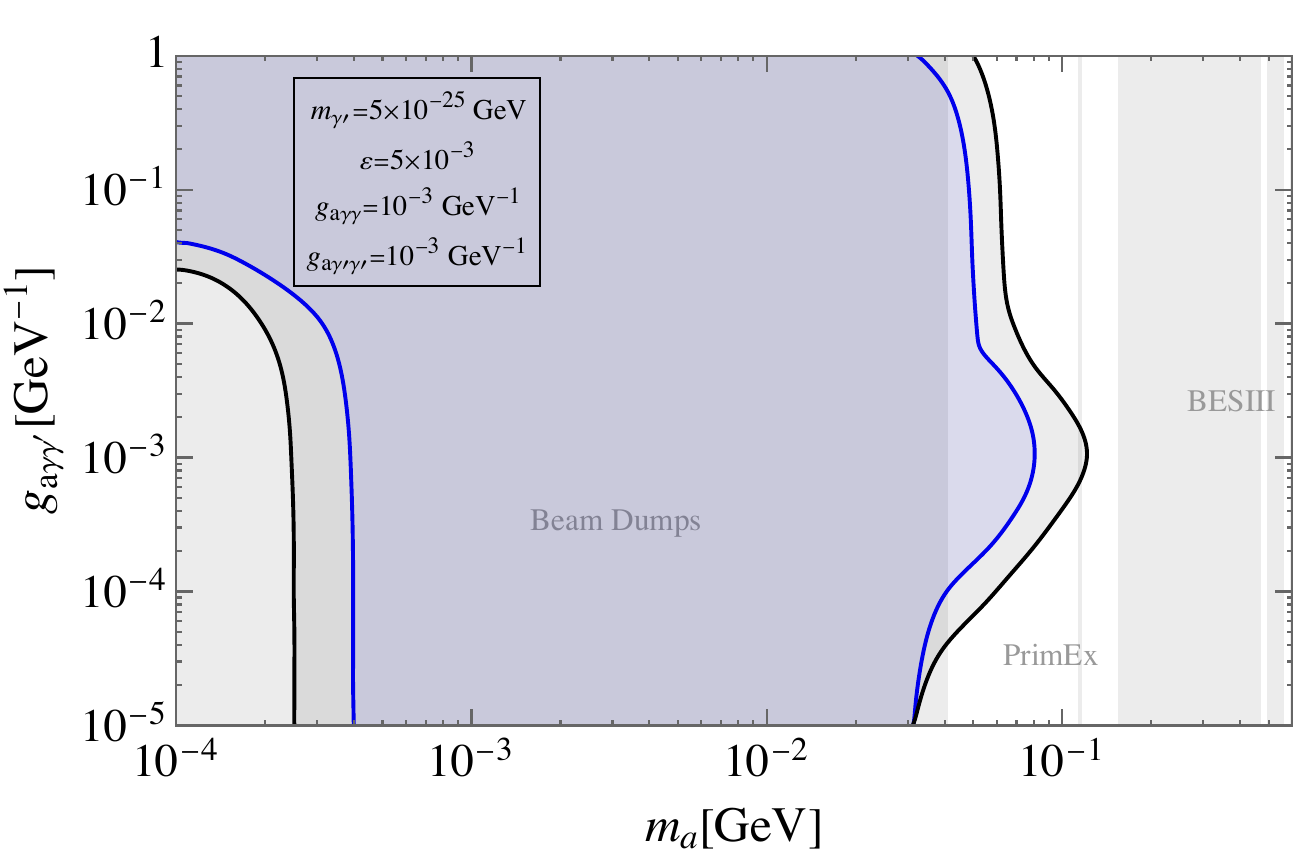}
    
    \includegraphics[width=0.3825\textwidth]{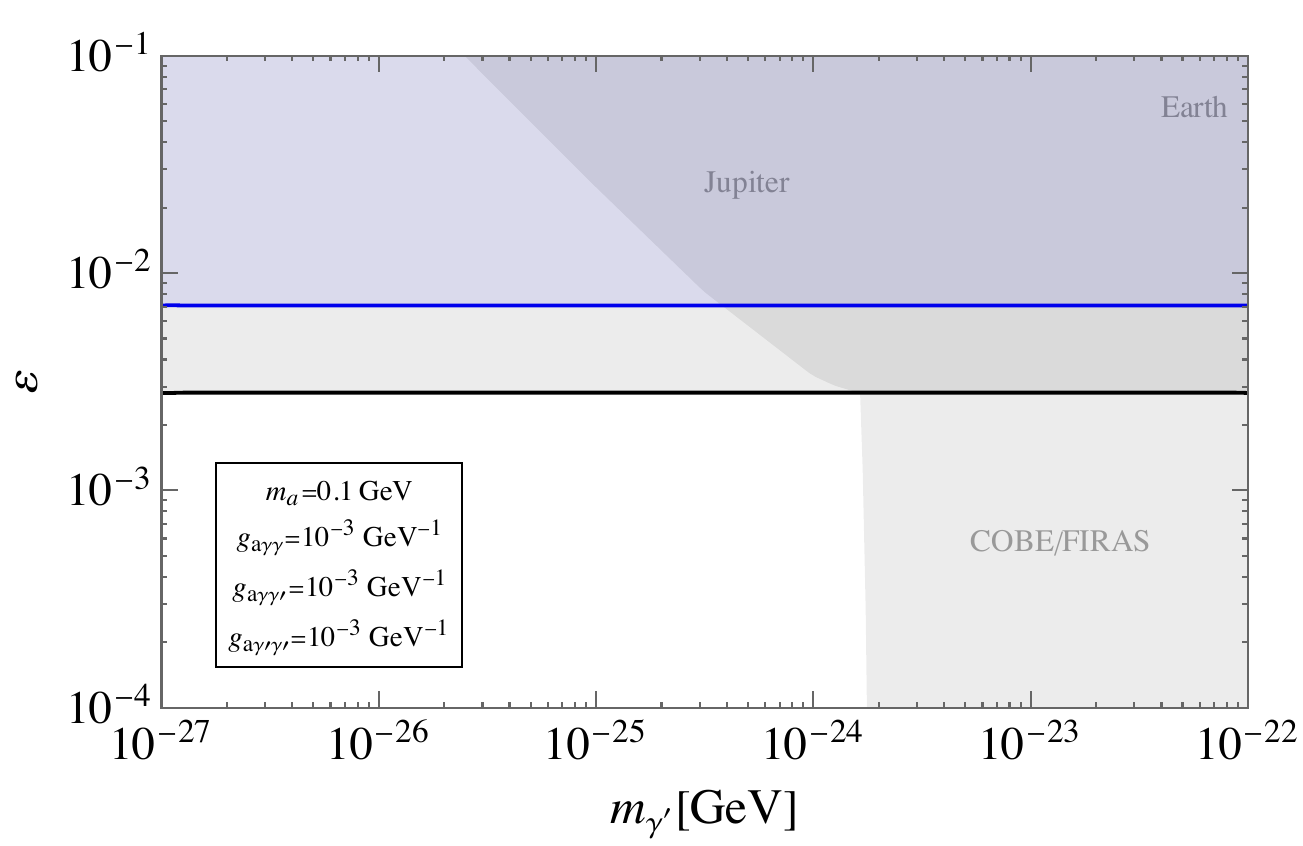}
    \hspace{0.075\textwidth}
    \includegraphics[width=0.3825\textwidth]{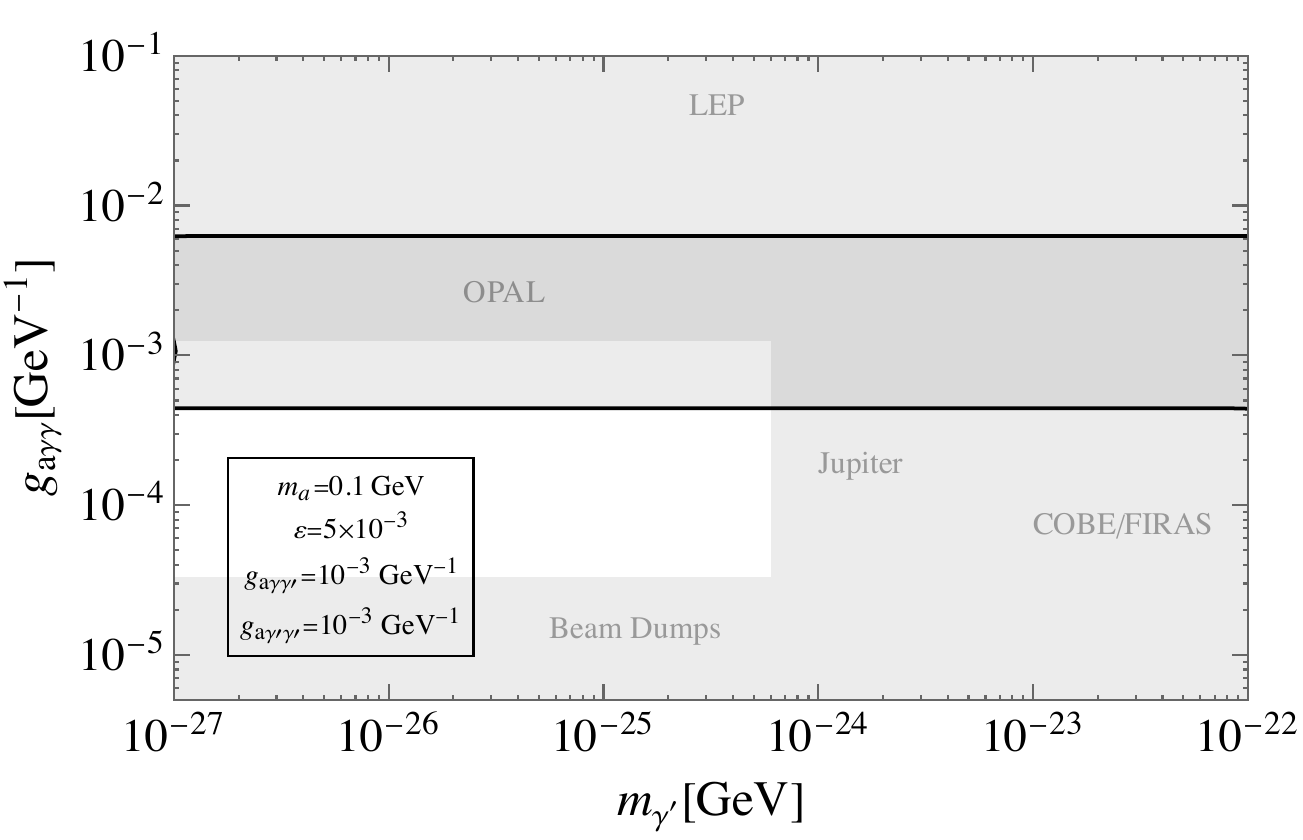}
    
    \includegraphics[width=0.3825\textwidth]{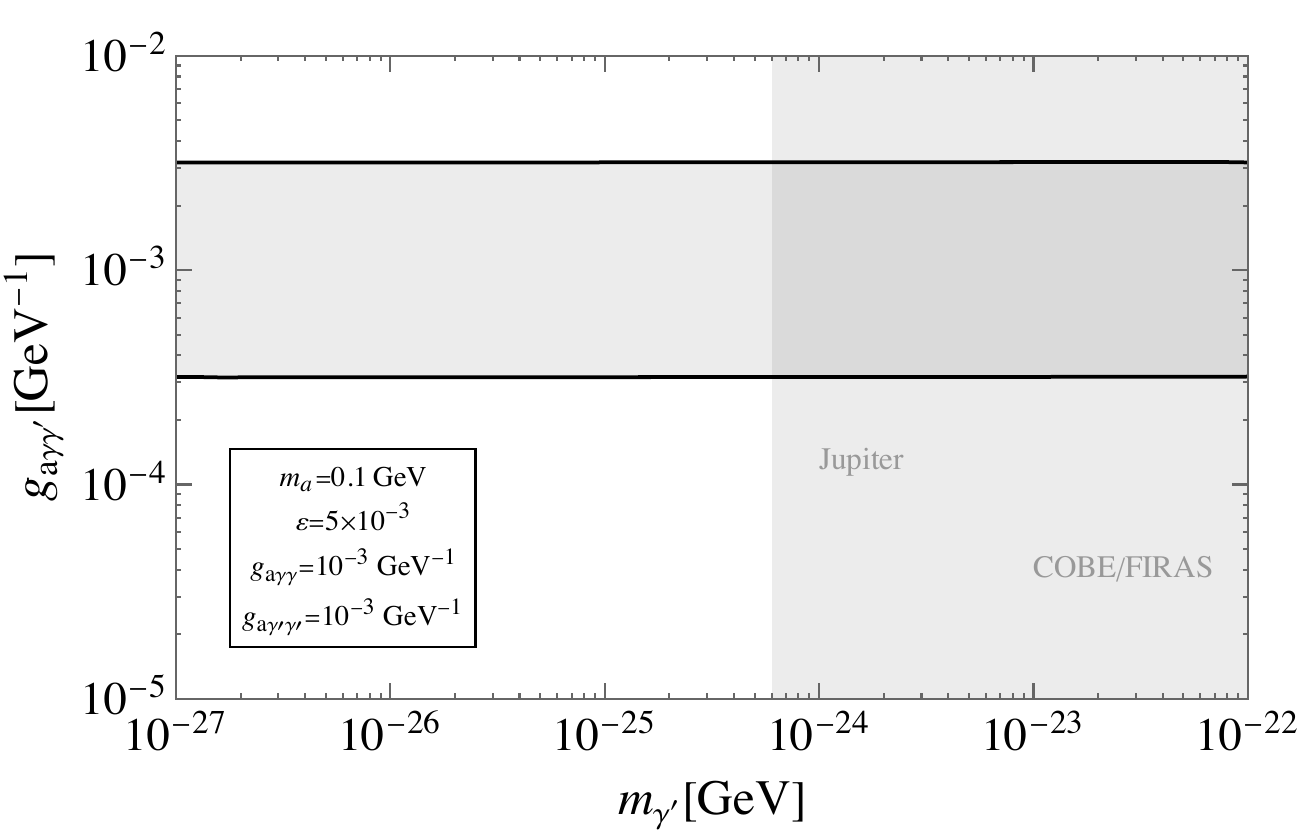}
    \hspace{0.075\textwidth}
    \includegraphics[width=0.3825\textwidth]{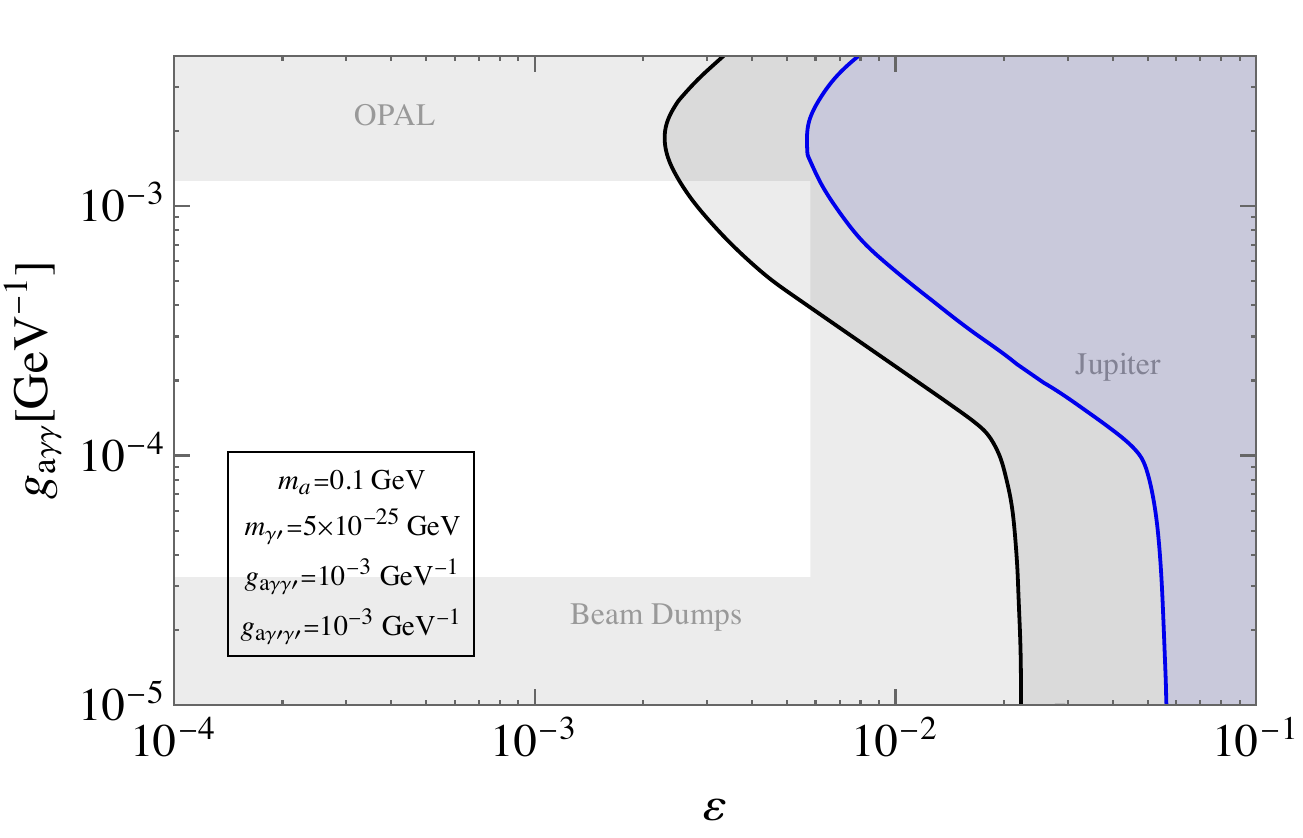}

    \includegraphics[width=0.3825\textwidth]{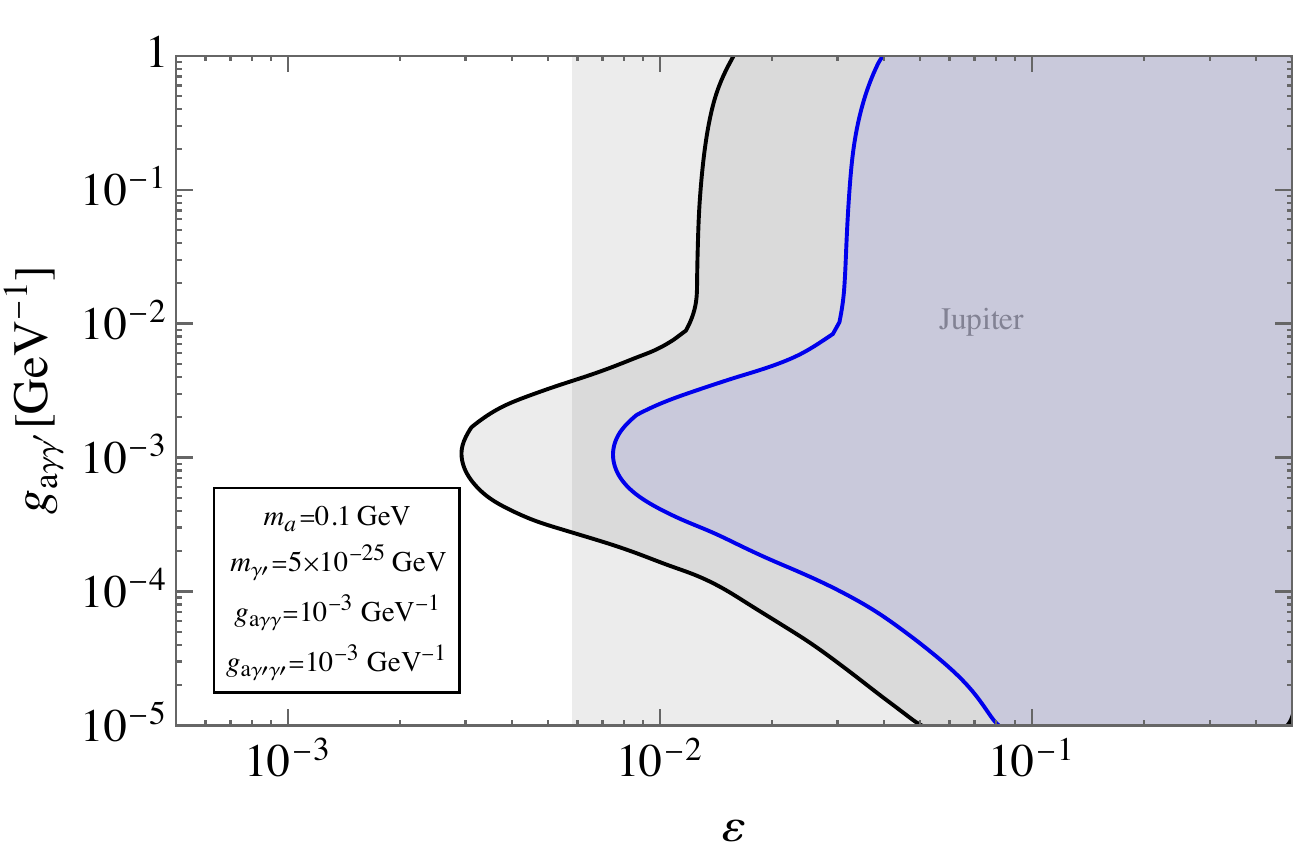}
    \hspace{0.075\textwidth}
    \includegraphics[width=0.3825\textwidth]{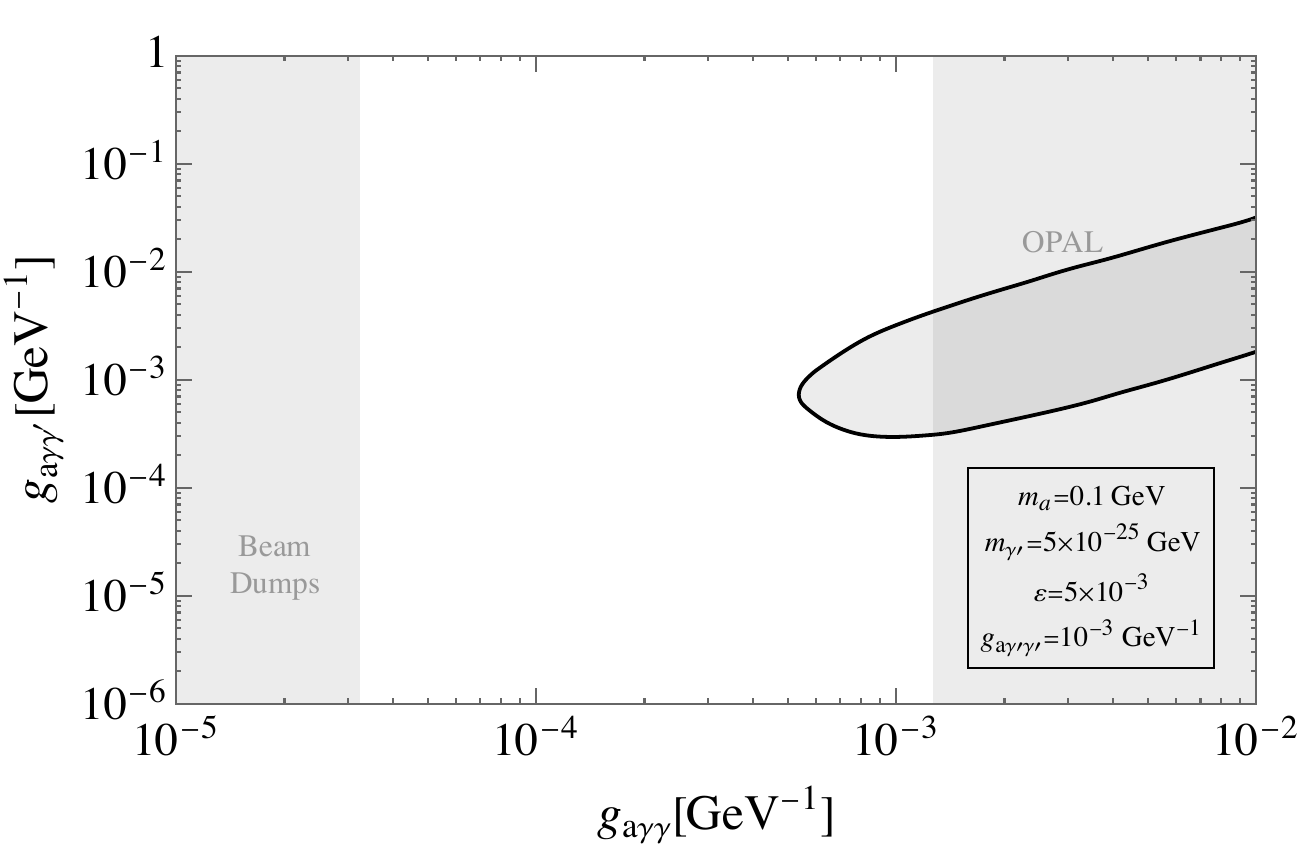}
    
    \caption{Projected reach of LUXE-NPOD in the dark axion parameter space when $m_{\gamma'}\ll m_a$.
    }
\label{fig:UltralightDPs}
\end{figure*}

\begin{figure}
    \centering
    \includegraphics[width=0.48\textwidth]{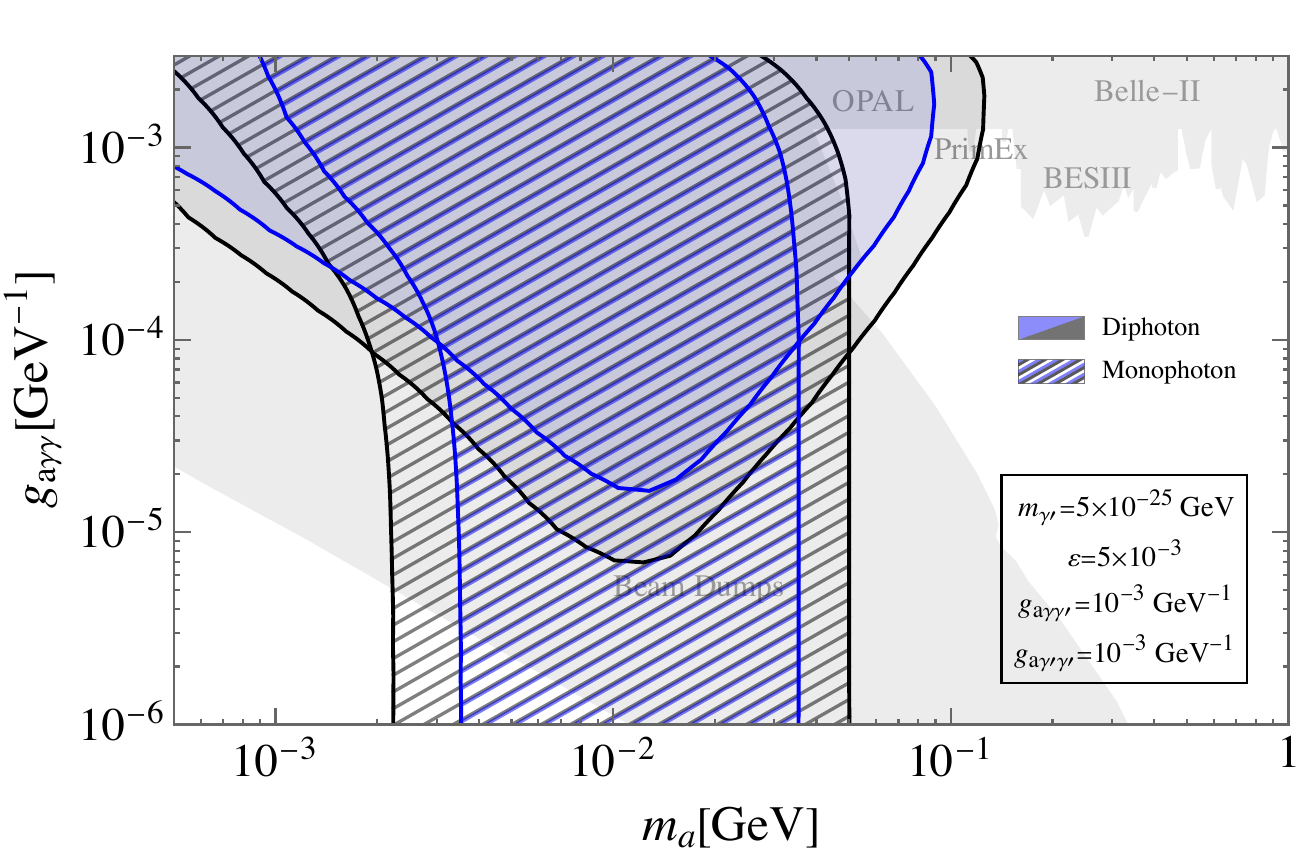}
    \caption{Projected reach of LUXE-NPOD in the $(m_a,g_{a\gamma\gamma})$ plane when $m_{\gamma'}\ll m_a$, separated into contributions from diphoton signals (solid semitransparent coloring) and monophoton signals (diagonal-hatch-line filling).}
\label{fig:Di_vs_Mono_signals}
\end{figure}

With the previous distinctions in mind, we reexamine the $\left(m_a,g_{a\gamma\gamma}\right)$ plane in the ultralight DP limit. For a given ALP mass $m_a$ and ALP-photon-photon coupling $g_{a\gamma\gamma}$, the possibility of $a\to\gamma'$ decays implies a shorter ALP lifetime compared to the ALP-only model (in which ALPs have just a single decay channel, into two photons). Therefore, the detectable volume outside the dump and before the detector corresponds to a downward skewed version of the standard-shaped beam dump projections (like those in Ref.~\cite{LUXENPOD} and \cref{fig:HeavyDPs} of this work). That is, ALPs we might hope to detect are ones with smaller $g_{a\gamma\gamma}$ relative to the previous case. This behavior can be seen in \cref{fig:UltralightDPs} (second row, left).
The existence of $a\to\gamma'$ decays means we lose a fraction of the diphoton signals we had in the ALP-only model, but this is partially compensated for by our inclusion of monophoton signals in the analysis. 
We observe this explicitly in \Cref{fig:Di_vs_Mono_signals}, where we separate LUXE-NPOD's sensitivity in the $(m_a,g_{a\gamma\gamma})$ plane into the two distinct decay modes that enable said reach---sensitivity from diphoton signals is depicted with a transparent coloring, and sensitivity from monophoton signals is depicted with diagonal hatch lines.
Interestingly, in the $g_{a\gamma\gamma}\lesssim\PeV^{-1}$ region of the plot, it is the ALP production described by \cref{eq:secNPprod_gamma'toa},
triggered by primary \emph{DPs} (rather than photons), that provides the most substantial contribution to the observed signal. Thus, we see that genuinely novel NP production processes facilitate the novel reach in this mass regime.

Let us also inspect the results obtained in the $\left(m_{\gamma'},\varepsilon\right)$ plane, shown in \cref{fig:UltralightDPs} (third row, left). It seems that LUXE-NPOD has the power to probe the kinetic mixing parameter $\varepsilon$ down to arbitrarily low DP masses, without losing any sensitivity. This is in stark contrast to virtually all existing limits on the $\left(m_{\gamma'},\varepsilon\right)$ parameters, which generically exhibit a maximum sensitivity at some mass scale below which their reach diminishes according to a power law (see, e.g., Fig.~1 in Ref.~\cite{DPLimitsLight_Handbook}). The generic behavior of these existing limits is well understood theoretically; for example, it is known that DPs modify the Coulomb potential $V(r)=\alpha/r+\varepsilon^2\alpha e^{-m_{\gamma'}r}/r$ (see Refs.~\cite{Paraphotons,DPCoulombModification}), resulting in an appropriate correction $\Delta V\sim \varepsilon^2m_{\gamma'}^2\alpha r/2$ when the DP is much lighter than the typical scale $1/r$ of the experiment. The reason we do not observe this power law in LUXE-NPOD is because we are not probing a model with just the DP as a new particle, but a system of DPs \emph{and} ALPs that mutually interact---enabling us to probe DP properties through ALP decays (and vice versa). Support for this claim comes from the fact that the dominant driver of LUXE-NPOD's reach in the $(m_{\gamma'},\varepsilon)$ plane is the chain $e_V^-\to e_V^-\gamma' \;\,\Rightarrow\;\, \gamma'N\to aN \;\,\Rightarrow\;\, a\to\gamma\gamma$. In the ultralight DP regime, the contribution of this process to the number of signals scales as
\begin{equation*}
    \!\underbrace{\varepsilon^2}_{\mtxt{\ensuremath{\gamma'} flux}}\!\!\times\underbrace{g_{a\gamma\gamma'}^2\!\left[1\!+\!\O\!\left(m_{\gamma'}^2/\abs{t}\right)\right]}_{\mtxt{\ensuremath{\gamma'\to a} conversion}}\times\!\!\!\underbrace{\prob_\mtxt{\ensuremath{a}-decay}}_{\!\!\mtxt{decay in volume}}\!\!\times\!\!\underbrace{\O(1)}_{\;\mtxt{\ensuremath{a\!\to\!\gamma\gamma} BR}}
\!\!\!\!\,.
\end{equation*}
Importantly, the expression above does not scale with $m_{\gamma'}$ to leading order ($\prob_\mtxt{\ensuremath{a}-decay}$ is only sensitive to the ALP parameters $m_a, g_{a\gamma\gamma}, g_{a\gamma\gamma'}, g_{a\gamma'\gamma'}$ in this regime), unlike the observables of dedicated DP-only searches. This is because the $\gamma'N\to aN$ conversion remains finite as $m_{\gamma'}$ decreases rather than vanishing as DP-only effects do---a direct consequence of the ALP-photon-DP coupling $g_{a\gamma\gamma'}$.

Finally, just like in the heavy-DP case, there are other pairs of parameters that can be constrained by LUXE-NPOD in the limit of ultralight DPs. \Cref{fig:UltralightDPs} shows a representative sample of the projected reach in each of the 10 parameter planes, in this regime. Although there are some planes to which LUXE's phase~0 is not sufficiently sensitive, we see that phase~1 of the experiment has the potential to probe new territory in all $10$ parameter subplanes. We observe the same trend seen in the heavy-DP case whereby $m_a$ and $g_{a\gamma\gamma}$ affect the ALP lifetime, but this time the results also noticeably depend on $\varepsilon$ and $g_{a\gamma\gamma'}$, as the DP is no longer decoupled from the ALP\@. Smaller $\varepsilon$ means fewer primary DPs are produced, leading to a smaller number of $\gamma'\to a$ Primakoff-like conversions in the dump and hence fewer detection events. Regarding $g_{a\gamma\gamma'}$, this parameter determines the magnitude of the $\gamma'\to a$ conversion cross section, so the number of such conversions (and consequently the number of detected signals) diminishes as $g_{a\gamma\gamma'}$ gets smaller.

%% file: sections/5_Conclusions.tex
In this work, we analyzed the implications of the dark axion model on the observed signals expected at the LUXE-NPOD experiment. The model, which incorporates DPs, ALPs, and their mutual interactions, can be effectively probed by this experiment due to its dual interaction points that allow for novel processes involving both particles.  

Our cascade simulations yielded a primary photon spectrum that closely matches previous works and confirmed that primary DP/ALP spectra are well approximated by simple coupling suppression. We showed that LUXE-NPOD is capable of probing novel regions in the six-dimensional parameter space of the dark axion model when the ALP mass falls within the $\numrange{10}{500}\,\MeV$ range. We are potentially able to place competitive constraints when the DP mass is either $\gtrsim\!\GeV$ or $\ll\!\eV$, the kinetic mixing is in the $\numrange{10^{-4}}{10^{-2}}$ range, and the ALP couplings to photons/DPs are $\O(\numrange{10^{-4}}{10^{-3}})\,\GeV^{-1}$. Notably, we found that the existence of a dark axion portal allows for the probing of kinetic mixings for arbitrarily low DP masses, unlike the generic behavior expected from DP-specific searches.

Our analysis demonstrated a systematic approach for constraining a model comprising multiple new particles and several free parameters. The spirit of this method can be replicated and applied to other models that include several NP processes and more than two parametric degrees of freedom.

It is important to note that throughout this work we did not allow for direct decays of DPs/ALPs into solely hidden sector particles. These could have significant effects on particle lifetimes and the expected number of observed signals, though we are currently unaware of a well-established method to quantify such decays based on existing observations.

%% file: sections/6_Acknowledgments.tex
The authors would like to thank Yotam Soreq for his collaboration and guidance throughout this project. They are also grateful to both Andrea Caputo and Edoardo Vitagliano for valuable insights regarding existing constraints on dark photon and axionlike particle parameters. The work of N.~N.\@ was supported by the Israel Science Foundation. B.~C.\@ is supported by the Natural Sciences and Engineering
Research Council of Canada through a Canada Graduate Scholarship --- Doctoral (Grant No. 599773) and by the U.S. Department of Energy, Office of Science, Office of High Energy Physics under Contract No.\@ DE-SC0012567.

%% file: sections/Appendix_A-Cascade_Equations_in_1D.tex
To obtain the spectra $dN_{Y}/dE_{Y}$ of particles $Y=\gamma,\gamma',a$ emanating from the primary interaction point, we solve the cascade equations for the particle spectra propagating through the laser pulse. To compute transition rates, we work in the framework of nonlinear QED, where the Gaussian laser pulse is treated as a slowly varying background field. The background field $A_\mu$ stimulates nonlinear Compton-like scattering (NCS) and nonlinear pair production (NPP). $A_\mu$ is taken to be circularly polarized with 4-momentum $k_\mu$, that is,
\begin{equation}
    A^\mu(x) = a_1^\mu \cos(kx) + a_2^\mu\sin(kx)
\,,
\end{equation}
where $a_1^2=a_2^2=A^2$ and $a_1\cdot a_2=a_1\cdot k=a_2\cdot k=0$. 

Electrons propagating through the laser pulse lose energy by emitting photons or NP particles via NCS or off-shell processes (see diagrams in \cref{fig:primNPprod}). The emitted photons, if of high enough energy, may then produce electron-positron pairs via NPP\@. NP particles may also produce electron-positron pairs (or other combinations of particles), but this effect is small and can be neglected. The interplay of these processes results in a cascade shower of particles. The intensity of the background field controls the rates of these processes and therefore the cascade dynamics. For $\mathcal{O}(1)$ values of the quantum nonlinearity parameter $\eta=e\abs{A}/m_e$ (here, $\abs{A}=\sqrt{\abs{A^2}}$ is the strength of the background field), such as those attainable at LUXE, an electron beam will lose most of its energy while propagating through the dump by producing hard photons and potentially NP particles. 

Since the energy of the electron beam $E_0$ is much larger than the electron mass ($E_0=16.5\,\GeV\gg m_e$), the cascade products are emitted through a narrow cone approximately collinear with the incident electrons. The evolution of the particle densities $I_X(E, t)$, where $X=e_V^-, e_V^+, \gamma, \gamma', a$, can therefore be effectively described by the 1D cascade-shower equations of QED, extended to include NP particle populations~\cite{1DCascadeEqs}
\begin{subequations}
\label{eq:Cascade}
\begin{align}
\label{eq:Cascade_Photons}
    &\frac{\partial I_\gamma(E_\gamma,t)}{\partial t}=
    \nonumber \\
    &\qquad -\int_{m_e}^{E_\gamma-m_e}\!d\tilde{E}_e\;\; I_\gamma(E_\gamma, t) \frac{dP_{\gamma\to ee}(\tilde{E}_e,E_\gamma,t)}{d\tilde{E}_e}  
    \nonumber \\
    &\qquad +\int_{m_e+E_\gamma}^{E_0}\!d\tilde{E}_e\;\; I_e(\tilde{E}_e, t) \frac{dP_{e\to e\gamma}(E_\gamma,\tilde{E}_e,t)}{d\tilde{E}_e}
    \nonumber \\
    &\qquad +\int_{m_e+E_\gamma+m_a}^{E_0}\!d\tilde{E}_e\;\; I_e(\tilde{E}_e,t)
    \nonumber \\
    &\qquad \quad\;\; \times\left[\left(\frac{dP_{e\to e\gamma a}(E_\gamma,\tilde{E}_a,\tilde{E}_e,t)}{d\tilde{E}_a}\right)_{\!\!\tilde{E}_a=\tilde{E}_e-m_e-E_\gamma} \right.
    \nonumber \\
    &\qquad \quad\;\; \quad\;\, \left.+\int_{m_a}^{\tilde{E}_e-m_e-E_\gamma}\!d\tilde{E}_a \frac{d^2P_{e\to e\gamma a}(E_\gamma,\tilde{E}_a,\tilde{E}_e,t)}{d\tilde{E}_ad\tilde{E}_e}\right]
\,,
\end{align}
\begin{align}
\label{eq:Cascade_Electrons}
    &\frac{\partial I_e(E_e,t)}{\partial t}= 
    \nonumber \\
    &\quad \int_{E_e+m_e}^{E_0}\!d\tilde{E}_\gamma \;\; I_\gamma(\tilde{E}_\gamma, t) \frac{dP_{\gamma\to ee}(E_e,\tilde{E}_\gamma,t)}{d\tilde{E}_\gamma} 
    \nonumber \\
    &\; +\!\int_{E_e}^{E_0}\!d\tilde{E}_e\;\; I_e(\tilde{E}_e,t) \frac{dP_{e\to e\gamma}(\tilde{E}_e-E_e,\tilde{E}_e,t)}{d\tilde{E}_e}  
    \nonumber \\
    &\; -\!\int_0^{E_e-m_e}\!d\tilde{E}_\gamma\;\; I_e(E_e, t) \frac{dP_{e\to e\gamma}(\tilde{E}_\gamma,E_e,t)}{d\tilde{E}_\gamma}
    \nonumber \\
    &\; +\!\int_{E_e+m_a}^{E_0}\!d\tilde{E}_e\;\; I_e(\tilde{E}_e,t)
    \nonumber \\
    &\quad\; \times\!\left[\!\left(\!\frac{dP_{e\to e\gamma a}(\tilde{E}_\gamma,\tilde{E}_e\!-\!E_e\!-\!\tilde{E}_\gamma,\tilde{E}_e,t)}{d\tilde{E}_\gamma}\!\right)_{\!\!\tilde{E}_\gamma=\tilde{E}_e\!-E_e\!-m_a}\right.
    \nonumber \\
    &\quad\; \quad \left.+\!\int_{0}^{\tilde{E}_e-E_e-m_a}\!\!d\tilde{E}_\gamma \frac{d^2P_{e\to e\gamma a}(\tilde{E}_\gamma,\tilde{E}_e\!-\!E_e\!-\!\tilde{E}_\gamma,\tilde{E}_e,t)}{d\tilde{E}_ed\tilde{E}_\gamma}\right]
    \nonumber \\
    &\; -\!\int_{0}^{E_e-m_e-m_a}\!d\tilde{E}_{\gamma}\;\; I_e(E_e,t)
    \nonumber \\
    &\quad\; \times\!\left[-\left(\frac{dP_{e\to e\gamma a}(\tilde{E}_{\gamma},\tilde{E}_a,E_e,t)}{d\tilde{E}_a}\right)_{\tilde{E}_a=E_e-m_e-\tilde{E}_{\gamma}}\right. 
    \nonumber \\
    &\quad\; \quad\quad \left.+\!\int_{m_a}^{E_e-m_e-\tilde{E}_{\gamma}}\!d\tilde{E}_a \frac{d^2P_{e\to e\gamma a}(\tilde{E}_{\gamma},\tilde{E}_a,E_e,t)}{d\tilde{E}_{\gamma}d\tilde{E}_a}\right]
    \nonumber \\
    &\; +\!\int_{E_e+m_{\gamma'}+m_a}^{E_0}\!d\tilde{E}_e\;\; I_e(\tilde{E}_e,t)
    \nonumber \\
    &\quad\; \times\!\left[\!\left(\!\frac{dP_{e\to e\gamma' a}(\tilde{E}_{\gamma'},\tilde{E}_e\!-\!E_e\!-\!\tilde{E}_{\gamma'},\tilde{E}_e,t)}{d\tilde{E}_{\gamma'}}\!\right)_{\!\!\tilde{E}_{\gamma'}=\tilde{E}_e\!-E_e\!-m_a}\right.
    \nonumber \\
    &\quad\; \quad \left.+\!\int_{m_{\gamma'}}^{\tilde{E}_e-E_e-m_a}\!\!d\tilde{E}_{\gamma'} \frac{d^2P_{e\to e\gamma' a}(\tilde{E}_\gamma,\tilde{E}_e\!-\!E_e\!-\!\tilde{E}_{\gamma'},\tilde{E}_e,t)}{d\tilde{E}_ed\tilde{E}_{\gamma'}}\right]
    \nonumber \\
    &\; -\!\int_{m_{\gamma'}}^{E_e-m_e-m_a}\!d\tilde{E}_{\gamma'}\;\; I_e(E_e,t)
    \nonumber \\
    &\quad\; \times\!\left[-\left(\frac{dP_{e\to e\gamma'a}(\tilde{E}_{\gamma'},\tilde{E}_a,E_e,t)}{d\tilde{E}_a}\right)_{\tilde{E}_a=E_e-m_e-\tilde{E}_{\gamma'}}\right. 
    \nonumber \\
    &\quad\; \quad\quad \left.+\!\int_{m_a}^{E_e-m_e-\tilde{E}_{\gamma'}}\!d\tilde{E}_a \frac{d^2P_{e\to e\gamma'a}(\tilde{E}_{\gamma'},\tilde{E}_a,E_e,t)}{d\tilde{E}_{\gamma'}d\tilde{E}_a}\right]
    \nonumber \\
    &\; +\!\int_{E_e+m_{\gamma'}}^{E_0}\!d\tilde{E}_e\;\; I_e(\tilde{E}_e,t) \frac{dP_{e\to e\gamma'}(\tilde{E}_e-E_e,\tilde{E}_e,t)}{d\tilde{E}_e} 
    \nonumber \\
    &\; -\!\int_{m_{\gamma'}}^{E_e-m_e}\!d\tilde{E}_{\gamma'}\;\; I_e(E_e, t) \frac{dP_{e\to e\gamma'}(\tilde{E}_{\gamma'},E_e,t)}{d\tilde{E}_{\gamma'}}
\,,
\end{align}
\begin{align}
\label{eq:Cascade_DPs}
    &\frac{\partial I_{\gamma'}(E_{\gamma'},t)}{\partial t}=
    \nonumber \\
    &\qquad \phantom{+}\;\, \int_{m_e+E_{\gamma'}}^{E_0}\!d\tilde{E}_e\;\; I_e(\tilde{E}_e, t)\frac{dP_{e\to e\gamma'}(E_{\gamma'},\tilde{E}_e,t)}{d\tilde{E}_e} 
    \nonumber \\
    &\qquad +\int_{m_e+E_{\gamma'}+m_a}^{E_0}\!d\tilde{E}_e\;\; I_e(\tilde{E}_e,t)
    \nonumber \\
    &\qquad \quad\;\; \times\!\! \left[\!\left(\!\frac{dP_{e\to e\gamma' a}(E_{\gamma'},\tilde{E}_a,\tilde{E}_e,t)}{d\tilde{E}_a}\!\right)_{\!\!\tilde{E}_a=\tilde{E}_e-m_e-E_{\gamma'}}\right.
    \nonumber \\
    &\qquad \quad\;\; \quad \left.+\!\int_{m_a}^{\tilde{E}_e\!-m_e\!-E_{\gamma'}}\!\!d\tilde{E}_a \frac{d^2P_{e\to e\gamma' a}(E_{\gamma'},\tilde{E}_a,\tilde{E}_e,t)}{d\tilde{E}_ad\tilde{E}_e}\!\right]
\,,
\end{align}
\begin{align}
\label{eq:Cascade_ALPs}
    &\frac{\partial I_a(E_a,t)}{\partial t}= 
    \nonumber \\
    &\qquad \int_{m_e+E_a}^{E_0}\!d\tilde{E}_e\;\; I_e(\tilde{E}_e, t)
    \nonumber \\
    &\qquad \quad\;\; \times\left[\left(\frac{dP_{e\to e\gamma a}(\tilde{E}_\gamma,E_a,\tilde{E}_e,t)}{d\tilde{E}_\gamma}\right)_{\tilde{E}_\gamma=\tilde{E}_e-m_e-E_a}\right.
    \nonumber \\
    &\qquad \quad\;\; \quad\;\;\; \left.+\int_{0}^{\tilde{E}_e-m_e-E_a}\frac{d^2P_{e\to e\gamma a}(\tilde{E}_\gamma,E_a,\tilde{E}_e,t)}{d\tilde{E}_\gamma d\tilde{E}_e}\right]
    \nonumber \\
    &\quad\; +\!\int_{m_e+m_{\gamma'}+E_a}^{E_0}\!d\tilde{E}_e\;\; I_e(\tilde{E}_e, t)
    \nonumber \\
    &\qquad \quad\;\; \times\left[\left(\frac{dP_{e\to e\gamma' a}(\tilde{E}_{\gamma'},E_a,\tilde{E}_e,t)}{d\tilde{E}_{\gamma'}}\right)_{\tilde{E}_{\gamma'}=\tilde{E}_e-m_e-E_a}\right.
    \nonumber \\
    &\qquad \quad\;\; \quad\;\;\, \left.+\int_{m_{\gamma'}}^{\tilde{E}_e-m_e-E_a}\frac{d^2P_{e\to e\gamma' a}(\tilde{E}_{\gamma'},E_a,\tilde{E}_e,t)}{d\tilde{E}_{\gamma'} d\tilde{E}_e}\right]
\,.
\end{align}
\end{subequations}
In the above, $P_{\gamma\to ee}(E_f,E_i,t)$ is the NPP probability for a photon with energy $E_i$ to produce an electron-positron pair with electron energy $E_f$, $P_{e\to e\gamma'}(E_f,E_i,t)$ is the NCS (or NCS-type) probability of an electron with energy $E_i$ to scatter and generate a photon (DP) with energy $E_f$, and $P_{e\to e\gamma^{(\prime)}a}(E_{f_{\gamma^{(\prime)}}},E_{f_a},E_i,t)$ is the probability of an electron with energy $E_i$ to undergo an off-shell emission process like in Figs.~\ref{fig:primNPprod}(b,c), producing a photon (DP) and ALP with energies $E_{f_{\gamma^{(\prime)}}}$ and $E_{f_a}$ respectively. The time dependence of these probabilities is an artifact of the time dependence of the background field felt by the particles.

The cascade equations can be understood as follows: the change in the electron intensity $I_e(E_e, t)$ at time $t$ is determined by the influx of electrons at energy $E_e$ due to more energetic electrons undergoing NCS/NCS-type scattering and off-shell emission processes, the efflux of electrons at energy $E_e$ undergoing similar processes, and the production of electrons at energy $E_e$ due to NPP. Likewise, the change in photon intensity $I_\gamma(E_\gamma, t)$ at a time $t$ is determined by the loss of photons at energy $E_\gamma$ due to NPP, and the production of photons at energy $E_\gamma$ due to electrons and positrons at energy $>E_\gamma$ undergoing NCS or off-shell emission. The change in DP and ALP intensities are determined by their production from electrons at higher energies. Note that here we have ignored the feedback of NCS and reverse off-shell processes from DPs, ALPs, and positrons due to their negligible impact on the dynamics.  

Implicit in the above equations is the locally monochromatic approximation (LMA), which says that the background field seen by the particles is slowly varying compared to the timescale of emission~\cite{LMAapproximation}. Also neglected above is the small amount of energy provided by the background field to stimulate NCS-type and NPP processes. For example, momentum conservation for the $n$th emission harmonic in NCS requires $q_0 + n\omega = q'_0 + \omega'$, where $q=p+e^2\abs{A}^2k/(2kp)$ (similarly $q'$) is the initial (final) effective momentum of an electron that is propagating through the background field with momentum $p$ ($p'$), and $\omega=k_0$ ($\omega'=k'_0$). Since we are interested in electron beams with energy $\sim\numrange{10}{100}\,\GeV$ in a $\numrange{10^2}{10^3}\,\nm$~($\numrange{1}{10}\,\eV$) laser, and the emission rates are exponentially suppressed in $n$, we can approximate the photon emission energy as $\omega'\approx q_0 - q'_0$. The limits of integration in \cref{eq:Cascade} are therefore taken to be the initial beam energy, rather than infinity.

The NPP, NCS, and off-shell emission rates are computed via nonperturbative QED calculations similar to those conducted in Appendix B of Ref.~\cite{LUXENPOD}. For NPP $\gamma(k')\to e^-_V(p)e^+_V(p')$, we obtain
\begin{equation}
    \begin{aligned}
    \frac{dP_{\gamma\rightarrow e^+e^-}}{dq_0} \!=\! \frac{2\alpha m_e^2}{\omega'k_-}\!\!\sum_{n>n_{0,\mtxt{PP}}} & \! \left[\left(1 + \frac{r^2}{2}\right)J_n^2 \right.
    \\
    & \; -\frac{\eta^2}{2}\!\left(\!J_n^2\!-\!\frac{J_{n-1}^2}{2}\!-\!\frac{J_{n+1}^2}{2}\!\right) 
    \\
    & \;\qquad\qquad\;\; \left.\times\left(\frac{kp}{kp'}+\frac{kp'}{kp}\right)\right]
\,,
\end{aligned}
\end{equation}
where $\alpha$ is the fine-structure constant, $n_{0,\mtxt{PP}}=(m_\gamma^2+2m_\gamma m_e(1+\eta^2))/(2kp)$, $r=m_\gamma/m_e$, and $J_n(z)$ are Bessel functions with argument 
\begin{equation}
    z= e|A|\sqrt{\frac{2n(kk')}{(kq)(kq')} - \frac{m_e^2(1 + \eta^2)(kk')^2}{(kq)^2(kq')^2} - \frac{m_\gamma^2}{(kp)(kp')}}
\,.
\end{equation}
For NCS $e^-_V(p)\to e^-_V(p')\gamma(k')$, we similarly have
\begin{equation}
\label{eq:NCSrate_photon}
\begin{aligned}
    \frac{dP_{e\to e\gamma}}{d\omega'} \!=\! \frac{2\alpha m_e^2}{q_0q_-} \!\sum_{n>n_{0,\mtxt{CS}}} & \! \left[\left(1 + \frac{r^2}{2}\right)J_n^2\right. 
    \\
    & \; +\frac{\eta^2}{2}\left(J_n^2-\frac{J_{n-1}^2}{2}-\frac{J_{n+1}^2}{2}\right) \\ 
    & \;\qquad\;\; \left.\times\left(\frac{kp}{kp'}+\frac{kp'}{kp}\right)\;\right]
\,,
\end{aligned}
\end{equation}
where $n_{0,\mtxt{CS}}=(m_X^2+2m_X m_e(1+\eta^2))/(2kp)$ is the kinematic limit for producing a particle of mass $m_X$ via NCS. The rate for the NCS-type process $e^-_V\to e^-_V\gamma'$ is straightforwardly obtained by substituting $\alpha \to \varepsilon^2\alpha$ and $m_\gamma \to m_{\gamma'}$ in \cref{eq:NCSrate_photon} and in the argument of the Bessel functions. For the off-shell emission of ALPs $e^-_V\to e^-_V\gamma a$, we refer to Eq.~(B23) in Ref.~\cite{LUXENPOD}. Lastly, we consider the simultaneous off-shell emission of DPs and ALPs $e^-_V\to e^-_V\gamma' a$; in the limit of light DPs, this rate is obtained from the ALP-only off-shell emission via the substitutions $g_{a\gamma\gamma}\to g_{a\gamma\gamma'}$ and $m_\gamma\to m_{\gamma'}$, while for non-negligible $m_{\gamma'}$ the final-state phase space is greatly reduced.

Plugging in the various rates $dP_{i\to f}/dE$, one can numerically solve Eqs.~(\ref{eq:Cascade}) with the initial conditions
\begin{equation}
\label{eq:initial_conditions}
\begin{aligned}
    I_e(E_e, 0) &= \delta(E_e-E_0)
\,,
    \\
    I_\gamma(E_\gamma, 0) &= I_{\gamma'}(E_{\gamma'}, 0) = I_a(E_a,0) = 0
\,.
\end{aligned}
\end{equation}
The differential $Y$ flux ($Y=\gamma,\gamma',a$) per incident electron is obtained by taking
\begin{equation}
    \frac{dN_Y}{dE_Y} = I_Y(E_Y,\infty)
\,.
\end{equation}
To accurately model the spectra produced at LUXE, we account for the spatial variance of the electron beam by discretizing its distribution into suitably fine bins and computing the spectra resulting from each bin independently. The cascade equations differ across bins beyond simple rescaling due to the differing field profile felt by each point. The beam and optical dump profiles are assumed to collide with intersecting centers of mass at $17.2^\circ$. The total spectrum $dN_Y/dE_Y$ is obtained by summing over the spectra produced by each bin, weighted by the beam intensity profile. We neglect dispersion effects and apply both the LMA and collinearity approximation.

\begin{figure}[t]
    \centering
    \includegraphics[width=0.48\textwidth]{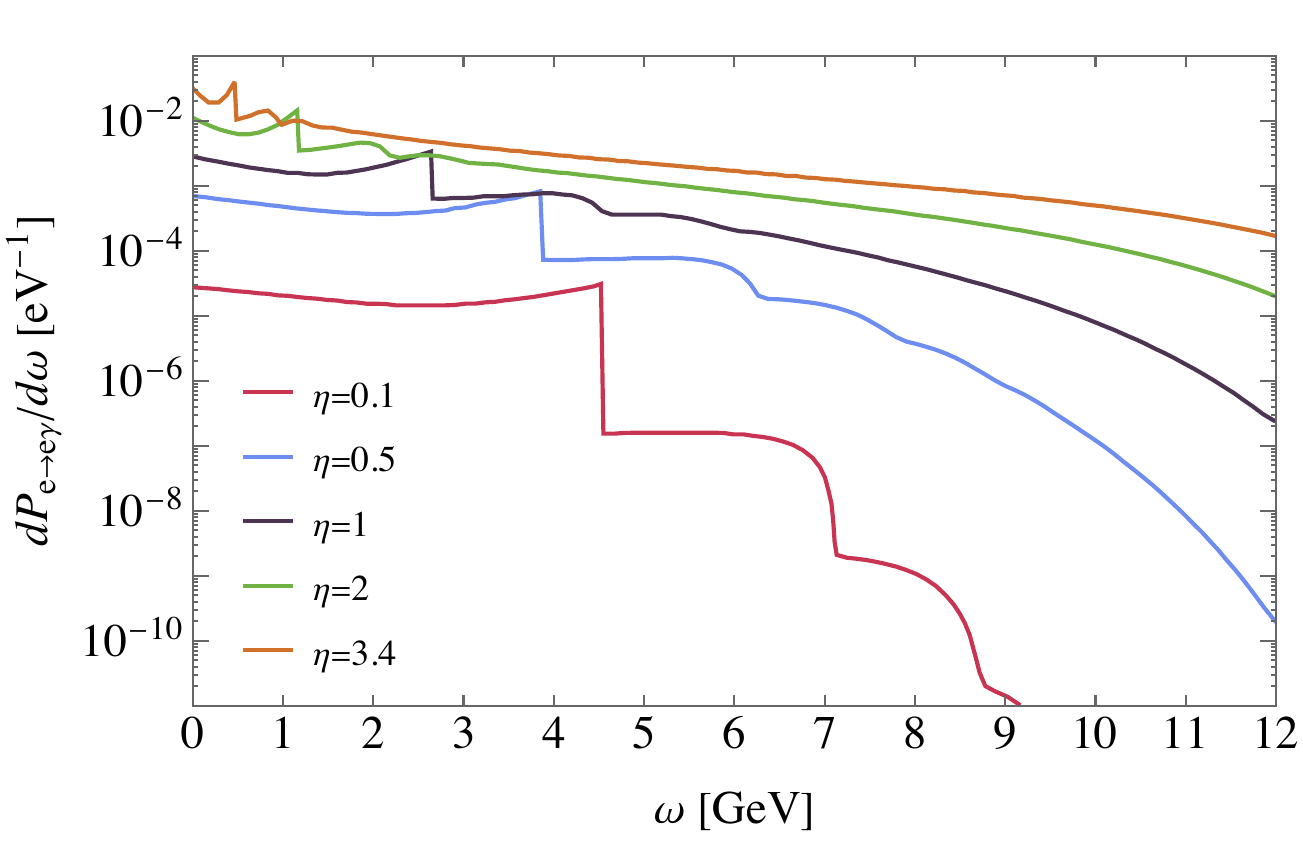}
    \includegraphics[width=0.48\textwidth]{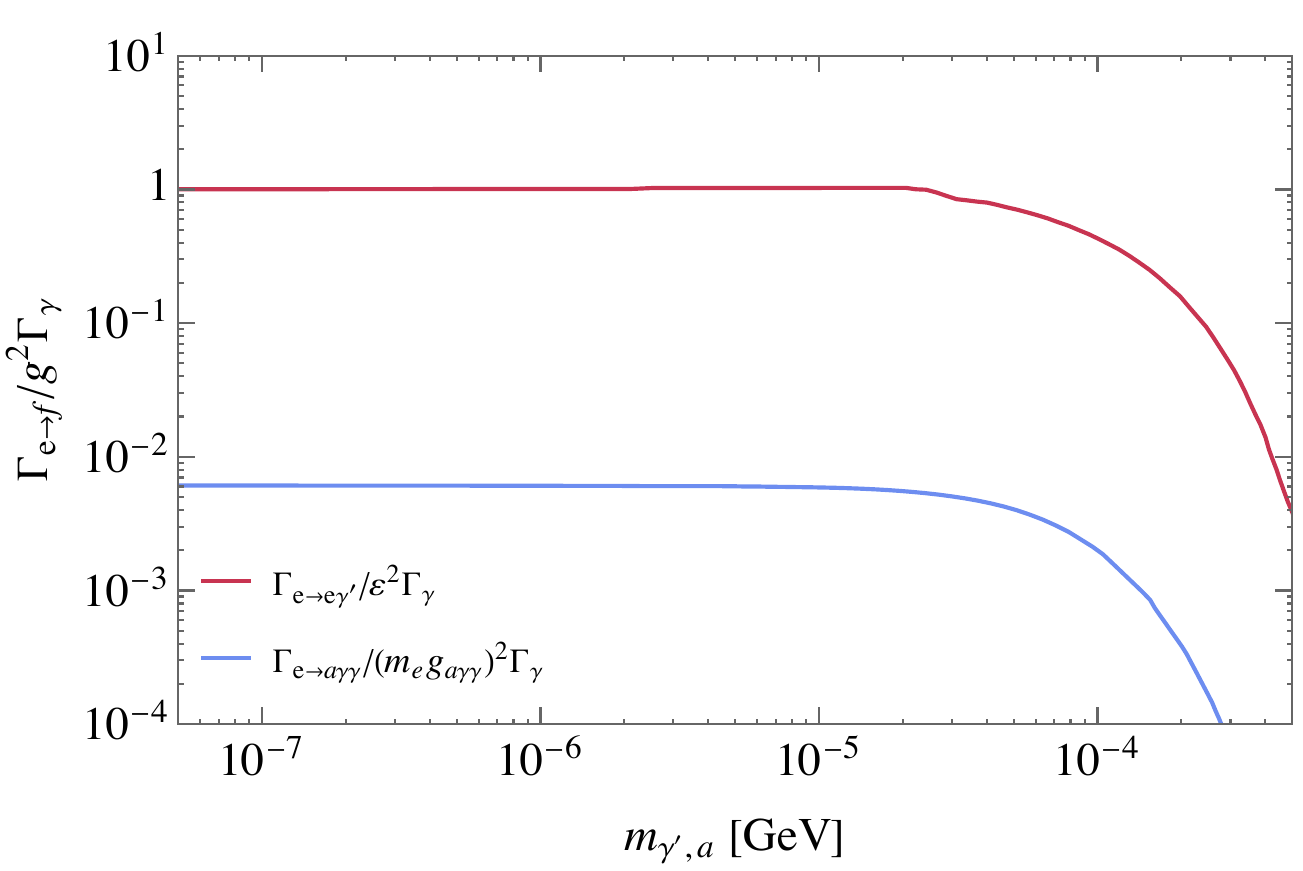}
    \caption{Rates for processes in strong-field QED.
    Top: the differential rate per unit energy for nonlinear Compton scattering from an incoming electron of energy 16.5~GeV for various field strengths/quantum nonlinearity parameters. Bottom: new physics production rates for a massive  dark photon (red) and an axionlike particle (blue) for varying NP particle mass, normalized by their respective couplings and rate of nonlinear Compton scattering $\Gamma_\gamma^{-1}\equiv\Gamma_{e\rightarrow e\gamma}^{-1}\approx 12~\text{fs}$.}
\label{fig:rates}
\end{figure}

\Cref{fig:rates} shows plots of the NCS, NCS-type, and off-shell emission rates for various parameters ($\eta,m_{\gamma'},m_a$).

%% file: sections/Appendix_B-Secondary_NP_Differential_Cross_Sections.tex
For secondary NP production processes, we perform an exact tree-level computation of the differential cross sections for all the possibilities appearing in \cref{eq:secNPprod}. The relevant Feynman diagrams contributing to each process are presented in \cref{fig:secNPprod}.

As seen in \cref{fig:secNPprod}, all secondary production avenues arise from elastic 2-to-2 scattering processes, where the incoming particles are a primary-produced particle ($\gamma$, $\gamma'$, or $a$) and a dump particle (dump electron $e^-$ or dump atom $N$), and the outgoing particles are a secondary-produced particle ($\gamma'$ or $a$) and a dump particle (same as incoming dump particle). 

We denote the incoming momenta by $k$, outgoing momenta by $p$, and indicate primary (secondary) particles by $Y$ ($X$) subscripts and dump particles by a $D$ subscript. Thus, for any of the secondary NP processes described above, the laboratory-frame 4-momenta are given by
\begin{equation}
\begin{aligned}[b]
    k_Y &= \left(E_Y,0,0,\sqrt{E_Y^2-m_Y^2}\right) \,, \\
    k_D &= \left(m_D,0,0,0\right) \,, \\
    p_X &= \left(E_X,\sin\theta\sqrt{E_X^2-m_X^2},0,\cos\theta\sqrt{E_X^2-m_X^2}\right) \,, \\
    p_D &= k_Y + k_D - p_X 
\,,
\end{aligned}
\end{equation}
where $\theta$ is the scattering angle. The transferred momentum is
\begin{equation}
    q = k_Y - p_X
\,.
\end{equation}
We use the standard Mandelstam variables
\begin{equation}
\begin{alignedat}[b]{2}
    s &= (k_Y + k_D)^2&& = m_Y^2 + m_D^2 + 2m_DE_Y \,, \\
    t &= q^2 = m_Y^2 +&& m_X^2 \\
    & &&\!- 2E_YE_X\sqrt{E_Y^2-m_Y^2}\sqrt{E_X^2-m_X^2}\cos\theta \,, \\
    u &= (k_D-p_X)^2&& = m_D^2 + m_X^2 - 2m_DE_X 
\,,
\end{alignedat}
\end{equation}
together with the identity $s+t+u=m_Y^2+m_X^2+2m_D^2$. Note that we are only considering elastic processes, so the incoming and outgoing dump particles are identical $k_D^2=p_D^2=m_D^2$.

With these definitions, the differential cross section $d\sigma_{Y\to X}/dt$ for any of the 2-to-2 processes we discussed is given by
\begin{equation}
    \frac{d\sigma_{Y\!\to\! X}}{dt} \!=\!
    \frac{\abs{\overline{\mathcal{M}}}^2}{16\pi\!\left(s^2\!+\!m_Y^4\!+\!m_D^4\!\!-\!2m_Y^{2}s\!-\!2m_D^2s\!-\!2m_Y^2m_D^2\right)}
\!\,,
\end{equation}
and the integration boundaries (assuming full $2\pi$ coverage of the forward direction in the center-of-mass frame) are
\begin{equation}
\label{eq:IntegrationLimits}
\begin{aligned}
    t_0 &= t_1+\frac{m_D\sqrt{E_Y^2-m_Y^2}}{m_Y^2+m_D^2+2m_DE_Y}\times \\
    &\!\!\!\!\!\sqrt{\!4m_D^2E_Y^2\!+\!4\left(m_Y^2\!\!-\!m_X^2\right)\!m_DE_Y\!+\!\left(m_Y^2\!\!-\!m_X^2\right)^2\!\!-\!4m_D^2m_X^2}
\,,
    \\
    t_1 &= \frac{-2m_D^2E_Y^2\!-\!\left(m_Y^2-m_X^2\right)\!m_DE_Y\!+\!m_D^2\!\left(m_Y^2\!+\!m_X^2\right)}{m_Y^2+m_D^2+2m_DE_Y}
\,.
\end{aligned}
\end{equation}
In the above, $\abs{\overline{\mathcal{M}}}^2$ is the spin/polarization-averaged squared amplitude, which is computed by evaluating the diagrams in \cref{fig:secNPprod}.

For processes involving $N$, we assume the dump atoms to be scalars with electric charge $Ze$ (here, $Z=74$ is the atomic number of tungsten) and model them using the normalized form factor
\begin{equation}
    F_0(t) = \frac{1}{Z}\sqrt{G_2^\mtxt{el}(t)+G_2^\mtxt{inel}(t)}
\,,
\end{equation}
where $G_2^\mtxt{el}$, $G_2^\mtxt{inel}$ are the elastic and inelastic atomic form factors (see Refs.~\cite{DPLimitEBeamDumps&FormFactors,FormFactors2}) defined by
\begin{align}
    G_2^\mtxt{el}(t) &= \left(\frac{a^2t}{1+a^2t}\right)^2 \left(\frac{1}{1+t/d}\right)^2 Z^2 \,, \\
    G_2^\mtxt{inel}(t) &= \left(\frac{{a'}^2t}{1+{a'}^2t}\right)^2 \left(\frac{1+\frac{t}{4m_p^2}\left(\mu_p^2-1\right)}{\left(1+\frac{t}{0.71\,\GeV^2}\right)^4}\right)^2 Z
\,,
\end{align}
with the parameters
\begin{equation}
\begin{aligned}
    a &= 111 Z^{-1/3} / m_e 
    \,, \\
    d &= 0.164\,\GeV^2 A_W^{-2/3}
    \,, \\
    a' &= 773 Z^{-2/3}/m_e
    \,, \\
    \mu_p &= 2.79
\end{aligned}
\end{equation}
($m_e$, $m_p$ are the electron and proton masses respectively).

The spin/polarization-averaged squared amplitudes are computed in Wolfram's \textit{Mathematica} software with the help of the \textsc{FeynCalc} package. We obtain the following differential cross sections:
\begin{subequations}
\begin{equation}
\begin{aligned}[b]
    &\frac{d\sigma_{\gamma\to\gamma'}}{dt}
    =
    \frac{2 \pi \alpha^2}{\left(s+t-m_e^2-m_{\gamma'}^2\right)^2}\times\left(s-m_e^2\right)^{-4}\times\varepsilon^2
    \\
    &\quad\times\!\smash[b]{\biggl[}\,4m_e^6\left(m_{\gamma'}^2\!-\!2s\right)+2m_e^8
    \\
    &\qquad\quad +m_e^4 \left(7m_{\gamma'}^4\!-\!4\left(3s\!+\!2t\right)m_{\gamma'}^2\!+\!12s^2\!+\!4st\!+\!3t^2\right)
    \\
    &\qquad\quad+s\left(s\!+\!t\!-\!m_{\gamma'}^2\right)\left(m_{\gamma'}^4\!-\!2m_{\gamma'}^2s\!+\!2s^2\!+\!2st\!+\!t^2\right)
    \\
    &\qquad\quad+m_e^2 \smash{\Bigl(}\left(12s^2\!+\!4st\!+\!3t^2\right)m_{\gamma'}^2\!-\!8s^3
    \\
    &\qquad\qquad\quad\quad\;
    -\!8s^2t\!-\!2st^2\!-\!t^3\!-\!\left(6s\!+\!5t\right)m_{\gamma'}^4\!+\!3m_{\gamma'}^6\smash[t]{\Bigr)\,\biggr]}
\,,
\end{aligned}
\end{equation}
\begin{equation}
\begin{aligned}[b]
    &\frac{d\sigma_{\gamma\to a}}{dt} 
    =
    \frac{8 \alpha Z^2 F_0^2(t)}{\left(4m_N^2-t\right)^2} \times\left(s-m_N^2\right)^{-2}
    \\
    &\quad\times\left( \frac{g_{a\gamma\gamma}^2}{4t^2} + \frac{\varepsilon g_{a\gamma\gamma} g_{a\gamma\gamma'}}{t\left(t-m_{\gamma'}^2\right)} + \frac{\varepsilon^2 g_{a\gamma\gamma'}^2}{\left(t-m_{\gamma'}^2\right)^2}\right) 
    \\
    &\quad\times m_N^4\left(m_a^2t\left(m_N^2\!+\!s\right)\!-\!m_a^4m_N^2\!-\!t\left(\left(m_N^2\!-\!s\right)^2\!+\!st\right)\right)
\,,
\end{aligned}
\end{equation}
\begin{equation}
\begin{aligned}[b]
    \frac{d\sigma_{\gamma'\to a}}{dt} 
    &=
    \frac{-16\alpha Z^2 F_0^2(t)}{3\left(4m_N^2-t\right)^2} 
    \\
    &\times\left(s^2\!+\!m_{\gamma'}^4\!+\!m_N^4\!-\!2m_{\gamma'}^2s\!-\!2m_N^2s\!-\!2m_N^2m_{\gamma'}^2\right)^{-1}
    \\
    &\times\left( \frac{g_{a\gamma\gamma'}^2}{t^2} + \frac{\varepsilon g_{a\gamma\gamma'} g_{a\gamma'\gamma'}}{t\left(t-m_{\gamma'}^2\right)} + \frac{\varepsilon^2 g_{a\gamma'\gamma'}^2}{4\left(t-m_{\gamma'}^2\right)^2}\right) 
    \\
    &\times m_N^4\smash[b]{\Bigl[}\,m_N^4t+m_a^4m_N^2
    \\
    &\qquad\quad\:\:-m_a^2\left(m_N^2\left(2m_{\gamma'}^2+t\right)+t\left(s-m_{\gamma'}^2\right)\right)
    \\
    &\qquad\quad\:\:+m_N^2\left(m_{\gamma'}^4-t\left(m_{\gamma'}^2+2s\right)\right)
    \\
    &\qquad\quad\:\:+st\left(s+t-m_{\gamma'}^2\right)\smash[t]{\Bigr]}
\,,
\end{aligned}
\end{equation}
\begin{equation}
\begin{aligned}[b]
    &\frac{d\sigma_{a \to \gamma'}}{dt} 
    =
    \frac{-16 \alpha Z^2 F_0^2(t)}{\left(4m_N^2-t\right)^2}
    \\
    &\quad\times\left(s^2+m_a^4+m_N^4-2m_a^2s-2m_N^2s-2m_N^2m_a^2\right)^{-1}
    \\
    &\quad\times 
    \left( \frac{g_{a\gamma\gamma'}^2}{t^2} + \frac{\varepsilon g_{a\gamma\gamma'} g_{a\gamma'\gamma'}}{t\left(t-m_{\gamma'}^2\right)} + \frac{\varepsilon^2 g_{a\gamma'\gamma'}^2}{4\left(t-m_{\gamma'}^2\right)^2}\right) 
    \\
    &\quad\times m_N^4\smash[b]{\Bigl[}\,m_N^4t\!+\!m_a^4m_N^2
    \\
    &\quad\qquad\quad\:\:-m_a^2\left(m_N^2\left(2m_{\gamma'}^2\!+\!t\right)\!+\!t\left(s\!-\!m_{\gamma'}^2\right)\right)
    \\
    &\quad\qquad\quad\:\:+\!m_N^2\left(m_{\gamma'}^4\!-\!t\left(m_{\gamma'}^2\!+\!2s\right)\right)\!+\!st\left(s\!+\!t\!-\!m_{\gamma'}^2\right)\smash[t]{\Bigr]}
\,.
\end{aligned}
\end{equation}
\end{subequations}
These are numerically integrated between the limits given in \cref{eq:IntegrationLimits} to achieve the result in \cref{eq:TotalSignals}.


%% file: main.bbl
%

%% file: main.bbl
\begin{thebibliography}{113}%
\makeatletter
\providecommand \@ifxundefined [1]{%
 \@ifx{#1\undefined}
}%
\providecommand \@ifnum [1]{%
 \ifnum #1\expandafter \@firstoftwo
 \else \expandafter \@secondoftwo
 \fi
}%
\providecommand \@ifx [1]{%
 \ifx #1\expandafter \@firstoftwo
 \else \expandafter \@secondoftwo
 \fi
}%
\providecommand \natexlab [1]{#1}%
\providecommand \enquote  [1]{``#1''}%
\providecommand \bibnamefont  [1]{#1}%
\providecommand \bibfnamefont [1]{#1}%
\providecommand \citenamefont [1]{#1}%
\providecommand \href@noop [0]{\@secondoftwo}%
\providecommand \href [0]{\begingroup \@sanitize@url \@href}%
\providecommand \@href[1]{\@@startlink{#1}\@@href}%
\providecommand \@@href[1]{\endgroup#1\@@endlink}%
\providecommand \@sanitize@url [0]{\catcode `\\12\catcode `\$12\catcode
  `\&12\catcode `\#12\catcode `\^12\catcode `\_12\catcode `\%12\relax}%
\providecommand \@@startlink[1]{}%
\providecommand \@@endlink[0]{}%
\providecommand \url  [0]{\begingroup\@sanitize@url \@url }%
\providecommand \@url [1]{\endgroup\@href {#1}{\urlprefix }}%
\providecommand \urlprefix  [0]{URL }%
\providecommand \Eprint [0]{\href }%
\providecommand \doibase [0]{https://doi.org/}%
\providecommand \selectlanguage [0]{\@gobble}%
\providecommand \bibinfo  [0]{\@secondoftwo}%
\providecommand \bibfield  [0]{\@secondoftwo}%
\providecommand \translation [1]{[#1]}%
\providecommand \BibitemOpen [0]{}%
\providecommand \bibitemStop [0]{}%
\providecommand \bibitemNoStop [0]{.\EOS\space}%
\providecommand \EOS [0]{\spacefactor3000\relax}%
\providecommand \BibitemShut  [1]{\csname bibitem#1\endcsname}%
\let\auto@bib@innerbib\@empty
\bibitem [{\citenamefont {Okun}(1982)}]{Paraphotons}%
  \BibitemOpen
  \bibfield  {author} {\bibinfo {author} {\bibfnamefont {L.~B.}\ \bibnamefont
  {Okun}},\ }\bibfield  {title} {\bibinfo {title} {{Limits on electrodynamics:
  paraphotons?}},\ }\href
  {https://jetp.ras.ru/cgi-bin/e/index/e/56/3/p502?a=list} {\bibfield
  {journal} {\bibinfo  {journal} {Soviet Physics JETP}\ }\textbf {\bibinfo
  {volume} {56}},\ \bibinfo {pages} {502} (\bibinfo {year} {1982})}\BibitemShut
  {NoStop}%
\bibitem [{\citenamefont {Holdom}(1986)}]{KineticMixing}%
  \BibitemOpen
  \bibfield  {author} {\bibinfo {author} {\bibfnamefont {B.}~\bibnamefont
  {Holdom}},\ }\bibfield  {title} {\bibinfo {title} {{Two $U(1)$'s and
  $\epsilon$ charge shifts}},\ }\href
  {https://doi.org/10.1016/0370-2693(86)91377-8} {\bibfield  {journal}
  {\bibinfo  {journal} {Physics Letters B}\ }\textbf {\bibinfo {volume}
  {166}},\ \bibinfo {pages} {196} (\bibinfo {year} {1986})}\BibitemShut
  {NoStop}%
\bibitem [{\citenamefont {Fabbrichesi}\ \emph {et~al.}(2021)\citenamefont
  {Fabbrichesi}, \citenamefont {Gabrielli},\ and\ \citenamefont
  {Lanfranchi}}]{DPBook}%
  \BibitemOpen
  \bibfield  {author} {\bibinfo {author} {\bibfnamefont {M.}~\bibnamefont
  {Fabbrichesi}}, \bibinfo {author} {\bibfnamefont {E.}~\bibnamefont
  {Gabrielli}},\ and\ \bibinfo {author} {\bibfnamefont {G.}~\bibnamefont
  {Lanfranchi}},\ }\href {https://doi.org/10.1007/978-3-030-62519-1} {\emph
  {\bibinfo {title} {{The Physics of the Dark Photon}}}}\ (\bibinfo
  {publisher} {Springer Cham},\ \bibinfo {year} {2021})\BibitemShut {NoStop}%
\bibitem [{\citenamefont {Peccei}\ and\ \citenamefont
  {Quinn}(1977{\natexlab{a}})}]{PecceiQuinnAxion}%
  \BibitemOpen
  \bibfield  {author} {\bibinfo {author} {\bibfnamefont {R.~D.}\ \bibnamefont
  {Peccei}}\ and\ \bibinfo {author} {\bibfnamefont {H.~R.}\ \bibnamefont
  {Quinn}},\ }\bibfield  {title} {\bibinfo {title} {{$CP$ Conservation in the
  Presence of Pseudoparticles}},\ }\href
  {https://doi.org/10.1103/PhysRevLett.38.1440} {\bibfield  {journal} {\bibinfo
   {journal} {Physical Review Letters}\ }\textbf {\bibinfo {volume} {38}},\
  \bibinfo {pages} {1440} (\bibinfo {year} {1977}{\natexlab{a}})}\BibitemShut
  {NoStop}%
\bibitem [{\citenamefont {Peccei}\ and\ \citenamefont
  {Quinn}(1977{\natexlab{b}})}]{PecceiQuinnAxion2}%
  \BibitemOpen
  \bibfield  {author} {\bibinfo {author} {\bibfnamefont {R.~D.}\ \bibnamefont
  {Peccei}}\ and\ \bibinfo {author} {\bibfnamefont {H.~R.}\ \bibnamefont
  {Quinn}},\ }\bibfield  {title} {\bibinfo {title} {{Constraints imposed by
  $CP$ conservation in the presence of pseudoparticles}},\ }\href
  {https://doi.org/10.1103/PhysRevD.16.1791} {\bibfield  {journal} {\bibinfo
  {journal} {Physical Review D}\ }\textbf {\bibinfo {volume} {16}},\ \bibinfo
  {pages} {1791} (\bibinfo {year} {1977}{\natexlab{b}})}\BibitemShut {NoStop}%
\bibitem [{\citenamefont {Wilczek}(1978)}]{WilczekAxion}%
  \BibitemOpen
  \bibfield  {author} {\bibinfo {author} {\bibfnamefont {F.}~\bibnamefont
  {Wilczek}},\ }\bibfield  {title} {\bibinfo {title} {{Problem of Strong $P$
  and $T$ Invariance in the Presence of Instantons}},\ }\href
  {https://doi.org/10.1103/PhysRevLett.40.279} {\bibfield  {journal} {\bibinfo
  {journal} {Physical Review Letters}\ }\textbf {\bibinfo {volume} {40}},\
  \bibinfo {pages} {279} (\bibinfo {year} {1978})}\BibitemShut {NoStop}%
\bibitem [{\citenamefont {Weinberg}(1978)}]{WeinbergAxion}%
  \BibitemOpen
  \bibfield  {author} {\bibinfo {author} {\bibfnamefont {S.}~\bibnamefont
  {Weinberg}},\ }\bibfield  {title} {\bibinfo {title} {{A New Light Boson?}},\
  }\href {https://doi.org/10.1103/PhysRevLett.40.223} {\bibfield  {journal}
  {\bibinfo  {journal} {Physical Review Letters}\ }\textbf {\bibinfo {volume}
  {40}},\ \bibinfo {pages} {223} (\bibinfo {year} {1978})}\BibitemShut
  {NoStop}%
\bibitem [{\citenamefont {Kim}(1979)}]{KSVZ}%
  \BibitemOpen
  \bibfield  {author} {\bibinfo {author} {\bibfnamefont {J.~E.}\ \bibnamefont
  {Kim}},\ }\bibfield  {title} {\bibinfo {title} {{Weak-Interaction Singlet and
  Strong $CP$ Invariance}},\ }\href
  {https://doi.org/10.1103/PhysRevLett.43.103} {\bibfield  {journal} {\bibinfo
  {journal} {Physical Review Letters}\ }\textbf {\bibinfo {volume} {43}},\
  \bibinfo {pages} {103} (\bibinfo {year} {1979})}\BibitemShut {NoStop}%
\bibitem [{\citenamefont {Shifman}\ \emph {et~al.}(1980)\citenamefont
  {Shifman}, \citenamefont {Vainshtein},\ and\ \citenamefont
  {Zakharov}}]{KSVZ2}%
  \BibitemOpen
  \bibfield  {author} {\bibinfo {author} {\bibfnamefont {M.~A.}\ \bibnamefont
  {Shifman}}, \bibinfo {author} {\bibfnamefont {A.~I.}\ \bibnamefont
  {Vainshtein}},\ and\ \bibinfo {author} {\bibfnamefont {V.~I.}\ \bibnamefont
  {Zakharov}},\ }\bibfield  {title} {\bibinfo {title} {{Can confinement ensure
  natural $CP$ invariance of strong interactions?}},\ }\href
  {https://doi.org/10.1016/0550-3213(80)90209-6} {\bibfield  {journal}
  {\bibinfo  {journal} {Nuclear Physics B}\ }\textbf {\bibinfo {volume}
  {166}},\ \bibinfo {pages} {493} (\bibinfo {year} {1980})}\BibitemShut
  {NoStop}%
\bibitem [{\citenamefont {Hook}(2019)}]{AxionsandALPsReview}%
  \BibitemOpen
  \bibfield  {author} {\bibinfo {author} {\bibfnamefont {A.}~\bibnamefont
  {Hook}},\ }\bibfield  {title} {\bibinfo {title} {{TASI Lectures on the Strong
  CP Problem and Axions}},\ }\href {https://doi.org/10.22323/1.333.0004}
  {\bibfield  {journal} {\bibinfo  {journal} {PoS}\ }\textbf {\bibinfo {volume}
  {TASI2018}},\ \bibinfo {pages} {004} (\bibinfo {year} {2019})}\BibitemShut
  {NoStop}%
\bibitem [{\citenamefont {Bartlett}\ and\ \citenamefont
  {L{\"o}gl}(1988)}]{DPLimitsCavendish}%
  \BibitemOpen
  \bibfield  {author} {\bibinfo {author} {\bibfnamefont {D.~F.}\ \bibnamefont
  {Bartlett}}\ and\ \bibinfo {author} {\bibfnamefont {S.}~\bibnamefont
  {L{\"o}gl}},\ }\bibfield  {title} {\bibinfo {title} {{Limits on an
  Electromagnetic Fifth Force}},\ }\href
  {https://doi.org/10.1103/PhysRevLett.61.2285} {\bibfield  {journal} {\bibinfo
   {journal} {Physical Review Letters}\ }\textbf {\bibinfo {volume} {61}},\
  \bibinfo {pages} {2285} (\bibinfo {year} {1988})}\BibitemShut {NoStop}%
\bibitem [{\citenamefont {Kroff}\ and\ \citenamefont
  {Malta}(2020)}]{DPLimitsPlimptonLawton&AtomicForceMicroscopy}%
  \BibitemOpen
  \bibfield  {author} {\bibinfo {author} {\bibfnamefont {D.}~\bibnamefont
  {Kroff}}\ and\ \bibinfo {author} {\bibfnamefont {P.~C.}\ \bibnamefont
  {Malta}},\ }\bibfield  {title} {\bibinfo {title} {{Constraining hidden
  photons via atomic force microscope measurements and the Plimpton-Lawton
  experiment}},\ }\href {https://doi.org/10.1103/PhysRevD.102.095015}
  {\bibfield  {journal} {\bibinfo  {journal} {Physical Review D}\ }\textbf
  {\bibinfo {volume} {102}},\ \bibinfo {pages} {095015} (\bibinfo {year}
  {2020})}\BibitemShut {NoStop}%
\bibitem [{\citenamefont {Jaeckel}\ and\ \citenamefont
  {Roy}(2010)}]{DPLimitSpectroscopy}%
  \BibitemOpen
  \bibfield  {author} {\bibinfo {author} {\bibfnamefont {J.}~\bibnamefont
  {Jaeckel}}\ and\ \bibinfo {author} {\bibfnamefont {S.}~\bibnamefont {Roy}},\
  }\bibfield  {title} {\bibinfo {title} {{Spectroscopy as a test of Coulomb's
  law: A probe of the hidden sector}},\ }\href
  {https://doi.org/10.1103/PhysRevD.82.125020} {\bibfield  {journal} {\bibinfo
  {journal} {Physical Review D}\ }\textbf {\bibinfo {volume} {82}},\ \bibinfo
  {pages} {125020} (\bibinfo {year} {2010})}\BibitemShut {NoStop}%
\bibitem [{\citenamefont {Marocco}(2021)}]{DPLimitsEarth}%
  \BibitemOpen
  \bibfield  {author} {\bibinfo {author} {\bibfnamefont {G.}~\bibnamefont
  {Marocco}},\ }\href@noop {} {\bibinfo {title} {{Dark photon limits from
  magnetic fields and astrophysical plasmas}}} (\bibinfo {year} {2021}),\
  \Eprint {https://arxiv.org/abs/2110.02875} {arXiv:2110.02875 [hep-ph]}
  \BibitemShut {NoStop}%
\bibitem [{\citenamefont {Yan}\ \emph {et~al.}(2024)\citenamefont {Yan},
  \citenamefont {Li},\ and\ \citenamefont {Fan}}]{DPLimitsJupiter}%
  \BibitemOpen
  \bibfield  {author} {\bibinfo {author} {\bibfnamefont {S.}~\bibnamefont
  {Yan}}, \bibinfo {author} {\bibfnamefont {L.}~\bibnamefont {Li}},\ and\
  \bibinfo {author} {\bibfnamefont {J.}~\bibnamefont {Fan}},\ }\bibfield
  {title} {\bibinfo {title} {{Constraints on photon mass and dark photon from
  the Jovian magnetic field}},\ }\href
  {https://doi.org/10.1007/JHEP06(2024)028} {\bibfield  {journal} {\bibinfo
  {journal} {Journal of High Energy Physics}\ }\textbf {\bibinfo {volume}
  {2024}},\ \bibinfo {pages} {028} (\bibinfo {year} {2024})}\BibitemShut
  {NoStop}%
\bibitem [{\citenamefont {McCarthy}\ \emph {et~al.}(2024)\citenamefont
  {McCarthy} \emph {et~al.}}]{COBE/FIRAS}%
  \BibitemOpen
  \bibfield  {author} {\bibinfo {author} {\bibfnamefont {F.}~\bibnamefont
  {McCarthy}} \emph {et~al.},\ }\bibfield  {title} {\bibinfo {title} {{Dark
  Photon Limits from Patchy Dark Screening of the Cosmic Microwave
  Background}},\ }\href {https://doi.org/10.1103/PhysRevLett.133.141003}
  {\bibfield  {journal} {\bibinfo  {journal} {Physical Review Letters}\
  }\textbf {\bibinfo {volume} {133}},\ \bibinfo {pages} {141003} (\bibinfo
  {year} {2024})}\BibitemShut {NoStop}%
\bibitem [{\citenamefont {Ehret}\ \emph {et~al.}(2010)\citenamefont {Ehret}
  \emph {et~al.}}]{DP/ALPLimitALPS}%
  \BibitemOpen
  \bibfield  {author} {\bibinfo {author} {\bibfnamefont {K.}~\bibnamefont
  {Ehret}} \emph {et~al.} (\bibinfo {collaboration} {ALPS Collaboration}),\
  }\bibfield  {title} {\bibinfo {title} {{New ALPS results on hidden-sector
  lightweights}},\ }\href {https://doi.org/10.1016/j.physletb.2010.04.066}
  {\bibfield  {journal} {\bibinfo  {journal} {Physics Letters B}\ }\textbf
  {\bibinfo {volume} {689}},\ \bibinfo {pages} {149} (\bibinfo {year}
  {2010})}\BibitemShut {NoStop}%
\bibitem [{\citenamefont {Inada}\ \emph {et~al.}(2013)\citenamefont {Inada}
  \emph {et~al.}}]{DPLimitSPring-8}%
  \BibitemOpen
  \bibfield  {author} {\bibinfo {author} {\bibfnamefont {T.}~\bibnamefont
  {Inada}} \emph {et~al.},\ }\bibfield  {title} {\bibinfo {title} {{Results of
  a search for paraphotons with intense X-ray beams at SPring-8}},\ }\href
  {https://doi.org/10.1016/j.physletb.2013.04.033} {\bibfield  {journal}
  {\bibinfo  {journal} {Physics Letters B}\ }\textbf {\bibinfo {volume}
  {722}},\ \bibinfo {pages} {301} (\bibinfo {year} {2013})}\BibitemShut
  {NoStop}%
\bibitem [{\citenamefont {Parker}\ \emph {et~al.}(2013)\citenamefont {Parker},
  \citenamefont {Hartnett}, \citenamefont {Povey},\ and\ \citenamefont
  {Tobar}}]{DPLimitsUWA}%
  \BibitemOpen
  \bibfield  {author} {\bibinfo {author} {\bibfnamefont {S.~R.}\ \bibnamefont
  {Parker}}, \bibinfo {author} {\bibfnamefont {J.~G.}\ \bibnamefont
  {Hartnett}}, \bibinfo {author} {\bibfnamefont {R.~G.}\ \bibnamefont
  {Povey}},\ and\ \bibinfo {author} {\bibfnamefont {M.~E.}\ \bibnamefont
  {Tobar}},\ }\bibfield  {title} {\bibinfo {title} {{Cryogenic resonant
  microwave cavity searches for hidden sector photons}},\ }\href
  {https://doi.org/10.1103/PhysRevD.88.112004} {\bibfield  {journal} {\bibinfo
  {journal} {Physical Review D}\ }\textbf {\bibinfo {volume} {88}},\ \bibinfo
  {pages} {112004} (\bibinfo {year} {2013})}\BibitemShut {NoStop}%
\bibitem [{\citenamefont {Wagner}\ \emph {et~al.}(2010)\citenamefont {Wagner}
  \emph {et~al.}}]{DPLimitsADMX}%
  \BibitemOpen
  \bibfield  {author} {\bibinfo {author} {\bibfnamefont {A.}~\bibnamefont
  {Wagner}} \emph {et~al.},\ }\bibfield  {title} {\bibinfo {title} {{Search for
  Hidden Sector Photons with the ADMX Detector}},\ }\href
  {https://doi.org/10.1103/PhysRevLett.105.171801} {\bibfield  {journal}
  {\bibinfo  {journal} {Physical Review Letters}\ }\textbf {\bibinfo {volume}
  {105}},\ \bibinfo {pages} {171801} (\bibinfo {year} {2010})}\BibitemShut
  {NoStop}%
\bibitem [{\citenamefont {Betz}\ \emph {et~al.}(2013)\citenamefont {Betz} \emph
  {et~al.}}]{DPLimitsCROWS}%
  \BibitemOpen
  \bibfield  {author} {\bibinfo {author} {\bibfnamefont {M.}~\bibnamefont
  {Betz}} \emph {et~al.},\ }\bibfield  {title} {\bibinfo {title} {{First
  results of the CERN Resonant Weakly Interacting Sub-eV Particle Search
  (CROWS)}},\ }\href {https://doi.org/10.1103/PhysRevD.88.075014} {\bibfield
  {journal} {\bibinfo  {journal} {Physical Review D}\ }\textbf {\bibinfo
  {volume} {88}},\ \bibinfo {pages} {075014} (\bibinfo {year}
  {2013})}\BibitemShut {NoStop}%
\bibitem [{\citenamefont {Redondo}(2008)}]{DPLimitsCAST}%
  \BibitemOpen
  \bibfield  {author} {\bibinfo {author} {\bibfnamefont {J.}~\bibnamefont
  {Redondo}},\ }\bibfield  {title} {\bibinfo {title} {{Helioscope bounds on
  hidden sector photons}},\ }\href
  {https://doi.org/10.1088/1475-7516/2008/07/008} {\bibfield  {journal}
  {\bibinfo  {journal} {Journal of Cosmology and Astroparticle Physics}\
  }\textbf {\bibinfo {volume} {2008}}\bibinfo  {number} { (07)},\ \bibinfo
  {pages} {008}}\BibitemShut {NoStop}%
\bibitem [{\citenamefont {Schwarz}\ \emph {et~al.}(2015)\citenamefont {Schwarz}
  \emph {et~al.}}]{DPLimitsSHIPS}%
  \BibitemOpen
\bibfield  {number} {  }\bibfield  {author} {\bibinfo {author} {\bibfnamefont
  {M.}~\bibnamefont {Schwarz}} \emph {et~al.},\ }\bibfield  {title} {\bibinfo
  {title} {{Results from the Solar Hidden Photon Search (SHIPS)}},\ }\href
  {https://doi.org/10.1088/1475-7516/2015/08/011} {\bibfield  {journal}
  {\bibinfo  {journal} {Journal of Cosmology and Astroparticle Physics}\
  }\textbf {\bibinfo {volume} {2015}}\bibinfo  {number} { (08)},\ \bibinfo
  {pages} {011}}\BibitemShut {NoStop}%
\bibitem [{\citenamefont {Danilov}\ \emph {et~al.}(2019)\citenamefont
  {Danilov}, \citenamefont {Demidov},\ and\ \citenamefont
  {Gorbunov}}]{DPLimitTEXONO}%
  \BibitemOpen
\bibfield  {number} {  }\bibfield  {author} {\bibinfo {author} {\bibfnamefont
  {M.}~\bibnamefont {Danilov}}, \bibinfo {author} {\bibfnamefont
  {S.}~\bibnamefont {Demidov}},\ and\ \bibinfo {author} {\bibfnamefont
  {D.}~\bibnamefont {Gorbunov}},\ }\bibfield  {title} {\bibinfo {title}
  {{Constraints on Hidden Photons Produced in Nuclear Reactors}},\ }\href
  {https://doi.org/10.1103/PhysRevLett.122.041801} {\bibfield  {journal}
  {\bibinfo  {journal} {Physical Review Letters}\ }\textbf {\bibinfo {volume}
  {122}},\ \bibinfo {pages} {041801} (\bibinfo {year} {2019})}\BibitemShut
  {NoStop}%
\bibitem [{\citenamefont {Gninenko}(2012)}]{DPLimitCHARM}%
  \BibitemOpen
  \bibfield  {author} {\bibinfo {author} {\bibfnamefont {S.~N.}\ \bibnamefont
  {Gninenko}},\ }\bibfield  {title} {\bibinfo {title} {{Constraints on Sub-GeV
  hidden sector gauge bosons from a search for heavy neutrino decays}},\ }\href
  {https://doi.org/10.1016/j.physletb.2012.06.002} {\bibfield  {journal}
  {\bibinfo  {journal} {Physics Letters B}\ }\textbf {\bibinfo {volume}
  {713}},\ \bibinfo {pages} {244} (\bibinfo {year} {2012})}\BibitemShut
  {NoStop}%
\bibitem [{\citenamefont {Bjorken}\ \emph {et~al.}(2009)\citenamefont
  {Bjorken}, \citenamefont {Essig}, \citenamefont {Schuster},\ and\
  \citenamefont {Toro}}]{DPLimitEBeamDumps&FormFactors}%
  \BibitemOpen
  \bibfield  {author} {\bibinfo {author} {\bibfnamefont {J.~D.}\ \bibnamefont
  {Bjorken}}, \bibinfo {author} {\bibfnamefont {R.}~\bibnamefont {Essig}},
  \bibinfo {author} {\bibfnamefont {P.}~\bibnamefont {Schuster}},\ and\
  \bibinfo {author} {\bibfnamefont {N.}~\bibnamefont {Toro}},\ }\bibfield
  {title} {\bibinfo {title} {{New fixed-target experiments to search for dark
  gauge forces}},\ }\href {https://doi.org/10.1103/PhysRevD.80.075018}
  {\bibfield  {journal} {\bibinfo  {journal} {Physical Review D}\ }\textbf
  {\bibinfo {volume} {80}},\ \bibinfo {pages} {075018} (\bibinfo {year}
  {2009})}\BibitemShut {NoStop}%
\bibitem [{\citenamefont {Bjorken}\ \emph {et~al.}(1988)\citenamefont {Bjorken}
  \emph {et~al.}}]{DPLimitE137}%
  \BibitemOpen
  \bibfield  {author} {\bibinfo {author} {\bibfnamefont {J.~D.}\ \bibnamefont
  {Bjorken}} \emph {et~al.},\ }\bibfield  {title} {\bibinfo {title} {{Search
  for neutral metastable penetrating particles produced in the SLAC beam
  dump}},\ }\href {https://doi.org/10.1103/PhysRevD.38.3375} {\bibfield
  {journal} {\bibinfo  {journal} {Physical Review D}\ }\textbf {\bibinfo
  {volume} {38}},\ \bibinfo {pages} {3375} (\bibinfo {year}
  {1988})}\BibitemShut {NoStop}%
\bibitem [{\citenamefont {Riordan}\ \emph
  {et~al.}(1987{\natexlab{a}})\citenamefont {Riordan} \emph
  {et~al.}}]{DPLimitE141}%
  \BibitemOpen
  \bibfield  {author} {\bibinfo {author} {\bibfnamefont {E.~M.}\ \bibnamefont
  {Riordan}} \emph {et~al.},\ }\bibfield  {title} {\bibinfo {title} {{Search
  for short-lived axions in an electron-beam-dump experiment}},\ }\href
  {https://doi.org/10.1103/PhysRevLett.59.755} {\bibfield  {journal} {\bibinfo
  {journal} {Physical Review Letters}\ }\textbf {\bibinfo {volume} {59}},\
  \bibinfo {pages} {755} (\bibinfo {year} {1987}{\natexlab{a}})}\BibitemShut
  {NoStop}%
\bibitem [{\citenamefont {Bross}\ \emph {et~al.}(1991)\citenamefont {Bross}
  \emph {et~al.}}]{DPLimitE774}%
  \BibitemOpen
  \bibfield  {author} {\bibinfo {author} {\bibfnamefont {A.}~\bibnamefont
  {Bross}} \emph {et~al.},\ }\bibfield  {title} {\bibinfo {title} {{Search for
  short-lived particles produced in an electron beam dump}},\ }\href
  {https://doi.org/10.1103/PhysRevLett.67.2942} {\bibfield  {journal} {\bibinfo
   {journal} {Physical Review Letters}\ }\textbf {\bibinfo {volume} {67}},\
  \bibinfo {pages} {2942} (\bibinfo {year} {1991})}\BibitemShut {NoStop}%
\bibitem [{\citenamefont {Endo}\ \emph {et~al.}(2012)\citenamefont {Endo},
  \citenamefont {Hamaguchi},\ and\ \citenamefont {Mishima}}]{DPLimitMuonG-2}%
  \BibitemOpen
  \bibfield  {author} {\bibinfo {author} {\bibfnamefont {M.}~\bibnamefont
  {Endo}}, \bibinfo {author} {\bibfnamefont {K.}~\bibnamefont {Hamaguchi}},\
  and\ \bibinfo {author} {\bibfnamefont {G.}~\bibnamefont {Mishima}},\
  }\bibfield  {title} {\bibinfo {title} {{Constraints on hidden photon models
  from electron $g-2$ and hydrogen spectroscopy}},\ }\href
  {https://doi.org/10.1103/PhysRevD.86.095029} {\bibfield  {journal} {\bibinfo
  {journal} {Physical Review D}\ }\textbf {\bibinfo {volume} {86}},\ \bibinfo
  {pages} {095029} (\bibinfo {year} {2012})}\BibitemShut {NoStop}%
\bibitem [{\citenamefont {Barducci}\ \emph {et~al.}(2021)\citenamefont
  {Barducci}, \citenamefont {Bertuzzo}, \citenamefont {Grilli~di Cortona},\
  and\ \citenamefont {Salla}}]{DPLimitLSND}%
  \BibitemOpen
  \bibfield  {author} {\bibinfo {author} {\bibfnamefont {D.}~\bibnamefont
  {Barducci}}, \bibinfo {author} {\bibfnamefont {E.}~\bibnamefont {Bertuzzo}},
  \bibinfo {author} {\bibfnamefont {G.}~\bibnamefont {Grilli~di Cortona}},\
  and\ \bibinfo {author} {\bibfnamefont {G.~M.}\ \bibnamefont {Salla}},\
  }\bibfield  {title} {\bibinfo {title} {{Dark photon bounds in the dark
  EFT}},\ }\href {https://doi.org/10.1007/JHEP12(2021)081} {\bibfield
  {journal} {\bibinfo  {journal} {Journal of High Energy Physics}\ }\textbf
  {\bibinfo {volume} {2021}},\ \bibinfo {pages} {081} (\bibinfo {year}
  {2021})}\BibitemShut {NoStop}%
\bibitem [{\citenamefont {Batley}\ \emph {et~al.}(2015)\citenamefont {Batley}
  \emph {et~al.}}]{DPLimitNA48/2}%
  \BibitemOpen
  \bibfield  {author} {\bibinfo {author} {\bibfnamefont {J.~R.}\ \bibnamefont
  {Batley}} \emph {et~al.} (\bibinfo {collaboration} {NA48/2 Collaboration}),\
  }\bibfield  {title} {\bibinfo {title} {{Search for the dark photon in $\pi^0$
  decays}},\ }\href@noop {} {\bibfield  {journal} {\bibinfo  {journal} {Physics
  Letters B}\ }\textbf {\bibinfo {volume} {746}},\ \bibinfo {pages} {178}
  (\bibinfo {year} {2015})}\BibitemShut {NoStop}%
\bibitem [{\citenamefont {Adlarson}\ \emph {et~al.}(2013)\citenamefont
  {Adlarson} \emph {et~al.}}]{DPLimitWASA}%
  \BibitemOpen
  \bibfield  {author} {\bibinfo {author} {\bibfnamefont {P.}~\bibnamefont
  {Adlarson}} \emph {et~al.} (\bibinfo {collaboration} {WASA-at-COSY
  Collaboration}),\ }\bibfield  {title} {\bibinfo {title} {{Search for a dark
  photon in the $\pi^0\to e^+e^-\gamma$ decay}},\ }\href
  {https://doi.org/10.1016/j.physletb.2013.08.055} {\bibfield  {journal}
  {\bibinfo  {journal} {Physics Letters B}\ }\textbf {\bibinfo {volume}
  {726}},\ \bibinfo {pages} {187} (\bibinfo {year} {2013})}\BibitemShut
  {NoStop}%
\bibitem [{\citenamefont {Abrahamyan}\ \emph {et~al.}(2011)\citenamefont
  {Abrahamyan} \emph {et~al.}}]{DPLimitAPEX}%
  \BibitemOpen
  \bibfield  {author} {\bibinfo {author} {\bibfnamefont {S.}~\bibnamefont
  {Abrahamyan}} \emph {et~al.},\ }\bibfield  {title} {\bibinfo {title} {{Search
  for a New Gauge Boson in Electron-Nucleus Fixed-Target Scattering by the APEX
  Experiment}},\ }\href {https://doi.org/10.1103/PhysRevLett.107.191804}
  {\bibfield  {journal} {\bibinfo  {journal} {Physical Review Letters}\
  }\textbf {\bibinfo {volume} {107}},\ \bibinfo {pages} {191804} (\bibinfo
  {year} {2011})}\BibitemShut {NoStop}%
\bibitem [{\citenamefont {Merkel}\ \emph {et~al.}(2011)\citenamefont {Merkel}
  \emph {et~al.}}]{DPLimitMAMI}%
  \BibitemOpen
  \bibfield  {author} {\bibinfo {author} {\bibfnamefont {H.}~\bibnamefont
  {Merkel}} \emph {et~al.} (\bibinfo {collaboration} {A1 Collaboration}),\
  }\bibfield  {title} {\bibinfo {title} {{Search for Light Gauge Bosons of the
  Dark Sector at the Mainz Microtron}},\ }\href
  {https://doi.org/10.1103/PhysRevLett.106.251802} {\bibfield  {journal}
  {\bibinfo  {journal} {Physical Review Letters}\ }\textbf {\bibinfo {volume}
  {106}},\ \bibinfo {pages} {251802} (\bibinfo {year} {2011})}\BibitemShut
  {NoStop}%
\bibitem [{\citenamefont {Lees}\ \emph {et~al.}(2014)\citenamefont {Lees} \emph
  {et~al.}}]{DPLimitBaBar}%
  \BibitemOpen
  \bibfield  {author} {\bibinfo {author} {\bibfnamefont {J.~P.}\ \bibnamefont
  {Lees}} \emph {et~al.},\ }\bibfield  {title} {\bibinfo {title} {{Search for a
  Dark Photon in $e^+e^-$ Collisions at BABAR}},\ }\href
  {https://doi.org/10.1103/PhysRevLett.113.201801} {\bibfield  {journal}
  {\bibinfo  {journal} {Physical Review Letters}\ }\textbf {\bibinfo {volume}
  {113}},\ \bibinfo {pages} {201801} (\bibinfo {year} {2014})}\BibitemShut
  {NoStop}%
\bibitem [{\citenamefont {Babusci}\ \emph {et~al.}(2013)\citenamefont {Babusci}
  \emph {et~al.}}]{DPLimitKLOE}%
  \BibitemOpen
  \bibfield  {author} {\bibinfo {author} {\bibfnamefont {D.}~\bibnamefont
  {Babusci}} \emph {et~al.} (\bibinfo {collaboration} {KLOE-2 Collaboration}),\
  }\bibfield  {title} {\bibinfo {title} {{Limit on the production of a light
  vector gauge boson in $\phi$ meson decays with the KLOE detector}},\ }\href
  {https://doi.org/10.1016/j.physletb.2013.01.067} {\bibfield  {journal}
  {\bibinfo  {journal} {Physics Letters B}\ }\textbf {\bibinfo {volume}
  {720}},\ \bibinfo {pages} {111} (\bibinfo {year} {2013})}\BibitemShut
  {NoStop}%
\bibitem [{\citenamefont {Abreu}\ \emph {et~al.}(2024)\citenamefont {Abreu}
  \emph {et~al.}}]{DPLimitFASER}%
  \BibitemOpen
  \bibfield  {author} {\bibinfo {author} {\bibfnamefont {H.}~\bibnamefont
  {Abreu}} \emph {et~al.} (\bibinfo {collaboration} {FASER Collaboration}),\
  }\bibfield  {title} {\bibinfo {title} {{Search for dark photons with the
  FASER detector at the LHC}},\ }\href
  {https://doi.org/10.1016/j.physletb.2023.138378} {\bibfield  {journal}
  {\bibinfo  {journal} {Physics Letters B}\ }\textbf {\bibinfo {volume}
  {848}},\ \bibinfo {pages} {138378} (\bibinfo {year} {2024})}\BibitemShut
  {NoStop}%
\bibitem [{\citenamefont {Andriamonje}\ \emph {et~al.}(2007)\citenamefont
  {Andriamonje} \emph {et~al.}}]{ALPLimitCAST}%
  \BibitemOpen
  \bibfield  {author} {\bibinfo {author} {\bibfnamefont {S.}~\bibnamefont
  {Andriamonje}} \emph {et~al.} (\bibinfo {collaboration} {CAST
  Collaboration}),\ }\bibfield  {title} {\bibinfo {title} {{An improved limit
  on the axion–photon coupling from the CAST experiment}},\ }\href
  {https://doi.org/10.1088/1475-7516/2007/04/010} {\bibfield  {journal}
  {\bibinfo  {journal} {Journal of Cosmology and Astroparticle Physics}\
  }\textbf {\bibinfo {volume} {2007}}\bibinfo  {number} { (04)},\ \bibinfo
  {pages} {010}}\BibitemShut {NoStop}%
\bibitem [{\citenamefont {Anastassopoulos}\ \emph {et~al.}(2017)\citenamefont
  {Anastassopoulos} \emph {et~al.}}]{ALPLimitCAST2}%
  \BibitemOpen
\bibfield  {number} {  }\bibfield  {author} {\bibinfo {author} {\bibfnamefont
  {V.}~\bibnamefont {Anastassopoulos}} \emph {et~al.} (\bibinfo {collaboration}
  {CAST Collaboration}),\ }\bibfield  {title} {\bibinfo {title} {{New CAST
  limit on the axion–photon interaction}},\ }\href
  {https://doi.org/10.1038/nphys4109} {\bibfield  {journal} {\bibinfo
  {journal} {Nature Physics}\ }\textbf {\bibinfo {volume} {13}},\ \bibinfo
  {pages} {584} (\bibinfo {year} {2017})}\BibitemShut {NoStop}%
\bibitem [{\citenamefont {Vinyoles}\ \emph {et~al.}(2015)\citenamefont
  {Vinyoles} \emph {et~al.}}]{ALPLimitSolarNu}%
  \BibitemOpen
  \bibfield  {author} {\bibinfo {author} {\bibfnamefont {N.}~\bibnamefont
  {Vinyoles}} \emph {et~al.},\ }\bibfield  {title} {\bibinfo {title} {{New
  axion and hidden photon constraints from a solar data global fit}},\ }\href
  {https://doi.org/10.1088/1475-7516/2015/10/015} {\bibfield  {journal}
  {\bibinfo  {journal} {Journal of Cosmology and Astroparticle Physics}\
  }\textbf {\bibinfo {volume} {2015}}\bibinfo  {number} { (10)},\ \bibinfo
  {pages} {015}}\BibitemShut {NoStop}%
\bibitem [{\citenamefont {Ayala}\ \emph {et~al.}(2014)\citenamefont {Ayala}
  \emph {et~al.}}]{ALPLimitGlobularClusters}%
  \BibitemOpen
\bibfield  {number} {  }\bibfield  {author} {\bibinfo {author} {\bibfnamefont
  {A.}~\bibnamefont {Ayala}} \emph {et~al.},\ }\bibfield  {title} {\bibinfo
  {title} {{Revisiting the Bound on Axion-Photon Coupling from Globular
  Clusters}},\ }\href {https://doi.org/10.1103/PhysRevLett.113.191302}
  {\bibfield  {journal} {\bibinfo  {journal} {Physical Review Letters}\
  }\textbf {\bibinfo {volume} {113}},\ \bibinfo {pages} {191302} (\bibinfo
  {year} {2014})}\BibitemShut {NoStop}%
\bibitem [{\citenamefont {Dolan}\ \emph {et~al.}(2022)\citenamefont {Dolan},
  \citenamefont {Hiskens},\ and\ \citenamefont
  {Volkas}}]{ALPLimitGlobularClusters2}%
  \BibitemOpen
  \bibfield  {author} {\bibinfo {author} {\bibfnamefont {M.~J.}\ \bibnamefont
  {Dolan}}, \bibinfo {author} {\bibfnamefont {F.~J.}\ \bibnamefont {Hiskens}},\
  and\ \bibinfo {author} {\bibfnamefont {R.~R.}\ \bibnamefont {Volkas}},\
  }\bibfield  {title} {\bibinfo {title} {{Advancing globular cluster
  constraints on the axion-photon coupling}},\ }\href
  {https://doi.org/10.1088/1475-7516/2022/10/096} {\bibfield  {journal}
  {\bibinfo  {journal} {Journal of Cosmology and Astroparticle Physics}\
  }\textbf {\bibinfo {volume} {2022}}\bibinfo  {number} { (10)},\ \bibinfo
  {pages} {096}}\BibitemShut {NoStop}%
\bibitem [{\citenamefont {Della~Valle}\ \emph {et~al.}(2016)\citenamefont
  {Della~Valle} \emph {et~al.}}]{ALPLimitPVLAS}%
  \BibitemOpen
\bibfield  {number} {  }\bibfield  {author} {\bibinfo {author} {\bibfnamefont
  {F.}~\bibnamefont {Della~Valle}} \emph {et~al.},\ }\bibfield  {title}
  {\bibinfo {title} {{The PVLAS experiment: measuring vacuum magnetic
  birefringence and dichroism with a birefringent Fabry--Perot cavity}},\
  }\href {https://doi.org/10.1140/epjc/s10052-015-3869-8} {\bibfield  {journal}
  {\bibinfo  {journal} {The European Physical Journal C}\ }\textbf {\bibinfo
  {volume} {76}},\ \bibinfo {pages} {24} (\bibinfo {year} {2016})}\BibitemShut
  {NoStop}%
\bibitem [{\citenamefont {Halliday}\ \emph {et~al.}(2025)\citenamefont
  {Halliday} \emph {et~al.}}]{ALPLimitEuXFL}%
  \BibitemOpen
  \bibfield  {author} {\bibinfo {author} {\bibfnamefont {J.~W.~D.}\
  \bibnamefont {Halliday}} \emph {et~al.},\ }\bibfield  {title} {\bibinfo
  {title} {{Bounds on Heavy Axions with an X-Ray Free Electron Laser}},\ }\href
  {https://doi.org/10.1103/PhysRevLett.134.055001} {\bibfield  {journal}
  {\bibinfo  {journal} {Physical Review Letters}\ }\textbf {\bibinfo {volume}
  {134}},\ \bibinfo {pages} {055001} (\bibinfo {year} {2025})}\BibitemShut
  {NoStop}%
\bibitem [{\citenamefont {Homma}\ \emph {et~al.}(2021)\citenamefont {Homma}
  \emph {et~al.}}]{ALPLimitSAPPHIRES}%
  \BibitemOpen
  \bibfield  {author} {\bibinfo {author} {\bibfnamefont {K.}~\bibnamefont
  {Homma}} \emph {et~al.} (\bibinfo {collaboration} {SAPPHIRES
  Collaboration}),\ }\bibfield  {title} {\bibinfo {title} {{Search for sub-eV
  axion-like resonance states via stimulated quasi-parallel laser collisions
  with the parameterization including fully asymmetric collisional geometry}},\
  }\href {https://doi.org/10.1007/JHEP12(2021)108} {\bibfield  {journal}
  {\bibinfo  {journal} {Journal of High Energy Physics}\ }\textbf {\bibinfo
  {volume} {2021}},\ \bibinfo {pages} {108} (\bibinfo {year}
  {2021})}\BibitemShut {NoStop}%
\bibitem [{\citenamefont {Dolan}\ \emph {et~al.}(2017)\citenamefont {Dolan}
  \emph
  {et~al.}}]{ALPLimitRevisedOldBeamDump/RevisedColliderInvisibleFinalStates}%
  \BibitemOpen
  \bibfield  {author} {\bibinfo {author} {\bibfnamefont {M.~J.}\ \bibnamefont
  {Dolan}} \emph {et~al.},\ }\bibfield  {title} {\bibinfo {title} {{Revised
  constraints and Belle II sensitivity for visible and invisible axion-like
  particles}},\ }\href {https://doi.org/10.1007/JHEP12(2017)094} {\bibfield
  {journal} {\bibinfo  {journal} {Journal of High Energy Physics}\ }\textbf
  {\bibinfo {volume} {2017}},\ \bibinfo {pages} {094} (\bibinfo {year}
  {2017})}\BibitemShut {NoStop}%
\bibitem [{\citenamefont {Astier}\ \emph {et~al.}(2000)\citenamefont {Astier}
  \emph {et~al.}}]{ALPLimitNOMAD}%
  \BibitemOpen
  \bibfield  {author} {\bibinfo {author} {\bibfnamefont {P.}~\bibnamefont
  {Astier}} \emph {et~al.} (\bibinfo {collaboration} {NOMAD Collaboration}),\
  }\bibfield  {title} {\bibinfo {title} {{Search for eV (pseudo)scalar
  penetrating particles in the SPS neutrino beam}},\ }\href
  {https://doi.org/10.1016/S0370-2693(00)00375-0} {\bibfield  {journal}
  {\bibinfo  {journal} {Physics Letters B}\ }\textbf {\bibinfo {volume}
  {479}},\ \bibinfo {pages} {371} (\bibinfo {year} {2000})}\BibitemShut
  {NoStop}%
\bibitem [{\citenamefont {Jaeckel}\ and\ \citenamefont
  {Spannowsky}(2016)}]{ALPLimitLEP}%
  \BibitemOpen
  \bibfield  {author} {\bibinfo {author} {\bibfnamefont {J.}~\bibnamefont
  {Jaeckel}}\ and\ \bibinfo {author} {\bibfnamefont {M.}~\bibnamefont
  {Spannowsky}},\ }\bibfield  {title} {\bibinfo {title} {{Probing MeV to 90 GeV
  axion-like particles with LEP and LHC}},\ }\href
  {https://doi.org/10.1016/j.physletb.2015.12.037} {\bibfield  {journal}
  {\bibinfo  {journal} {Physics Letters B}\ }\textbf {\bibinfo {volume}
  {753}},\ \bibinfo {pages} {482} (\bibinfo {year} {2016})}\BibitemShut
  {NoStop}%
\bibitem [{\citenamefont {Abudin{\'e}n}\ \emph {et~al.}(2020)\citenamefont
  {Abudin{\'e}n} \emph {et~al.}}]{ALPLimitBelleII}%
  \BibitemOpen
  \bibfield  {author} {\bibinfo {author} {\bibfnamefont {F.}~\bibnamefont
  {Abudin{\'e}n}} \emph {et~al.} (\bibinfo {collaboration} {Belle II
  Collaboration}),\ }\bibfield  {title} {\bibinfo {title} {{Search for
  Axionlike Particles Produced in $e^+e^-$ Collisions at Belle II}},\ }\href
  {https://doi.org/10.1103/PhysRevLett.125.161806} {\bibfield  {journal}
  {\bibinfo  {journal} {Physical Review Letters}\ }\textbf {\bibinfo {volume}
  {125}},\ \bibinfo {pages} {161806} (\bibinfo {year} {2020})}\BibitemShut
  {NoStop}%
\bibitem [{\citenamefont {Ablikim}\ \emph {et~al.}(2023)\citenamefont {Ablikim}
  \emph {et~al.}}]{ALPLimitBESIII}%
  \BibitemOpen
  \bibfield  {author} {\bibinfo {author} {\bibfnamefont {M.}~\bibnamefont
  {Ablikim}} \emph {et~al.} (\bibinfo {collaboration} {BESIII Collaboration}),\
  }\bibfield  {title} {\bibinfo {title} {{Search for an axion-like particle in
  radiative $J/\psi$ decays}},\ }\href
  {https://doi.org/10.1016/j.physletb.2023.137698} {\bibfield  {journal}
  {\bibinfo  {journal} {Physics Letters B}\ }\textbf {\bibinfo {volume}
  {838}},\ \bibinfo {pages} {137698} (\bibinfo {year} {2023})}\BibitemShut
  {NoStop}%
\bibitem [{\citenamefont {Knapen}\ \emph {et~al.}(2017)\citenamefont {Knapen},
  \citenamefont {Lin}, \citenamefont {Lou},\ and\ \citenamefont
  {Melia}}]{ALPLimitOPAL/LHCpp}%
  \BibitemOpen
  \bibfield  {author} {\bibinfo {author} {\bibfnamefont {S.}~\bibnamefont
  {Knapen}}, \bibinfo {author} {\bibfnamefont {T.}~\bibnamefont {Lin}},
  \bibinfo {author} {\bibfnamefont {H.~K.}\ \bibnamefont {Lou}},\ and\ \bibinfo
  {author} {\bibfnamefont {T.}~\bibnamefont {Melia}},\ }\bibfield  {title}
  {\bibinfo {title} {{Searching for Axionlike Particles with Ultraperipheral
  Heavy-Ion Collisions}},\ }\href
  {https://doi.org/10.1103/PhysRevLett.118.171801} {\bibfield  {journal}
  {\bibinfo  {journal} {Physical Review Letters}\ }\textbf {\bibinfo {volume}
  {118}},\ \bibinfo {pages} {171801} (\bibinfo {year} {2017})}\BibitemShut
  {NoStop}%
\bibitem [{\citenamefont {Sirunyan}\ \emph {et~al.}(2019)\citenamefont
  {Sirunyan} \emph {et~al.}}]{ALPLimitCMS}%
  \BibitemOpen
  \bibfield  {author} {\bibinfo {author} {\bibfnamefont {A.~M.}\ \bibnamefont
  {Sirunyan}} \emph {et~al.} (\bibinfo {collaboration} {CMS Collaboration}),\
  }\bibfield  {title} {\bibinfo {title} {{Evidence for light-by-light
  scattering and searches for axion-like particles in ultraperipheral PbPb
  collisions at $\sqrt{s_\text{NN}}=5.02\,\text{TeV}$}},\ }\href
  {https://doi.org/10.1016/j.physletb.2019.134826} {\bibfield  {journal}
  {\bibinfo  {journal} {Physics Letters B}\ }\textbf {\bibinfo {volume}
  {797}},\ \bibinfo {pages} {134826} (\bibinfo {year} {2019})}\BibitemShut
  {NoStop}%
\bibitem [{\citenamefont {Aad}\ \emph {et~al.}(2021)\citenamefont {Aad} \emph
  {et~al.}}]{ALPLimitATLAS}%
  \BibitemOpen
  \bibfield  {author} {\bibinfo {author} {\bibfnamefont {G.}~\bibnamefont
  {Aad}} \emph {et~al.} (\bibinfo {collaboration} {ATLAS Collaboration}),\
  }\bibfield  {title} {\bibinfo {title} {{Measurement of light-by-light
  scattering and search for axion-like particles with $2.2\text{nb}^{-1}$ of
  $\text{Pb}+\text{Pb}$ data with the ATLAS detector}},\ }\href
  {https://doi.org/10.1007/JHEP03(2021)243} {\bibfield  {journal} {\bibinfo
  {journal} {Journal of High Energy Physics}\ }\textbf {\bibinfo {volume}
  {2021}},\ \bibinfo {pages} {243} (\bibinfo {year} {2021})}\BibitemShut
  {NoStop}%
\bibitem [{\citenamefont {Bl{\"u}mlein}\ \emph {et~al.}(1992)\citenamefont
  {Bl{\"u}mlein} \emph {et~al.}}]{ALPLimitOldBeamDump1}%
  \BibitemOpen
  \bibfield  {author} {\bibinfo {author} {\bibfnamefont {J.}~\bibnamefont
  {Bl{\"u}mlein}} \emph {et~al.},\ }\bibfield  {title} {\bibinfo {title}
  {{Limits on the mass of light (pseudo)scalar particles from Bethe-Heitler
  $e^+e^-$ and $\mu^+\mu^-$ pair production in a proton-iron beam dump
  experiment}},\ }\href {https://doi.org/10.1142/S0217751X9200171X} {\bibfield
  {journal} {\bibinfo  {journal} {International Journal of Modern Physics A}\
  }\textbf {\bibinfo {volume} {7}},\ \bibinfo {pages} {3835} (\bibinfo {year}
  {1992})}\BibitemShut {NoStop}%
\bibitem [{\citenamefont {Riordan}\ \emph
  {et~al.}(1987{\natexlab{b}})\citenamefont {Riordan} \emph
  {et~al.}}]{ALPLimitOldBeamDump2}%
  \BibitemOpen
  \bibfield  {author} {\bibinfo {author} {\bibfnamefont {E.~M.}\ \bibnamefont
  {Riordan}} \emph {et~al.},\ }\bibfield  {title} {\bibinfo {title} {{Search
  for short-lived axions in an electron-beam-dump experiment}},\ }\href
  {https://doi.org/10.1103/PhysRevLett.59.755} {\bibfield  {journal} {\bibinfo
  {journal} {Physical Review Letters}\ }\textbf {\bibinfo {volume} {59}},\
  \bibinfo {pages} {755} (\bibinfo {year} {1987}{\natexlab{b}})}\BibitemShut
  {NoStop}%
\bibitem [{\citenamefont {Banerjee}\ \emph {et~al.}(2020)\citenamefont
  {Banerjee} \emph {et~al.}}]{ALPLimitNA64}%
  \BibitemOpen
  \bibfield  {author} {\bibinfo {author} {\bibfnamefont {D.}~\bibnamefont
  {Banerjee}} \emph {et~al.} (\bibinfo {collaboration} {NA64 Collaboration}),\
  }\bibfield  {title} {\bibinfo {title} {{Search for Axionlike and Scalar
  Particles with the NA64 Experiment}},\ }\href
  {https://doi.org/10.1103/PhysRevLett.125.081801} {\bibfield  {journal}
  {\bibinfo  {journal} {Physical Review Letters}\ }\textbf {\bibinfo {volume}
  {125}},\ \bibinfo {pages} {081801} (\bibinfo {year} {2020})}\BibitemShut
  {NoStop}%
\bibitem [{\citenamefont {Bergsma}\ \emph {et~al.}(1985)\citenamefont {Bergsma}
  \emph {et~al.}}]{ALPLimitCHARM}%
  \BibitemOpen
  \bibfield  {author} {\bibinfo {author} {\bibfnamefont {F.}~\bibnamefont
  {Bergsma}} \emph {et~al.} (\bibinfo {collaboration} {CHARM Collaboration}),\
  }\bibfield  {title} {\bibinfo {title} {{Search for axion-like particle
  production in $400\,\text{GeV}$ proton-copper interactions}},\ }\href
  {https://doi.org/10.1016/0370-2693(85)90400-9} {\bibfield  {journal}
  {\bibinfo  {journal} {Physics Letters B}\ }\textbf {\bibinfo {volume}
  {157}},\ \bibinfo {pages} {458} (\bibinfo {year} {1985})}\BibitemShut
  {NoStop}%
\bibitem [{\citenamefont {Capozzi}\ \emph {et~al.}(2023)\citenamefont {Capozzi}
  \emph {et~al.}}]{ALPLimitMiniBooNE}%
  \BibitemOpen
  \bibfield  {author} {\bibinfo {author} {\bibfnamefont {F.}~\bibnamefont
  {Capozzi}} \emph {et~al.},\ }\bibfield  {title} {\bibinfo {title} {{New
  constraints on ALP couplings to electrons and photons from ArgoNeuT and the
  MiniBooNE beam dump}},\ }\href {https://doi.org/10.1103/PhysRevD.108.075019}
  {\bibfield  {journal} {\bibinfo  {journal} {Physical Review D}\ }\textbf
  {\bibinfo {volume} {108}},\ \bibinfo {pages} {075019} (\bibinfo {year}
  {2023})}\BibitemShut {NoStop}%
\bibitem [{\citenamefont {Aloni}\ \emph {et~al.}(2019)\citenamefont {Aloni},
  \citenamefont {Fanelli}, \citenamefont {Soreq},\ and\ \citenamefont
  {Williams}}]{ALPLimitPrimEx}%
  \BibitemOpen
  \bibfield  {author} {\bibinfo {author} {\bibfnamefont {D.}~\bibnamefont
  {Aloni}}, \bibinfo {author} {\bibfnamefont {C.}~\bibnamefont {Fanelli}},
  \bibinfo {author} {\bibfnamefont {Y.}~\bibnamefont {Soreq}},\ and\ \bibinfo
  {author} {\bibfnamefont {M.}~\bibnamefont {Williams}},\ }\bibfield  {title}
  {\bibinfo {title} {{Photoproduction of Axionlike Particles}},\ }\href
  {https://doi.org/10.1103/PhysRevLett.123.071801} {\bibfield  {journal}
  {\bibinfo  {journal} {Physical Review Letters}\ }\textbf {\bibinfo {volume}
  {123}},\ \bibinfo {pages} {071801} (\bibinfo {year} {2019})}\BibitemShut
  {NoStop}%
\bibitem [{\citenamefont {Kaneta}\ \emph {et~al.}(2017)\citenamefont {Kaneta},
  \citenamefont {Lee},\ and\ \citenamefont {Yun}}]{DarkAxionPortal}%
  \BibitemOpen
  \bibfield  {author} {\bibinfo {author} {\bibfnamefont {K.}~\bibnamefont
  {Kaneta}}, \bibinfo {author} {\bibfnamefont {H.~S.}\ \bibnamefont {Lee}},\
  and\ \bibinfo {author} {\bibfnamefont {S.}~\bibnamefont {Yun}},\ }\bibfield
  {title} {\bibinfo {title} {{Portal Connecting Dark Photons and Axions}},\
  }\href {https://doi.org/10.1103/PhysRevLett.118.101802} {\bibfield  {journal}
  {\bibinfo  {journal} {Physical Review Letters}\ }\textbf {\bibinfo {volume}
  {118}},\ \bibinfo {pages} {101802} (\bibinfo {year} {2017})}\BibitemShut
  {NoStop}%
\bibitem [{\citenamefont {deNiverville}\ \emph {et~al.}(2018)\citenamefont
  {deNiverville}, \citenamefont {Lee},\ and\ \citenamefont {Seo}}]{DAP2}%
  \BibitemOpen
  \bibfield  {author} {\bibinfo {author} {\bibfnamefont {P.}~\bibnamefont
  {deNiverville}}, \bibinfo {author} {\bibfnamefont {H.-S.}\ \bibnamefont
  {Lee}},\ and\ \bibinfo {author} {\bibfnamefont {M.-S.}\ \bibnamefont {Seo}},\
  }\bibfield  {title} {\bibinfo {title} {{Implications of the dark axion portal
  for the muon $g-2$, $B$ factories, fixed target neutrino experiments, and
  beam dumps}},\ }\href {https://doi.org/10.1103/PhysRevD.98.115011} {\bibfield
   {journal} {\bibinfo  {journal} {Physical Review D}\ }\textbf {\bibinfo
  {volume} {98}},\ \bibinfo {pages} {115011} (\bibinfo {year}
  {2018})}\BibitemShut {NoStop}%
\bibitem [{\citenamefont {deNiverville}\ and\ \citenamefont
  {Lee}(2019)}]{DAP3}%
  \BibitemOpen
  \bibfield  {author} {\bibinfo {author} {\bibfnamefont {P.}~\bibnamefont
  {deNiverville}}\ and\ \bibinfo {author} {\bibfnamefont {H.~S.}\ \bibnamefont
  {Lee}},\ }\bibfield  {title} {\bibinfo {title} {{Implications of the dark
  axion portal for SHiP and FASER and the advantages of monophoton signals}},\
  }\href {https://doi.org/10.1103/PhysRevD.100.055017} {\bibfield  {journal}
  {\bibinfo  {journal} {Physical Review D}\ }\textbf {\bibinfo {volume}
  {100}},\ \bibinfo {pages} {055017} (\bibinfo {year} {2019})}\BibitemShut
  {NoStop}%
\bibitem [{\citenamefont {Kirpichnikov}\ \emph {et~al.}(2020)\citenamefont
  {Kirpichnikov}, \citenamefont {Lyubovitskij},\ and\ \citenamefont
  {Zhevlakov}}]{DAP4}%
  \BibitemOpen
  \bibfield  {author} {\bibinfo {author} {\bibfnamefont {D.~V.}\ \bibnamefont
  {Kirpichnikov}}, \bibinfo {author} {\bibfnamefont {V.~E.}\ \bibnamefont
  {Lyubovitskij}},\ and\ \bibinfo {author} {\bibfnamefont {A.~S.}\ \bibnamefont
  {Zhevlakov}},\ }\bibfield  {title} {\bibinfo {title} {{Implication of hidden
  sub-GeV bosons for the $(g-2)_\mu$, ${}^8\text{Be}\text{--}{}^4\text{He}$
  anomaly, proton charge radius, EDM of fermions, and dark axion portal}},\
  }\href {https://doi.org/10.1103/PhysRevD.102.095024} {\bibfield  {journal}
  {\bibinfo  {journal} {Physical Review D}\ }\textbf {\bibinfo {volume}
  {102}},\ \bibinfo {pages} {095024} (\bibinfo {year} {2020})}\BibitemShut
  {NoStop}%
\bibitem [{\citenamefont {deNiverville}\ \emph {et~al.}(2021)\citenamefont
  {deNiverville}, \citenamefont {Lee},\ and\ \citenamefont {Lee}}]{DAP5}%
  \BibitemOpen
  \bibfield  {author} {\bibinfo {author} {\bibfnamefont {P.}~\bibnamefont
  {deNiverville}}, \bibinfo {author} {\bibfnamefont {H.~S.}\ \bibnamefont
  {Lee}},\ and\ \bibinfo {author} {\bibfnamefont {Y.~M.}\ \bibnamefont {Lee}},\
  }\bibfield  {title} {\bibinfo {title} {{New searches at reactor experiments
  based on the dark axion portal}},\ }\href
  {https://doi.org/10.1103/PhysRevD.103.075006} {\bibfield  {journal} {\bibinfo
   {journal} {Physical Review D}\ }\textbf {\bibinfo {volume} {103}},\ \bibinfo
  {pages} {075006} (\bibinfo {year} {2021})}\BibitemShut {NoStop}%
\bibitem [{\citenamefont {Hook}\ \emph {et~al.}(2021)\citenamefont {Hook},
  \citenamefont {Marques-Tavares},\ and\ \citenamefont {Ristow}}]{DAP6}%
  \BibitemOpen
  \bibfield  {author} {\bibinfo {author} {\bibfnamefont {A.}~\bibnamefont
  {Hook}}, \bibinfo {author} {\bibfnamefont {G.}~\bibnamefont
  {Marques-Tavares}},\ and\ \bibinfo {author} {\bibfnamefont {C.}~\bibnamefont
  {Ristow}},\ }\bibfield  {title} {\bibinfo {title} {{Supernova constraints on
  an axion-photon-dark photon interaction}},\ }\href
  {https://doi.org/10.1007/JHEP06(2021)167} {\bibfield  {journal} {\bibinfo
  {journal} {Journal of High Energy Physics}\ }\textbf {\bibinfo {volume}
  {2021}},\ \bibinfo {pages} {167} (\bibinfo {year} {2021})}\BibitemShut
  {NoStop}%
\bibitem [{\citenamefont {Ge}\ \emph {et~al.}(2021)\citenamefont {Ge},
  \citenamefont {Ma},\ and\ \citenamefont {Pasquini}}]{DAP7}%
  \BibitemOpen
  \bibfield  {author} {\bibinfo {author} {\bibfnamefont {S.~F.}\ \bibnamefont
  {Ge}}, \bibinfo {author} {\bibfnamefont {X.~D.}\ \bibnamefont {Ma}},\ and\
  \bibinfo {author} {\bibfnamefont {P.}~\bibnamefont {Pasquini}},\ }\bibfield
  {title} {\bibinfo {title} {{Probing the dark axion portal with muon anomalous
  magnetic moment}},\ }\href {https://doi.org/10.1140/epjc/s10052-021-09571-1}
  {\bibfield  {journal} {\bibinfo  {journal} {The European Physical Journal C}\
  }\textbf {\bibinfo {volume} {81}},\ \bibinfo {pages} {787} (\bibinfo {year}
  {2021})}\BibitemShut {NoStop}%
\bibitem [{\citenamefont {Guti{\'e}rrez}\ \emph {et~al.}(2021)\citenamefont
  {Guti{\'e}rrez} \emph {et~al.}}]{DAP8}%
  \BibitemOpen
  \bibfield  {author} {\bibinfo {author} {\bibfnamefont {J.~C.}\ \bibnamefont
  {Guti{\'e}rrez}} \emph {et~al.},\ }\href@noop {} {\bibinfo {title}
  {{Cosmology and direct detection of the Dark Axion Portal}}} (\bibinfo {year}
  {2021}),\ \Eprint {https://arxiv.org/abs/2112.11387} {arXiv:2112.11387
  [hep-ph]} \BibitemShut {NoStop}%
\bibitem [{\citenamefont {Zhevlakov}\ \emph {et~al.}(2022)\citenamefont
  {Zhevlakov}, \citenamefont {Kirpichnikov},\ and\ \citenamefont
  {Lyubovitskij}}]{DAP9}%
  \BibitemOpen
  \bibfield  {author} {\bibinfo {author} {\bibfnamefont {A.~S.}\ \bibnamefont
  {Zhevlakov}}, \bibinfo {author} {\bibfnamefont {D.~V.}\ \bibnamefont
  {Kirpichnikov}},\ and\ \bibinfo {author} {\bibfnamefont {V.~E.}\ \bibnamefont
  {Lyubovitskij}},\ }\bibfield  {title} {\bibinfo {title} {{Implication of the
  dark axion portal for the EDM of fermions and dark matter probing with
  $\text{NA64}e$, $\text{NA64}\mu$, LDMX, $M^3$, and BABAR}},\ }\href
  {https://doi.org/10.1103/PhysRevD.106.035018} {\bibfield  {journal} {\bibinfo
   {journal} {Physical Review D}\ }\textbf {\bibinfo {volume} {106}},\ \bibinfo
  {pages} {035018} (\bibinfo {year} {2022})}\BibitemShut {NoStop}%
\bibitem [{\citenamefont {Carenza}\ \emph {et~al.}(2023)\citenamefont
  {Carenza}, \citenamefont {Lucente},\ and\ \citenamefont
  {Vitagliano}}]{DAP10}%
  \BibitemOpen
  \bibfield  {author} {\bibinfo {author} {\bibfnamefont {P.}~\bibnamefont
  {Carenza}}, \bibinfo {author} {\bibfnamefont {G.}~\bibnamefont {Lucente}},\
  and\ \bibinfo {author} {\bibfnamefont {E.}~\bibnamefont {Vitagliano}},\
  }\bibfield  {title} {\bibinfo {title} {{Probing the blue axion with cosmic
  optical background anisotropies}},\ }\href
  {https://doi.org/10.1103/PhysRevD.107.083032} {\bibfield  {journal} {\bibinfo
   {journal} {Physical Review D}\ }\textbf {\bibinfo {volume} {107}},\ \bibinfo
  {pages} {083032} (\bibinfo {year} {2023})}\BibitemShut {NoStop}%
\bibitem [{\citenamefont {Jod{\l}owski}(2023)}]{DAP11}%
  \BibitemOpen
  \bibfield  {author} {\bibinfo {author} {\bibfnamefont {K.}~\bibnamefont
  {Jod{\l}owski}},\ }\bibfield  {title} {\bibinfo {title} {{Probing some photon
  portals to new physics at intensity frontier experiments}},\ }\href
  {https://doi.org/10.1103/PhysRevD.108.115017} {\bibfield  {journal} {\bibinfo
   {journal} {Physical Review D}\ }\textbf {\bibinfo {volume} {108}},\ \bibinfo
  {pages} {115017} (\bibinfo {year} {2023})}\BibitemShut {NoStop}%
\bibitem [{\citenamefont {Hong}\ \emph {et~al.}(2024)\citenamefont {Hong},
  \citenamefont {Min}, \citenamefont {Son},\ and\ \citenamefont {You}}]{DAP12}%
  \BibitemOpen
  \bibfield  {author} {\bibinfo {author} {\bibfnamefont {H.}~\bibnamefont
  {Hong}}, \bibinfo {author} {\bibfnamefont {U.}~\bibnamefont {Min}}, \bibinfo
  {author} {\bibfnamefont {M.}~\bibnamefont {Son}},\ and\ \bibinfo {author}
  {\bibfnamefont {T.}~\bibnamefont {You}},\ }\bibfield  {title} {\bibinfo
  {title} {{A cosmic window on the dark axion portal}},\ }\href
  {https://doi.org/10.1007/JHEP03(2024)155} {\bibfield  {journal} {\bibinfo
  {journal} {Journal of High Energy Physics}\ }\textbf {\bibinfo {volume}
  {2024}},\ \bibinfo {pages} {155} (\bibinfo {year} {2024})}\BibitemShut
  {NoStop}%
\bibitem [{\citenamefont {Hook}\ \emph {et~al.}(2024)\citenamefont {Hook},
  \citenamefont {Marques-Tavares},\ and\ \citenamefont {Ristow}}]{DAP13}%
  \BibitemOpen
  \bibfield  {author} {\bibinfo {author} {\bibfnamefont {A.}~\bibnamefont
  {Hook}}, \bibinfo {author} {\bibfnamefont {G.}~\bibnamefont
  {Marques-Tavares}},\ and\ \bibinfo {author} {\bibfnamefont {C.}~\bibnamefont
  {Ristow}},\ }\bibfield  {title} {\bibinfo {title} {{CMB spectral distortions
  from an axion-dark photon-photon interaction}},\ }\href
  {https://doi.org/10.1007/JHEP05(2024)086} {\bibfield  {journal} {\bibinfo
  {journal} {Journal of High Energy Physics}\ }\textbf {\bibinfo {volume}
  {2024}},\ \bibinfo {pages} {086} (\bibinfo {year} {2024})}\BibitemShut
  {NoStop}%
\bibitem [{\citenamefont {Jod{\l}owski}(2025)}]{DAP14}%
  \BibitemOpen
  \bibfield  {author} {\bibinfo {author} {\bibfnamefont {K.}~\bibnamefont
  {Jod{\l}owski}},\ }\bibfield  {title} {\bibinfo {title} {{Dark axion portal
  at $Z$ boson factories}},\ }\href {https://doi.org/10.1007/JHEP08(2025)022}
  {\bibfield  {journal} {\bibinfo  {journal} {Journal of High Energy Physics}\
  }\textbf {\bibinfo {volume} {2025}},\ \bibinfo {pages} {022} (\bibinfo {year}
  {2025})}\BibitemShut {NoStop}%
\bibitem [{\citenamefont {Arias}\ \emph {et~al.}(2025)\citenamefont {Arias},
  \citenamefont {S\'{a}ez},\ and\ \citenamefont {Jaeckel}}]{DAP15}%
  \BibitemOpen
  \bibfield  {author} {\bibinfo {author} {\bibfnamefont {P.}~\bibnamefont
  {Arias}}, \bibinfo {author} {\bibfnamefont {B.~D.}\ \bibnamefont
  {S\'{a}ez}},\ and\ \bibinfo {author} {\bibfnamefont {J.}~\bibnamefont
  {Jaeckel}},\ }\bibfield  {title} {\bibinfo {title} {{Freezing-in the pure
  dark axion portal}},\ }\href {https://doi.org/10.1088/1475-7516/2025/06/060}
  {\bibfield  {journal} {\bibinfo  {journal} {Journal of Cosmology and
  Astroparticle Physics}\ }\textbf {\bibinfo {volume} {2025}}\bibinfo  {number}
  { (06)},\ \bibinfo {pages} {060}}\BibitemShut {NoStop}%
\bibitem [{\citenamefont {Arias}\ \emph {et~al.}(2026)\citenamefont {Arias}
  \emph {et~al.}}]{DAP16}%
  \BibitemOpen
\bibfield  {number} {  }\bibfield  {author} {\bibinfo {author} {\bibfnamefont
  {P.}~\bibnamefont {Arias}} \emph {et~al.},\ }\bibfield  {title} {\bibinfo
  {title} {{Probing displaced (dark)photons from low reheating freeze-in at the
  LHC}},\ }\href {https://doi.org/10.1007/JHEP01(2026)135} {\bibfield
  {journal} {\bibinfo  {journal} {Journal of High Energy Physics}\ }\textbf
  {\bibinfo {volume} {2026}},\ \bibinfo {pages} {135} (\bibinfo {year}
  {2026})}\BibitemShut {NoStop}%
\bibitem [{\citenamefont {Bai}\ \emph {et~al.}(2022)\citenamefont {Bai} \emph
  {et~al.}}]{LUXENPOD}%
  \BibitemOpen
  \bibfield  {author} {\bibinfo {author} {\bibfnamefont {Z.}~\bibnamefont
  {Bai}} \emph {et~al.},\ }\bibfield  {title} {\bibinfo {title} {{New physics
  searches with an optical dump at LUXE}},\ }\href
  {https://doi.org/10.1103/PhysRevD.106.115034} {\bibfield  {journal} {\bibinfo
   {journal} {Physical Review D}\ }\textbf {\bibinfo {volume} {106}},\ \bibinfo
  {pages} {115034} (\bibinfo {year} {2022})}\BibitemShut {NoStop}%
\bibitem [{\citenamefont {Soto}\ \emph {et~al.}(2025)\citenamefont {Soto} \emph
  {et~al.}}]{LUXENPOD2}%
  \BibitemOpen
  \bibfield  {author} {\bibinfo {author} {\bibfnamefont {M.~A.}\ \bibnamefont
  {Soto}} \emph {et~al.},\ }\bibfield  {title} {\bibinfo {title} {{Layout
  optimization for the LUXE-NPOD experiment}},\ }\href
  {https://doi.org/10.1103/n2nh-7v5b} {\bibfield  {journal} {\bibinfo
  {journal} {Physical Review D}\ }\textbf {\bibinfo {volume} {112}},\ \bibinfo
  {pages} {112014} (\bibinfo {year} {2025})}\BibitemShut {NoStop}%
\bibitem [{\citenamefont {Abramowicz}\ \emph {et~al.}(2021)\citenamefont
  {Abramowicz} \emph {et~al.}}]{LUXECDR}%
  \BibitemOpen
  \bibfield  {author} {\bibinfo {author} {\bibfnamefont {H.}~\bibnamefont
  {Abramowicz}} \emph {et~al.},\ }\bibfield  {title} {\bibinfo {title}
  {{Conceptual design report for the LUXE experiment}},\ }\href
  {https://doi.org/10.1140/epjs/s11734-021-00249-z} {\bibfield  {journal}
  {\bibinfo  {journal} {The European Physical Journal Special Topics}\ }\textbf
  {\bibinfo {volume} {230}},\ \bibinfo {pages} {2445} (\bibinfo {year}
  {2021})}\BibitemShut {NoStop}%
\bibitem [{\citenamefont {Abramowicz}\ \emph {et~al.}(2024)\citenamefont
  {Abramowicz} \emph {et~al.}}]{LUXETDR}%
  \BibitemOpen
  \bibfield  {author} {\bibinfo {author} {\bibfnamefont {H.}~\bibnamefont
  {Abramowicz}} \emph {et~al.} (\bibinfo {collaboration} {LUXE
  Collaboration}),\ }\bibfield  {title} {\bibinfo {title} {{Technical Design
  Report for the LUXE experiment}},\ }\href
  {https://doi.org/10.1140/epjs/s11734-024-01164-9} {\bibfield  {journal}
  {\bibinfo  {journal} {The European Physical Journal Special Topics}\ }\textbf
  {\bibinfo {volume} {233}},\ \bibinfo {pages} {1709–1974} (\bibinfo {year}
  {2024})}\BibitemShut {NoStop}%
\bibitem [{\citenamefont {Abramowicz}\ \emph {et~al.}(2025)\citenamefont
  {Abramowicz} \emph {et~al.}}]{LUXE}%
  \BibitemOpen
  \bibfield  {author} {\bibinfo {author} {\bibfnamefont {H.}~\bibnamefont
  {Abramowicz}} \emph {et~al.},\ }\href@noop {} {\bibinfo {title} {{Input to
  the ESPPU: The LUXE Experiment}}} (\bibinfo {year} {2025}),\ \Eprint
  {https://arxiv.org/abs/2504.00873} {arXiv:2504.00873 [hep-ex]} \BibitemShut
  {NoStop}%
\bibitem [{\citenamefont {Heisenberg}\ and\ \citenamefont
  {Euler}(1936)}]{LightByLightScattering}%
  \BibitemOpen
  \bibfield  {author} {\bibinfo {author} {\bibfnamefont {W.}~\bibnamefont
  {Heisenberg}}\ and\ \bibinfo {author} {\bibfnamefont {H.}~\bibnamefont
  {Euler}},\ }\bibfield  {title} {\bibinfo {title} {{Folgerungen aus der
  Diracschen Theorie des Positrons}},\ }\href
  {https://doi.org/10.1007/BF01343663} {\bibfield  {journal} {\bibinfo
  {journal} {Zeitschrift f{\"u}r Physik}\ }\textbf {\bibinfo {volume} {98}},\
  \bibinfo {pages} {714} (\bibinfo {year} {1936})}\BibitemShut {NoStop}%
\bibitem [{\citenamefont {Karplus}\ and\ \citenamefont
  {Neuman}(1951)}]{LightByLightScattering2}%
  \BibitemOpen
  \bibfield  {author} {\bibinfo {author} {\bibfnamefont {R.}~\bibnamefont
  {Karplus}}\ and\ \bibinfo {author} {\bibfnamefont {M.}~\bibnamefont
  {Neuman}},\ }\bibfield  {title} {\bibinfo {title} {{The Scattering of Light
  by Light}},\ }\href {https://doi.org/10.1103/PhysRev.83.776} {\bibfield
  {journal} {\bibinfo  {journal} {Physical Review}\ }\textbf {\bibinfo {volume}
  {83}},\ \bibinfo {pages} {776} (\bibinfo {year} {1951})}\BibitemShut
  {NoStop}%
\bibitem [{\citenamefont {Schwinger}(1951)}]{SchwingerPairProduction}%
  \BibitemOpen
  \bibfield  {author} {\bibinfo {author} {\bibfnamefont {J.}~\bibnamefont
  {Schwinger}},\ }\bibfield  {title} {\bibinfo {title} {{On Gauge Invariance
  and Vacuum Polarization}},\ }\href {https://doi.org/10.1103/PhysRev.82.664}
  {\bibfield  {journal} {\bibinfo  {journal} {Physical Review}\ }\textbf
  {\bibinfo {volume} {82}},\ \bibinfo {pages} {664} (\bibinfo {year}
  {1951})}\BibitemShut {NoStop}%
\bibitem [{\citenamefont {Reiss}(1962)}]{NonLinearBreitWheeler}%
  \BibitemOpen
  \bibfield  {author} {\bibinfo {author} {\bibfnamefont {H.~R.}\ \bibnamefont
  {Reiss}},\ }\bibfield  {title} {\bibinfo {title} {{Absorption of Light by
  Light}},\ }\href {https://doi.org/10.1063/1.1703787} {\bibfield  {journal}
  {\bibinfo  {journal} {Journal of Mathematical Physics}\ }\textbf {\bibinfo
  {volume} {3}},\ \bibinfo {pages} {59} (\bibinfo {year} {1962})}\BibitemShut
  {NoStop}%
\bibitem [{\citenamefont {Nikishov}\ and\ \citenamefont
  {Ritus}(1964)}]{NonlinearCompton}%
  \BibitemOpen
  \bibfield  {author} {\bibinfo {author} {\bibfnamefont {A.~I.}\ \bibnamefont
  {Nikishov}}\ and\ \bibinfo {author} {\bibfnamefont {V.~I.}\ \bibnamefont
  {Ritus}},\ }\bibfield  {title} {\bibinfo {title} {{Quantum Processes in the
  Field of a Plane Electromagnetic Wave and in a Constant Field. I}},\ }\href
  {https://jetp.ras.ru/cgi-bin/e/index/e/19/2/p529?a=list} {\bibfield
  {journal} {\bibinfo  {journal} {Soviet Physics JETP}\ }\textbf {\bibinfo
  {volume} {19}},\ \bibinfo {pages} {529} (\bibinfo {year} {1964})}\BibitemShut
  {NoStop}%
\bibitem [{\citenamefont {Brown}\ and\ \citenamefont
  {Kibble}(1964)}]{NonlinearCompton2}%
  \BibitemOpen
  \bibfield  {author} {\bibinfo {author} {\bibfnamefont {L.~S.}\ \bibnamefont
  {Brown}}\ and\ \bibinfo {author} {\bibfnamefont {T.~W.~B.}\ \bibnamefont
  {Kibble}},\ }\bibfield  {title} {\bibinfo {title} {{Interaction of Intense
  Laser Beams with Electrons}},\ }\href
  {https://doi.org/10.1103/PhysRev.133.A705} {\bibfield  {journal} {\bibinfo
  {journal} {Physical Review}\ }\textbf {\bibinfo {volume} {133}},\ \bibinfo
  {pages} {A705} (\bibinfo {year} {1964})}\BibitemShut {NoStop}%
\bibitem [{\citenamefont {Wolkow}(1935)}]{VolkovStates}%
  \BibitemOpen
  \bibfield  {author} {\bibinfo {author} {\bibfnamefont {D.~M.}\ \bibnamefont
  {Wolkow}},\ }\bibfield  {title} {\bibinfo {title} {{{\"U}ber eine Klasse von
  L{\"o}sungen der Diracschen Gleichung}},\ }\href
  {https://doi.org/10.1007/BF01331022} {\bibfield  {journal} {\bibinfo
  {journal} {Zeitschrift f{\"u}r Physik}\ }\textbf {\bibinfo {volume} {94}},\
  \bibinfo {pages} {250} (\bibinfo {year} {1935})}\BibitemShut {NoStop}%
\bibitem [{\citenamefont {Workman}\ \emph {et~al.}(2022)\citenamefont {Workman}
  \emph {et~al.}}]{PDGreview}%
  \BibitemOpen
  \bibfield  {author} {\bibinfo {author} {\bibfnamefont {R.~L.}\ \bibnamefont
  {Workman}} \emph {et~al.} (\bibinfo {collaboration} {Particle Data Group}),\
  }\bibfield  {title} {\bibinfo {title} {{Review of Particle Physics}},\ }\href
  {https://doi.org/10.1093/ptep/ptac097} {\bibfield  {journal} {\bibinfo
  {journal} {Progress of Theoretical and Experimental Physics}\ }\textbf
  {\bibinfo {volume} {2022}},\ \bibinfo {pages} {083C01} (\bibinfo {year}
  {2022})}\BibitemShut {NoStop}%
\bibitem [{\citenamefont {Tsai}\ and\ \citenamefont
  {Whitis}(1966)}]{CascadeEqs}%
  \BibitemOpen
  \bibfield  {author} {\bibinfo {author} {\bibfnamefont {Y.~S.}\ \bibnamefont
  {Tsai}}\ and\ \bibinfo {author} {\bibfnamefont {V.}~\bibnamefont {Whitis}},\
  }\bibfield  {title} {\bibinfo {title} {{Thick-Target Bremsstrahlung and
  Target Considerations for Secondary-Particle Production by Electrons}},\
  }\href {https://doi.org/10.1103/PhysRev.149.1248} {\bibfield  {journal}
  {\bibinfo  {journal} {Physical Review}\ }\textbf {\bibinfo {volume} {149}},\
  \bibinfo {pages} {1248} (\bibinfo {year} {1966})}\BibitemShut {NoStop}%
\bibitem [{\citenamefont {Heinzl}\ \emph {et~al.}(2020)\citenamefont {Heinzl},
  \citenamefont {King},\ and\ \citenamefont {MacLeod}}]{LMAapproximation}%
  \BibitemOpen
  \bibfield  {author} {\bibinfo {author} {\bibfnamefont {T.}~\bibnamefont
  {Heinzl}}, \bibinfo {author} {\bibfnamefont {B.}~\bibnamefont {King}},\ and\
  \bibinfo {author} {\bibfnamefont {A.~J.}\ \bibnamefont {MacLeod}},\
  }\bibfield  {title} {\bibinfo {title} {{Locally monochromatic approximation
  to QED in intense laser fields}},\ }\href
  {https://doi.org/10.1103/PhysRevA.102.063110} {\bibfield  {journal} {\bibinfo
   {journal} {Physical Review A}\ }\textbf {\bibinfo {volume} {102}},\ \bibinfo
  {pages} {063110} (\bibinfo {year} {2020})}\BibitemShut {NoStop}%
\bibitem [{\citenamefont {Bonivento}(2018)}]{AngularResolution}%
  \BibitemOpen
  \bibfield  {author} {\bibinfo {author} {\bibfnamefont {W.~M.}\ \bibnamefont
  {Bonivento}},\ }\bibfield  {title} {\bibinfo {title} {{Studies for the
  electro-magnetic calorimeter SplitCal for the SHiP experiment at CERN with
  shower direction reconstruction capability}},\ }\href
  {https://doi.org/10.1088/1748-0221/13/02/C02041} {\bibfield  {journal}
  {\bibinfo  {journal} {{Journal of Instrumentation}}\ }\textbf {\bibinfo
  {volume} {13}},\ \bibinfo {pages} {C02041} (\bibinfo {year}
  {2018})}\BibitemShut {NoStop}%
\bibitem [{\citenamefont {Bornheim}(2015)}]{Detector1}%
  \BibitemOpen
  \bibfield  {author} {\bibinfo {author} {\bibfnamefont {A.}~\bibnamefont
  {Bornheim}},\ }\bibfield  {title} {\bibinfo {title} {{Timing performance of
  the CMS electromagnetic calorimeter and prospects for the future}},\ }\href
  {https://doi.org/10.22323/1.213.0021} {\bibfield  {journal} {\bibinfo
  {journal} {PoS}\ }\textbf {\bibinfo {volume} {TIPP2014}},\ \bibinfo {pages}
  {021} (\bibinfo {year} {2015})}\BibitemShut {NoStop}%
\bibitem [{\citenamefont {Lobanov}(2020)}]{Detector2}%
  \BibitemOpen
  \bibfield  {author} {\bibinfo {author} {\bibfnamefont {A.}~\bibnamefont
  {Lobanov}},\ }\bibfield  {title} {\bibinfo {title} {{Precision timing
  calorimetry with the CMS HGCAL}},\ }\href
  {https://doi.org/10.1088/1748-0221/15/07/C07003} {\bibfield  {journal}
  {\bibinfo  {journal} {{Journal of Instrumentation}}\ }\textbf {\bibinfo
  {volume} {15}},\ \bibinfo {pages} {C07003} (\bibinfo {year}
  {2020})}\BibitemShut {NoStop}%
\bibitem [{\citenamefont {Martinazzoli}(2020)}]{Detector3}%
  \BibitemOpen
  \bibfield  {author} {\bibinfo {author} {\bibfnamefont {L.}~\bibnamefont
  {Martinazzoli}},\ }\bibfield  {title} {\bibinfo {title} {{Crystal Fibers for
  the LHCb Calorimeter Upgrade}},\ }\href
  {https://doi.org/10.1109/TNS.2020.2975570} {\bibfield  {journal} {\bibinfo
  {journal} {IEEE Transactions on Nuclear Science}\ }\textbf {\bibinfo {volume}
  {67}},\ \bibinfo {pages} {1003} (\bibinfo {year} {2020})}\BibitemShut
  {NoStop}%
\bibitem [{\citenamefont {Liu}\ \emph {et~al.}(2020)\citenamefont {Liu},
  \citenamefont {Jiang},\ and\ \citenamefont {Wang}}]{Detector4}%
  \BibitemOpen
  \bibfield  {author} {\bibinfo {author} {\bibfnamefont {Y.}~\bibnamefont
  {Liu}}, \bibinfo {author} {\bibfnamefont {J.}~\bibnamefont {Jiang}},\ and\
  \bibinfo {author} {\bibfnamefont {Y.}~\bibnamefont {Wang}},\ }\bibfield
  {title} {\bibinfo {title} {{High-granularity crystal calorimetry: conceptual
  designs and first studies}},\ }\href
  {https://doi.org/10.1088/1748-0221/15/04/C04056} {\bibfield  {journal}
  {\bibinfo  {journal} {{Journal of Instrumentation}}\ }\textbf {\bibinfo
  {volume} {15}},\ \bibinfo {pages} {C04056} (\bibinfo {year}
  {2020})}\BibitemShut {NoStop}%
\bibitem [{\citenamefont {Ilten}\ \emph {et~al.}(2018)\citenamefont {Ilten},
  \citenamefont {Soreq}, \citenamefont {Williams},\ and\ \citenamefont
  {Xue}}]{DPLimitsHeavy_DarkCast}%
  \BibitemOpen
  \bibfield  {author} {\bibinfo {author} {\bibfnamefont {P.}~\bibnamefont
  {Ilten}}, \bibinfo {author} {\bibfnamefont {Y.}~\bibnamefont {Soreq}},
  \bibinfo {author} {\bibfnamefont {M.}~\bibnamefont {Williams}},\ and\
  \bibinfo {author} {\bibfnamefont {W.}~\bibnamefont {Xue}},\ }\bibfield
  {title} {\bibinfo {title} {{Serendipity in dark photon searches}},\ }\href
  {https://doi.org/10.1007/JHEP06(2018)004} {\bibfield  {journal} {\bibinfo
  {journal} {Journal of High Energy Physics}\ }\textbf {\bibinfo {volume}
  {2018}},\ \bibinfo {pages} {004} (\bibinfo {year} {2018})}\BibitemShut
  {NoStop}%
\bibitem [{\citenamefont {Caputo}\ \emph {et~al.}(2021)\citenamefont {Caputo},
  \citenamefont {Millar}, \citenamefont {O’Hare},\ and\ \citenamefont
  {Vitagliano}}]{DPLimitsLight_Handbook}%
  \BibitemOpen
  \bibfield  {author} {\bibinfo {author} {\bibfnamefont {A.}~\bibnamefont
  {Caputo}}, \bibinfo {author} {\bibfnamefont {A.~J.}\ \bibnamefont {Millar}},
  \bibinfo {author} {\bibfnamefont {C.~A.~J.}\ \bibnamefont {O’Hare}},\ and\
  \bibinfo {author} {\bibfnamefont {E.}~\bibnamefont {Vitagliano}},\ }\bibfield
   {title} {\bibinfo {title} {{Dark photon limits: A handbook}},\ }\href
  {https://doi.org/10.1103/PhysRevD.104.095029} {\bibfield  {journal} {\bibinfo
   {journal} {Physical Review D}\ }\textbf {\bibinfo {volume} {104}},\ \bibinfo
  {pages} {095029} (\bibinfo {year} {2021})}\BibitemShut {NoStop}%
\bibitem [{\citenamefont {O'Hare}(2020)}]{CairanOhareGithub}%
  \BibitemOpen
  \bibfield  {author} {\bibinfo {author} {\bibfnamefont {C.}~\bibnamefont
  {O'Hare}},\ }\href {https://doi.org/10.5281/zenodo.3932430} {\bibinfo {title}
  {{cajohare/AxionLimits: AxionLimits}}},\ \bibinfo {howpublished}
  {\url{https://cajohare.github.io/AxionLimits/}} (\bibinfo {year}
  {2020})\BibitemShut {NoStop}%
\bibitem [{\citenamefont {Ning}\ and\ \citenamefont
  {Safdi}(2025)}]{ALPLimitAstro1}%
  \BibitemOpen
  \bibfield  {author} {\bibinfo {author} {\bibfnamefont {O.}~\bibnamefont
  {Ning}}\ and\ \bibinfo {author} {\bibfnamefont {B.~R.}\ \bibnamefont
  {Safdi}},\ }\bibfield  {title} {\bibinfo {title} {{Leading Axion-Photon
  Sensitivity with NuSTAR Observations of M82 and M87}},\ }\href
  {https://doi.org/10.1103/PhysRevLett.134.171003} {\bibfield  {journal}
  {\bibinfo  {journal} {Physical Review Letters}\ }\textbf {\bibinfo {volume}
  {134}},\ \bibinfo {pages} {171003} (\bibinfo {year} {2025})}\BibitemShut
  {NoStop}%
\bibitem [{\citenamefont {Cand\'on}\ \emph {et~al.}(2025)\citenamefont
  {Cand\'on} \emph {et~al.}}]{ALPLimitAstro2}%
  \BibitemOpen
  \bibfield  {author} {\bibinfo {author} {\bibfnamefont {F.~R.}\ \bibnamefont
  {Cand\'on}} \emph {et~al.},\ }\bibfield  {title} {\bibinfo {title} {{NuSTAR
  Bounds on Radiatively Decaying Particles from M82}},\ }\href
  {https://doi.org/10.1103/PhysRevLett.134.171004} {\bibfield  {journal}
  {\bibinfo  {journal} {Physical Review Letters}\ }\textbf {\bibinfo {volume}
  {134}},\ \bibinfo {pages} {171004} (\bibinfo {year} {2025})}\BibitemShut
  {NoStop}%
\bibitem [{\citenamefont {Dolan}\ \emph {et~al.}(2021)\citenamefont {Dolan},
  \citenamefont {Hiskens},\ and\ \citenamefont {Volkas}}]{ALPLimitAstro3}%
  \BibitemOpen
  \bibfield  {author} {\bibinfo {author} {\bibfnamefont {M.~J.}\ \bibnamefont
  {Dolan}}, \bibinfo {author} {\bibfnamefont {F.~J.}\ \bibnamefont {Hiskens}},\
  and\ \bibinfo {author} {\bibfnamefont {R.~R.}\ \bibnamefont {Volkas}},\
  }\bibfield  {title} {\bibinfo {title} {{Constraining axion-like particles
  using the white dwarf initial-final mass relation}},\ }\href
  {https://doi.org/10.1088/1475-7516/2021/09/010} {\bibfield  {journal}
  {\bibinfo  {journal} {Journal of Cosmology and Astroparticle Physics}\
  }\textbf {\bibinfo {volume} {2021}}\bibinfo  {number} { (09)},\ \bibinfo
  {pages} {010}}\BibitemShut {NoStop}%
\bibitem [{\citenamefont {Caputo}\ \emph
  {et~al.}(2022{\natexlab{a}})\citenamefont {Caputo} \emph
  {et~al.}}]{ALPLimitAstro4}%
  \BibitemOpen
\bibfield  {number} {  }\bibfield  {author} {\bibinfo {author} {\bibfnamefont
  {A.}~\bibnamefont {Caputo}} \emph {et~al.},\ }\bibfield  {title} {\bibinfo
  {title} {{Low-Energy Supernovae Severely Constrain Radiative Particle
  Decays}},\ }\href {https://doi.org/10.1103/PhysRevLett.128.221103} {\bibfield
   {journal} {\bibinfo  {journal} {Physical Review Letters}\ }\textbf {\bibinfo
  {volume} {128}},\ \bibinfo {pages} {221103} (\bibinfo {year}
  {2022}{\natexlab{a}})}\BibitemShut {NoStop}%
\bibitem [{\citenamefont {Lucente}\ \emph {et~al.}(2020)\citenamefont {Lucente}
  \emph {et~al.}}]{ALPLimitAstro5}%
  \BibitemOpen
  \bibfield  {author} {\bibinfo {author} {\bibfnamefont {G.}~\bibnamefont
  {Lucente}} \emph {et~al.},\ }\bibfield  {title} {\bibinfo {title} {{Heavy
  axion-like particles and core-collapse supernovae: constraints and impact on
  the explosion mechanism}},\ }\href
  {https://doi.org/10.1088/1475-7516/2020/12/008} {\bibfield  {journal}
  {\bibinfo  {journal} {Journal of Cosmology and Astroparticle Physics}\
  }\textbf {\bibinfo {volume} {2020}}\bibinfo  {number} { (12)},\ \bibinfo
  {pages} {008}}\BibitemShut {NoStop}%
\bibitem [{\citenamefont {Caputo}\ \emph
  {et~al.}(2022{\natexlab{b}})\citenamefont {Caputo}, \citenamefont {Raffelt},\
  and\ \citenamefont {Vitagliano}}]{ALPLimitAstro6}%
  \BibitemOpen
\bibfield  {number} {  }\bibfield  {author} {\bibinfo {author} {\bibfnamefont
  {A.}~\bibnamefont {Caputo}}, \bibinfo {author} {\bibfnamefont
  {G.}~\bibnamefont {Raffelt}},\ and\ \bibinfo {author} {\bibfnamefont
  {E.}~\bibnamefont {Vitagliano}},\ }\bibfield  {title} {\bibinfo {title}
  {{Muonic boson limits: Supernova redux}},\ }\href
  {https://doi.org/10.1103/PhysRevD.105.035022} {\bibfield  {journal} {\bibinfo
   {journal} {Physical Review D}\ }\textbf {\bibinfo {volume} {105}},\ \bibinfo
  {pages} {035022} (\bibinfo {year} {2022}{\natexlab{b}})}\BibitemShut
  {NoStop}%
\bibitem [{\citenamefont {Diamond}\ \emph {et~al.}(2023)\citenamefont {Diamond}
  \emph {et~al.}}]{ALPLimitAstro7}%
  \BibitemOpen
  \bibfield  {author} {\bibinfo {author} {\bibfnamefont {M.}~\bibnamefont
  {Diamond}} \emph {et~al.},\ }\bibfield  {title} {\bibinfo {title}
  {{Axion-sourced fireballs from supernovae}},\ }\href
  {https://doi.org/10.1103/PhysRevD.107.103029} {\bibfield  {journal} {\bibinfo
   {journal} {Physical Review D}\ }\textbf {\bibinfo {volume} {107}},\ \bibinfo
  {pages} {103029} (\bibinfo {year} {2023})}\BibitemShut {NoStop}%
\bibitem [{\citenamefont {Fiorillo}\ \emph {et~al.}(2025)\citenamefont
  {Fiorillo}, \citenamefont {Pitik},\ and\ \citenamefont
  {Vitagliano}}]{ALPLimitAstro8}%
  \BibitemOpen
  \bibfield  {author} {\bibinfo {author} {\bibfnamefont {D.~F.~G.}\
  \bibnamefont {Fiorillo}}, \bibinfo {author} {\bibfnamefont {T.}~\bibnamefont
  {Pitik}},\ and\ \bibinfo {author} {\bibfnamefont {E.}~\bibnamefont
  {Vitagliano}},\ }\bibfield  {title} {\bibinfo {title} {{Energy Transfer by
  Feebly Interacting Particles in Supernovae: The Trapping Regime}},\ }\href
  {https://doi.org/10.1103/cz94-dqxt} {\bibfield  {journal} {\bibinfo
  {journal} {Physical Review Letters}\ }\textbf {\bibinfo {volume} {135}},\
  \bibinfo {pages} {071005} (\bibinfo {year} {2025})}\BibitemShut {NoStop}%
\bibitem [{\citenamefont {Bhupal~Dev}\ \emph {et~al.}(2024)\citenamefont
  {Bhupal~Dev} \emph {et~al.}}]{ALPLimitAstro9}%
  \BibitemOpen
  \bibfield  {author} {\bibinfo {author} {\bibfnamefont {P.~S.~B.}\
  \bibnamefont {Bhupal~Dev}} \emph {et~al.},\ }\bibfield  {title} {\bibinfo
  {title} {{First Constraints on the Photon Coupling of Axionlike Particles
  from Multimessenger Studies of the Neutron Star Merger GW170817}},\ }\href
  {https://doi.org/10.1103/PhysRevLett.132.101003} {\bibfield  {journal}
  {\bibinfo  {journal} {Physical Review Letters}\ }\textbf {\bibinfo {volume}
  {132}},\ \bibinfo {pages} {101003} (\bibinfo {year} {2024})}\BibitemShut
  {NoStop}%
\bibitem [{\citenamefont {Diamond}\ \emph {et~al.}(2024)\citenamefont {Diamond}
  \emph {et~al.}}]{ALPLimitAstro10}%
  \BibitemOpen
  \bibfield  {author} {\bibinfo {author} {\bibfnamefont {M.}~\bibnamefont
  {Diamond}} \emph {et~al.},\ }\bibfield  {title} {\bibinfo {title}
  {{Multimessenger Constraints on Radiatively Decaying Axions from GW170817}},\
  }\href {https://doi.org/10.1103/PhysRevLett.132.101004} {\bibfield  {journal}
  {\bibinfo  {journal} {Physical Review Letters}\ }\textbf {\bibinfo {volume}
  {132}},\ \bibinfo {pages} {101004} (\bibinfo {year} {2024})}\BibitemShut
  {NoStop}%
\bibitem [{\citenamefont {DeRocco}\ \emph {et~al.}(2020)\citenamefont
  {DeRocco}, \citenamefont {Graham},\ and\ \citenamefont
  {Rajendran}}]{ChameleonModels}%
  \BibitemOpen
  \bibfield  {author} {\bibinfo {author} {\bibfnamefont {W.}~\bibnamefont
  {DeRocco}}, \bibinfo {author} {\bibfnamefont {P.~W.}\ \bibnamefont
  {Graham}},\ and\ \bibinfo {author} {\bibfnamefont {S.}~\bibnamefont
  {Rajendran}},\ }\bibfield  {title} {\bibinfo {title} {{Exploring the
  robustness of stellar cooling constraints on light particles}},\ }\href
  {https://doi.org/10.1103/PhysRevD.102.075015} {\bibfield  {journal} {\bibinfo
   {journal} {Physical Review D}\ }\textbf {\bibinfo {volume} {102}},\ \bibinfo
  {pages} {075015} (\bibinfo {year} {2020})}\BibitemShut {NoStop}%
\bibitem [{\citenamefont {Popov}(1999)}]{DPCoulombModification}%
  \BibitemOpen
  \bibfield  {author} {\bibinfo {author} {\bibfnamefont {V.}~\bibnamefont
  {Popov}},\ }\bibfield  {title} {\bibinfo {title} {{On the experimental search
  for photon mixing}},\ }\href@noop {} {\bibfield  {journal} {\bibinfo
  {journal} {Turkish Journal of Physics}\ }\textbf {\bibinfo {volume} {23}},\
  \bibinfo {pages} {943} (\bibinfo {year} {1999})}\BibitemShut {NoStop}%
\bibitem [{\citenamefont {Landau}\ and\ \citenamefont
  {Rumer}(1938)}]{1DCascadeEqs}%
  \BibitemOpen
  \bibfield  {author} {\bibinfo {author} {\bibfnamefont {L.~D.}\ \bibnamefont
  {Landau}}\ and\ \bibinfo {author} {\bibfnamefont {G.}~\bibnamefont {Rumer}},\
  }\bibfield  {title} {\bibinfo {title} {{The cascade theory of electronic
  showers}},\ }\href {https://doi.org/10.1098/rspa.1938.0088} {\bibfield
  {journal} {\bibinfo  {journal} {Proceedings of the Royal Society of London.
  A. Mathematical and Physical Sciences}\ }\textbf {\bibinfo {volume} {166}},\
  \bibinfo {pages} {213} (\bibinfo {year} {1938})}\BibitemShut {NoStop}%
\bibitem [{\citenamefont {Tsai}(1974)}]{FormFactors2}%
  \BibitemOpen
  \bibfield  {author} {\bibinfo {author} {\bibfnamefont {Y.~S.}\ \bibnamefont
  {Tsai}},\ }\bibfield  {title} {\bibinfo {title} {{Pair production and
  bremsstrahlung of charged leptons}},\ }\href
  {https://doi.org/10.1103/RevModPhys.46.815} {\bibfield  {journal} {\bibinfo
  {journal} {Reviews of Modern Physics}\ }\textbf {\bibinfo {volume} {46}},\
  \bibinfo {pages} {815} (\bibinfo {year} {1974})}\BibitemShut {NoStop}%
\end{thebibliography}
